\documentclass{aa}    

\usepackage{graphicx}
\usepackage{txfonts}

\usepackage{natbib}
\usepackage{dcolumn,url,longtable}
\usepackage{amssymb}
\usepackage{color}

\newcolumntype{d}{D{.}{.}{-1}}

\newcommand{\radd}{\text{rad}\,\text{d}^{-2}}
\newcommand{\degd}{\text{deg}\,\text{d}^{-2}}
\newcommand{\dgr}{^\circ}
\newcommand{\dd}{\mathrm{d}}
\newcommand{\domdt}{\dd\omega/\dd t}
\newcommand{\dadt}{\dd a/\dd t}

\usepackage[colorlinks=true,urlcolor=blue,citecolor=blue,linkcolor=blue,bookmarks=true]{hyperref}

\usepackage{amsmath}
\usepackage{gensymb}

\def\sig{J\,m$^{-2}$\,s$^{-0.5}$\,K$^{-1}$}
\begin{document} 

  \title{Secular change in the spin states of asteroids due to radiation and gravitation torques. New detections and updates of the YORP effect}

  \titlerunning{Secular change in the spin states of asteroids due to radiation and gravitation torques}

  \author{  J.~\v{D}urech		      \inst{1}     \and
            D.~Vokrouhlick\'y       \inst{1}    \and
            P.~Pravec               \inst{2}    \and   
            Yu.~Krugly              \inst{3,12}    \and
            D.~Polishook            \inst{4}    \and
            J.~Hanu\v{s}            \inst{1}    \and
            F.~Marchis              \inst{16}   \and 
            A.~Ro{\.z}ek            \inst{9}    \and
            C.~Snodgrass            \inst{9}    \and
            L.~Alegre               \inst{9}    \and
            Z.~Donchev              \inst{6}    \and
            Sh.~A.~Ehgamberdiev     \inst{15}   \and
            P.~Fatka                \inst{2}    \and
            N.~M.~Gaftonyuk         \inst{13}   \and
            A.~Gal\'ad              \inst{11}   \and
            K.~Hornoch              \inst{2}    \and
            R.~Ya.~Inasaridze       \inst{5,8}  \and
            E.~Khalouei             \inst{10}   \and
            H.~Ku\v{c}\'akov\'a     \inst{2}    \and
            P.~Ku\v{s}nir\'ak       \inst{2}    \and
            J.~Oey                  \inst{7}    \and
            D.~P.~Pray              \inst{17}   \and
            A.~Sergeev              \inst{3,14} \and
            I.~Slyusarev            \inst{3}
	 }

  \institute{Charles University, Faculty of Mathematics and Physics, Institute of Astronomy, V Hole\v{s}ovi\v{c}k\'ach 2, 180\,00 Prague, Czech Republic\\
             \email{durech@sirrah.troja.mff.cuni.cz} \and
             Astronomical Institute, Academy of Sciences of the Czech Republic, Fri\v{c}ova~1, 251\,65 Ond\v{r}ejov, Czech Republic \and
             Institute of Astronomy of V. N. Karazin Kharkiv National University, Sumska Str. 35, Kharkiv 61022, Ukraine            \and
             Faculty of Physics, Weizmann Institute of Science, 234 Herzl St., Rehovot 7610001, Israel                              \and
             E. Kharadze Georgian National Astrophysical Observatory, Abastumani, Georgia                                           \and
             Institute of Astronomy and NAO, Bulgarian Academy of Sciences, 72 Tsarigradsko Chaussee Blvd., BG-1784 Sofia, Bulgaria \and
             Blue Mountains Observatory, 94 Rawson Pde. Leura, NSW 2780, Australia                                                  \and
             Samtskhe-Javakheti State University, Rustaveli Street 113, Akhaltsikhe 0080, Georgia                                   \and
             Institute for Astronomy, University of Edinburgh, Royal Observatory, Edinburgh, EH9 3HJ, UK                            \and
             Astronomy Research Center, Research Institute of Basic Sciences, Seoul National University, 1 Gwanak-ro, Gwanak-gu, Seoul 08826, Korea \and
             Modra Observatory, Department of Astronomy, Physics of the Earth, and Meteorology, FMPI UK, Bratislava SK-84248, Slovakia \and
             Astronomical Observatory Institute, Faculty of Physics, Adam Mickiewicz University, S{\l}oneczna 36, 60-286 Pozna\'n, Poland \and
             Crimean Astrophysical Observatory, Simeiz, Crimea, Ukraine                                                             \and
             Université Côte d’Azur, Observatoire de la Côte d’Azur, CNRS, Laboratoire Lagrange, France                             \and
             Ulugh Beg Astronomical Institute, Astronomicheskaya Str. 33, Tashkent 100052, Uzbekistan                               \and
             Carl Sagan Center at the SETI Institute, 189 Bernardo Av., Mountain View, CA 94043, USA                                \and
             Sugarloaf Mountain Observatory, South Deerfield, MA, USA                                                               
	     }

  \date{Received ?; accepted ?}

  \abstract
% context heading (optional)
{The rotation state of small asteroids is affected in the long term by perturbing
 torques of gravitational and radiative origin (the YORP effect). The former can
 be detected by a change in the spin-axis orientation in the inertial space; the latter 
 manifests itself by a quadratic increase in the rotation phase.}
% aims heading (mandatory)
{Direct observational evidence of the YORP effect is the primary goal of our
 work. This includes both the YORP detection for new objects and an improvement in the accuracy of previously known detections.}
% methods heading (mandatory)
{We carried out photometric observations of five near-Earth asteroids: (1862)~Apollo,
 (2100)~Ra-Shalom, (85989) 1999~JD6, (138852) 2000~WN10, and (161989)~Cacus. Then we
 applied the light-curve inversion method to all available data to determine the spin
 state and a convex shape model for each of the five studied asteroids. The YORP effect was modeled as a linear change of the
 rotation frequency $\upsilon \equiv \domdt$. In the case of (2100)~Ra-Shalom, the analysis
 required that the spin-axis precession due to the solar gravitational torque also be included.}
% results heading (mandatory)
{We obtained two new detections of the YORP effect: (i) $\upsilon = (2.9 \pm 2.0)\times
 10^{-9}\,\radd$ for (2100)~Ra-Shalom, and (ii) $\upsilon = (5.5\pm 0.7)\times 10^{-8}\,
 \radd$ for (138852) 2000~WN10. The analysis of Ra-Shalom also reveals a precession of the
 spin axis with a precession constant $\alpha \sim 3000\,\arcsec$~yr$^{-1}$. This is the first
 such detection from Earth-bound photometric data. For the other two asteroids, we improved
 the accuracy of the previously reported YORP detection: (i) $\upsilon = (4.94 \pm 0.09)\times 10^{-8}\,\radd$ for
 (1862)~Apollo, and (ii) $\upsilon = (1.86\pm 0.09)\times 10^{-8}\, \radd$ for (161989)~Cacus.
 With this value, Apollo has the most precisely determined YORP effect so far. Despite
 the recent report of a detected YORP effect for (85989) 1999~JD6, we show that the
 model without YORP cannot be rejected statistically. Therefore, the detection of the YORP
 effect for this asteroid requires future observations. In several of our targets, the
 currently available observations do not provide enough constraints on the shape model
 (even at large scales) to compute the theoretical YORP effect with sufficient precision. Nevertheless,
 the interpretation of the detected signal as the YORP effect is fairly plausible. The spin-axis precession constant of Ra-Shalom determined from observations matches the theoretically expected value.}
% conclusions heading (optional)
{The total number of asteroids with a YORP detection has increased to 12. In all
 cases, the rotation frequency increases in time. The analysis of a rich photometric data set
 of irregularly shaped asteroids may require inclusion of spin-axis precession in future studies.}

  \keywords{Minor planets, asteroids: general, Methods: data analysis, Techniques: photometric}

  \maketitle

  \section{Introduction}

 Photometric observations of asteroids can be used to determine their rotation state
(in most cases, a unique rotation period and spin-axis direction) and shape. In contrast to
the shape, which is most often only a convex approximation due to the limited information content
of the disk-integrated light curves, the spin state can be determined very precisely. 
This precision increases with increasing number of available observations. The ability
to accurately constrain the direction of the spin axis principally stems from a number of
independent viewing geometries, defined by the observer (usually Earth), the asteroid, and 
the Sun, that are represented by the data. In the case of a few exceptional near-Earth asteroids,
a sufficient number of observations may be achieved in one or two apparitions \citep{Mon.ea:20, Kwi.ea:21}. More
often, only data accumulated over years or decades eventually help to determine the spin-axis direction to within a few degrees of accuracy. The precision of the rotation-period determination
mainly depends on the time span covered by observations. With light curves that are observed over several
decades and a typical rotation period of some hours, the rotation-period precision
can reach about $0.01$\,s.

The simplest lowest-energy rotation model, characterized by a fixed direction of the spin
axis in the inertial space and by a constant rotation frequency, is adequate to fit the observations
of most asteroids. This model is about as sophisticated as the Keplerian
description of the asteroid heliocentric motion on a fixed ellipse, however.
As data spanning sufficiently long intervals of time become available and help determine the spin state very
accurately, the basic model may need generalization by effects of even tiny torques, especially if they accumulate
over time into a strong perturbation. 
The relevant torques originate (i) in the
solar gravitational field and (ii) in the solar radiation. 

The first torque (i) is a well-known effect
that has been empirically known and was later theoretically analyzed by astronomers for centuries in the case of the rotation of Earth (the difference is only that two centers, the Sun and the Moon, act together to produce the resulting
lunisolar precession of the Earth axis). The explanation of the effect was first given by
Isaac Newton \citep[e.g.,][Sec. 23]{chandra} and was later mathematically mastered by Jean d'Alembert
\citep{alem1749}. While it is obvious for the Earth, it took centuries before the solar-induced
gravitational precession was discussed in the context of asteroid rotation. The pioneering works
in this respect came from the Uppsala school during the 1990s \citep[e.g.][]{sko1996,sko1997,
sko1998,sko2002}. Still, in spite of significant improvements in mathematical modeling,
the effect was elusive to direct detection. Only when very precise measurements from the Dawn spacecraft
visiting (4) Vesta became available was the asteroidal precession observationally
determined \citep[e.g.,][]{kono2014}. However, the accuracy was not impressive, to the point that some consider the Vesta case inconclusive \citep[see, e.g.,][]{archie2018}. The reason is that
the effect is rather small for a roundish shape like that of Vesta, only $\simeq 0.28^\circ$~cy$^{-1}$,
and Dawn stayed near Vesta for only about a year. The detection of asteroidal spin precession using
much less precise Earth-bound observations can only be successful when the effect at a properly
chosen target is much stronger than was found for Vesta. Ideally, this would require a slowly rotating,
highly irregular body on a near-Earth orbit for which the effect may increase to a fraction of a degree
in a year. Still, a good-quality data set spanning a long period of time is needed. The first hint
that it might be possible to detect the asteroidal spin precession has recently been given by \citet{Dur.ea:22}
for (1620) Geographos and (1685) Toro. However, the first convincing detection awaited the
current paper. For the first time, we provide the detection of the spin-axis precession for
(2100) Ra-Shalom. This object is nearly an optimum target for this goal: it resides on an Aten orbit,
it rotates slowly, it has an irregular shape and obliquity away from the unfavorable values of
$0^\circ$, $90^\circ$ and $180^\circ$, and the available numerous photometric data span a long
interval of $44$ years. The details are given in Sec.~\ref{the:rasha}.

The second torque (ii) is a novel process without relevance for planets, satellites, or even large
asteroids. It has to do with radiation pressure that is imparted on small, irregularly shaped asteroids, and it
consists of two components: a smaller torque due to directly reflected sunlight in the optical band,
and a more important component due to the recoil of the thermally emitted radiation by the asteroid itself.
The analysis of this phenomenon has undergone an impressive revival in the last two decades after the
pioneering work of \citet{rub2000}. Rubincam also coined the acronym YORP, which stands for
Yarkovsky-O'Keefe-Radzievski-Paddack effect. Many of the important applications of YORP in planetary
science can be found in the review by \citet{vetal2015}. While the gravitational torque only affects
the direction of the spin axis, the YORP effect triggers secular perturbation of both the spin-axis direction 
and the spin frequency. Both are tiny effects, so that it is not surprising that the direct detection
of YORP proved to be a tricky task. \citet{ito2004} pointed out that the only foreseeable way
to detect the YORP effect is the secular change in the rotation frequency, which fortuitously
produces a quadratic perturbation in the rotation phase. This is directly observable with photometric
data.  \citet{yorp12007,yorp22007,apollo2007} indeed followed this method and obtained the
first YORP detections for the small near-Earth asteroids (54509) YORP and (1862) Apollo. Since then, the observations of
a handful of other asteroids allowed us to detect the YORP torque directly, but the sample still remains
very limited. Most importantly, the available results are still insufficient to solve the quantitative challenges posed by the theory of the YORP effect \citep[see][for a review and
detailed discussion]{vetal2015}. This also motivated our work: We aimed to increase the sample of asteroids with YORP detections, to determine their physical properties, and to compare the observed YORP with the theory.

Building partly on our previous research and also reporting on new targets, we present
evidence for the YORP effect in four near-Earth asteroids: (i) An improved accuracy of the
YORP effect detection for the two bodies (1862) Apollo and (161989) Cacus, and (ii)
a determination of the YORP effect for the first time for another two bodies, (2100) Ra-Shalom and
(138852) 2000~WN10. Several aspects related to these detections are notable. The YORP effect
detection in (1862) Apollo now has the highest signal-to-noise ratio, with a very good prospect
of further improvements. In the case of (138852) 2000~WN10, the YORP effect is detected for a 
prograde-rotating
body for the first time. Finally, (2100) Ra-Shalom represents a difficult analysis in which the
spin-axis precession is detected for the first time together with the YORP effect on a body
with the lowest rotation frequency so far. Additionally, we analyzed available data for
asteroid (85989) 1999~JD6, for which the YORP effect has previously been reported by \citet{tian2022}. 
However, the addition of new observations to the data set of this object shows that the YORP effect has not
yet been detected at a statistically significant level. Nevertheless, we find it interesting that while a no-YORP solution is statistically admissible at a level of about
one sigma for (85989) 1999~JD6, most solutions require the
rotational frequency to be decelerated. If this is confirmed in the future, this would be a breakthrough case in the context of the YORP effect theory because all YORP detections so far have shown an acceleration of the rotation frequency (see Sect.~\ref{sec:discussion}).

   \begin{figure}[t]
        \includegraphics[trim=0.5cm 4cm 1cm 4cm, clip, width=\columnwidth]{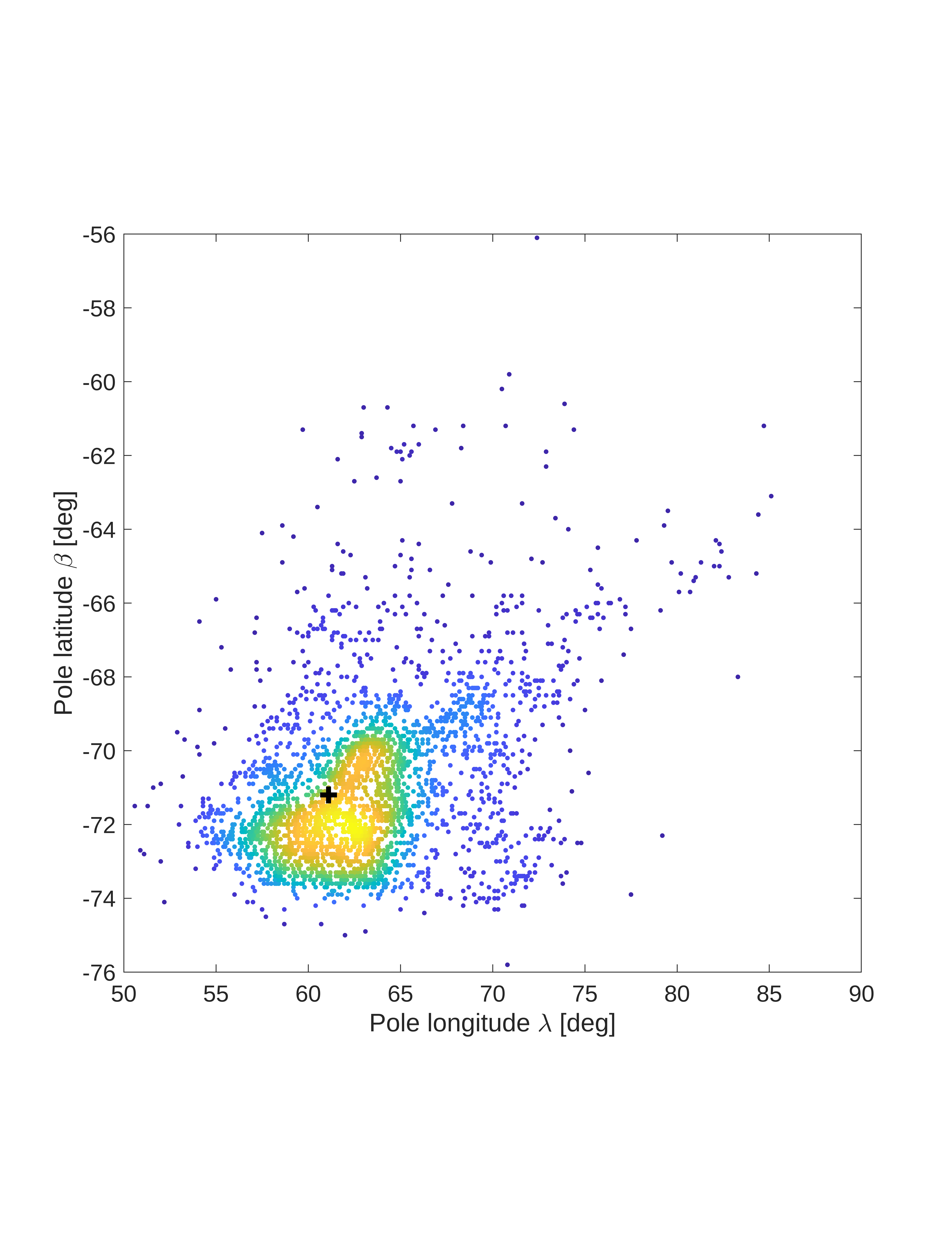}
        \includegraphics[trim=0.5cm 4cm 1cm 4cm, clip, width=\columnwidth]{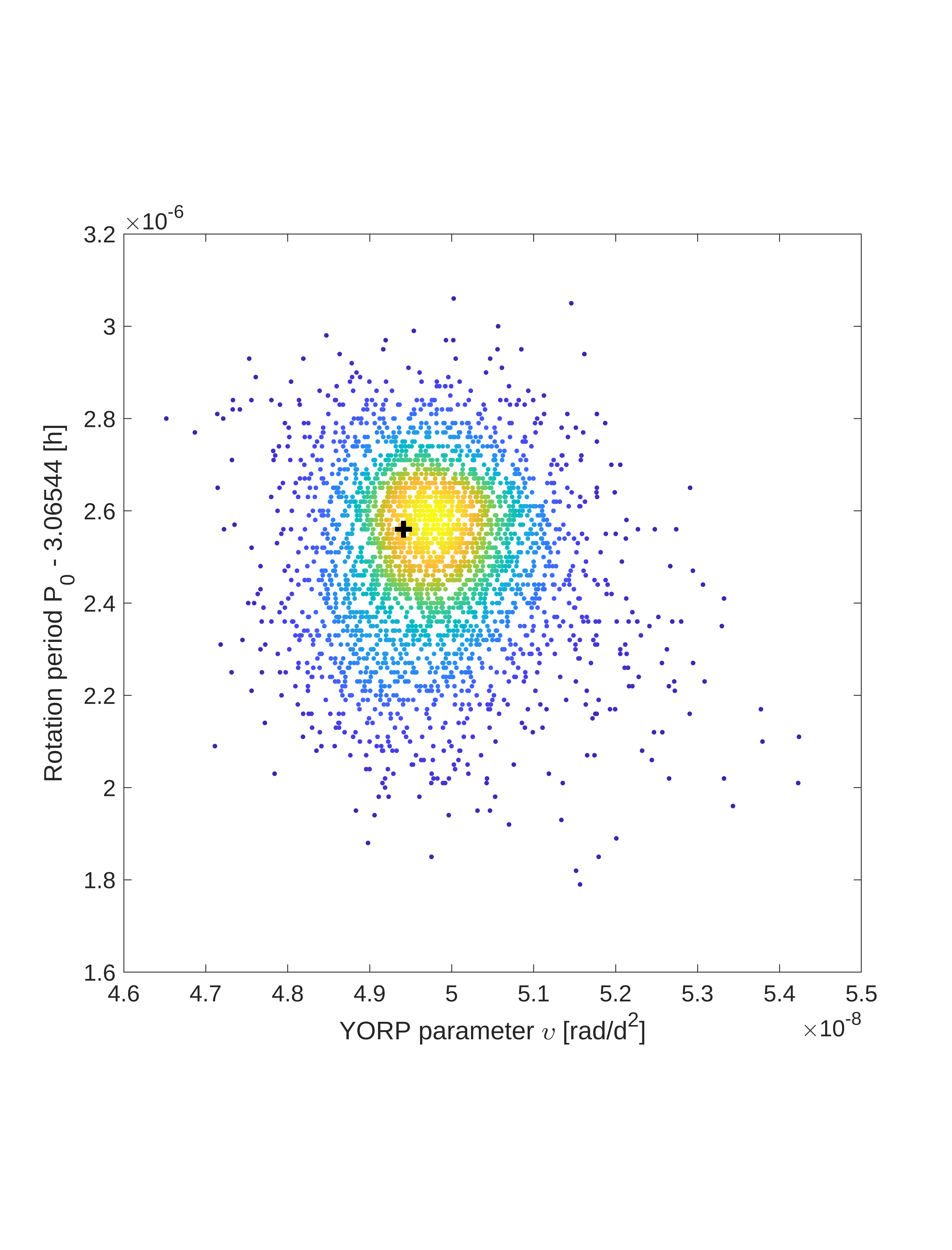}
        \caption{Bootstrap distribution of the pole direction in the ecliptic longitude $\lambda$ and latitude $\beta$ ({\it top\/}), the YORP parameter $\upsilon$, and the rotation period $P_0$ ({\it bottom\/}) for asteroid (1862)~Apollo (the latter corresponding to epoch JD$_0$ 2452276.0). The color-coding corresponds to the density of points. The black cross marks the best-fit solution based on the full original (not bootstrapped) data set.}
        \label{fig:BS_1862}
    \end{figure}

The paper is organized as follows. In Sec.~\ref{sec:models} we describe the available observation data sets, the reconstructed models, and the uncertainty in the rotation parameters obtained using our light-curve inversion method. We compare these values with
the theoretical prediction in Sec.~\ref{sec:theory}, and we discuss the broader context of our results in Sec.~\ref{sec:discussion}.
   
\section{New photometric observations and model reconstruction} \label{sec:models}

    This section describes new photometric observations and model reconstruction for each asteroid in our sample. The final model is clearly based not only on our new data, but also uses previously published observations whenever available. The observatories contributing to our campaign are listed in Table~\ref{tab:telescopes}. Observations carried out with the Danish Telescope at La Silla (DK154) and the Ond\v{r}ejov Observatory telescope (D65) were calibrated, the other photometry was relative without all-sky calibration. Observing times were converted from the reported UTC into TDB. 

    The new observations are summarized in the tables in the appendix. For simplicity, we only list individual nights, not individual light curves when more of them (due to observations in different filters or calibration issues) were observed in a single night with the same telescope.
    
    To reconstruct the physical model, we used the light-curve inversion method of \citet{Kaa.ea:01} and \citet{Kaa.Tor:01}, in which the angular frequency $\omega$ evolves linearly in time as $\omega = \omega_0 + \dot\omega t$, where the rate $\upsilon \equiv \dot\omega =  \text{d}\omega / \text{d} t$ and the initial value $\omega_0$ are free parameters of the optimization (a comment on the validity of this approach can be found in Sec.~\ref{sec:discussion}). In the following text, we call $\upsilon$ a YORP parameter, although the connection between the change in the rotation rate detected from light curves and the YORP effect is discussed below in Sect.~\ref{sec:theory}. Instead of $\omega_0$, we report the sidereal rotation period $P_0 = 2\pi / \omega_0$. Because the rotation period evolves with time, it is necessary to also report the epoch JD$_0$, for which $P_0$ is given. In our previous works \citep[e.g.,][]{Dur.ea:08, Dur.ea:18a, Dur.ea:22}, we set JD$_0$ to the epoch of the first observation. However, this causes a correlation between $\upsilon$ and $P_0$, a shorter initial rotation period (at the beginning of the observing data set) and a smaller YORP produce about the same evolution in the rotation phase angle as a longer initial period that evolves faster due to a larger YORP parameter $\upsilon$. 
    The rotation phase angle $\varphi$ evolves over time $t$ from some initial value $\varphi_0$ as
    \begin{equation}
        \varphi(t) = \varphi_0 + \frac{2\pi}{P_0} \left(t - \text{JD}_0\right) + \frac{1}{2} \upsilon \left(t - \text{JD}_0\right)^2\,.
    \end{equation}
    With JD$_0$ at the beginning of the observing data set, the second term is always positive, so that by decreasing $P_0$, the phase $\varphi$ increases for all observations, and this can partly be balanced by decreasing $\upsilon$, from which a positive correlation between $P_0$ and $\upsilon$ arises. However, when JD$_0$ is somewhere in the middle of the observing interval, the second term is negative for half of the observations, and a small change in $P_0$ leads to an increase in $\varphi$ for half of the observations and a decrease for the remaining half, which cannot be compensated for by changing $\upsilon$.
    To avoid a correlation between $\upsilon$ and $P_0$, we now set JD$_0$ somewhere close to the center of the interval covered by observations.

     \begin{figure}[t]
        \includegraphics[width=\columnwidth]{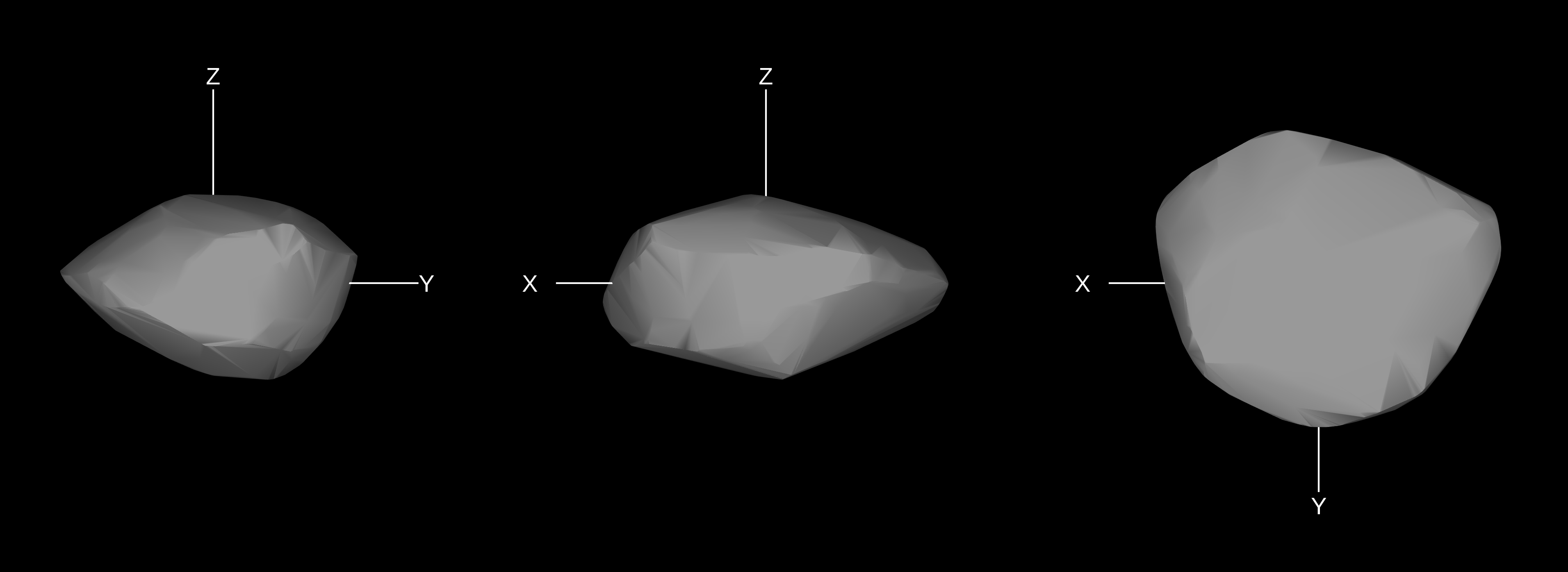}
        \caption{Convex shape model of asteroid (1862) Apollo shown from the equatorial level ({\it left} and {\it center\/}, $90^\circ$ apart) and pole-on ({\it right\/}).}
        \label{fig:model_1862}
    \end{figure}
        \begin{figure*}
        \includegraphics[width=\textwidth]{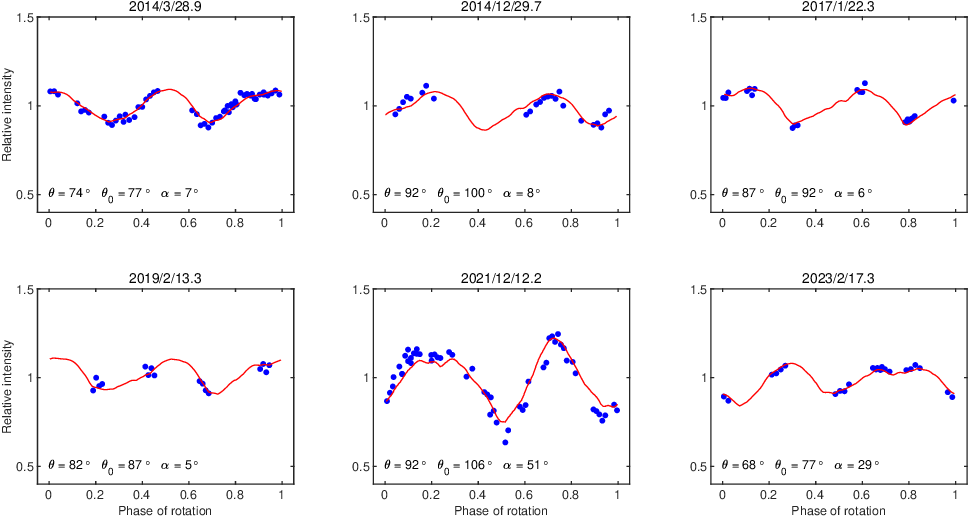}
        \caption{Example light curves of (1862) Apollo. The epoch of observation is shown by the label at the top of each panel.
        The blue points are the observed data, and the red curve is our best model including YORP.
        The geometry of the observation is described by the aspect angle $\theta$, the solar aspect angle $\theta_0$, and the solar phase angle $\alpha$.}
        \label{fig:lc_fit_1862}
    \end{figure*}
    
    To estimate the uncertainties in the spin and YORP parameters, we created 3000 bootstrap data sets and repeated the inversion. The bootstrap sample was a random selection of the same number of light curves; in each selected light curve, we bootstrapped its points. The parameter uncertainties reported below are the standard deviations of the parameter distributions.

    We also included the precession effect of the spin axis due to solar torque for asteroids (2100)~Ra-Shalom and (85989) 1999~JD6, for which this effect is strong enough to be taken into account
    \citep[see][for details]{Dur.ea:22}.
  
    \subsection{(1862) Apollo} \label{obs:apollo}

    Apollo was one of the two asteroids, together with (54509) YORP, for which the YORP effect was detected for the first time \citep{apollo2007, yorp22007, yorp12007}. \cite{Dur.ea:08} published an update of the original detection and derived a YORP acceleration of $(5.5 \pm 1.2) \times 10^{-8}\radd$ from 27 years of observations from 1980--2007.
    
    We observed Apollo during five more apparitions in 2014, 2017, 2019, 2021, and 2023. We also used other published observations from 2014 \citep{War:14k} and 2021 \citep{War.Ste:22c} that are available in the ALCDEF database\footnote{\url{https://alcdef.org}} \citep{War.ea:09}. The full data set now consists of 87 light curves covering the time 1980--2023. The new light curves are listed in Table~\ref{tab:aspect_1862}. The observations taken with DK154 and D65 were absolutely calibrated in the Johnson-Cousins VR photometric system. We determined a color index $(V - R) = 0.428 \pm 0.010$\,mag, which is compatible with its Q-type spectral classification.

    The new model we derived has the following spin parameters: an ecliptic longitude $\lambda = (61.1 \pm 4.6)^\circ$, an ecliptic latitude $\beta = (-71.2 \pm 2.2)^\circ$, a sidereal rotation period $P_0 = (3.065\,422\,6 \pm 0.000\,000\,2)$\,h (given at JD$_0$ 2452276.0 epoch), and a secular change in the rotation rate $\upsilon = (4.94 \pm 0.09)\times 10^{-8}\,\radd$. The uncertainties of these spin parameters were estimated from the distribution of the bootstrap solutions shown in Fig.~\ref{fig:BS_1862}. The shape model is shown in Fig.~\ref{fig:model_1862}. The difference in the rotation phase accumulated over 43 years between a constant-period model with $P_0$ and our best-fit model is $\sim 340^\circ$, almost one full rotation. In Fig.~\ref{fig:lc_fit_1862} we show the match between the data and model for the six new light curves as an example.

    \begin{figure}
        \includegraphics[width=\columnwidth]{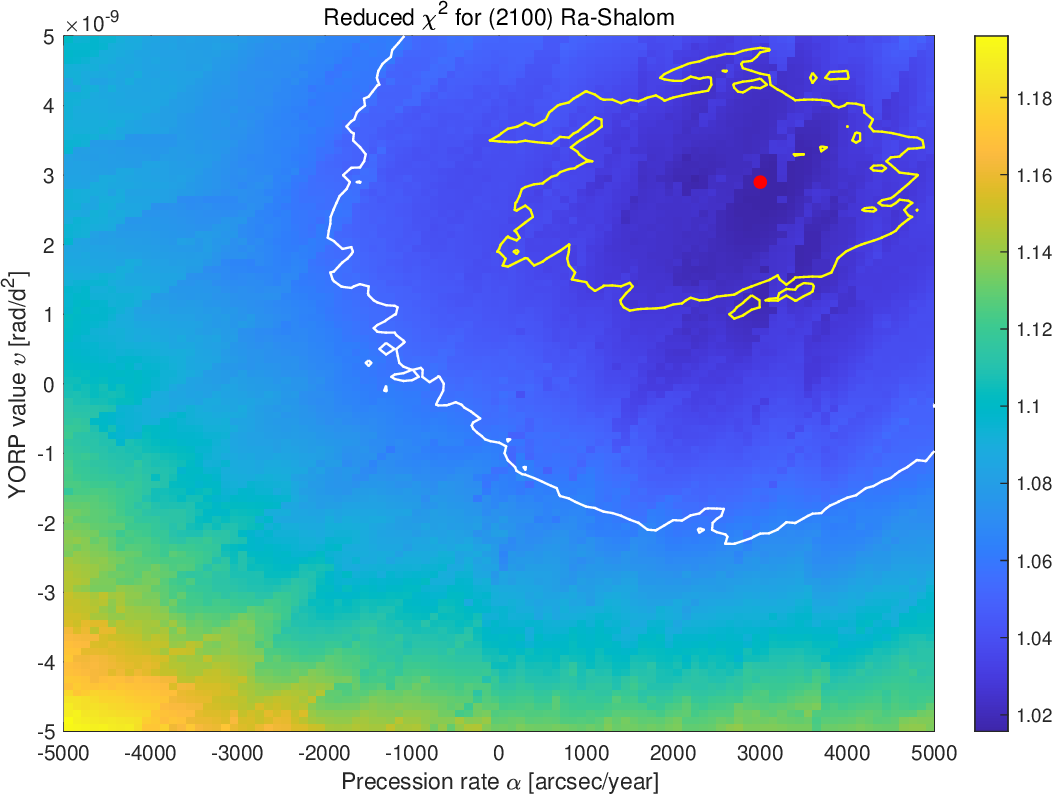}
        \caption{Color map of the reduced $\chi^2$ values of the light-curve fit for asteroid Ra-Shalom for different values of the precession rate $\alpha$ and the YORP strength $\upsilon$. The red point marks the lowest $\chi^2$ value, the inner yellow contour is the estimated $1\sigma$ uncertainty limit, and the outer white contour is a $3\sigma$ boundary.}
        \label{fig:precession_YORP_2100}
    \end{figure}

    \begin{figure}[t]
        \includegraphics[width=\columnwidth]{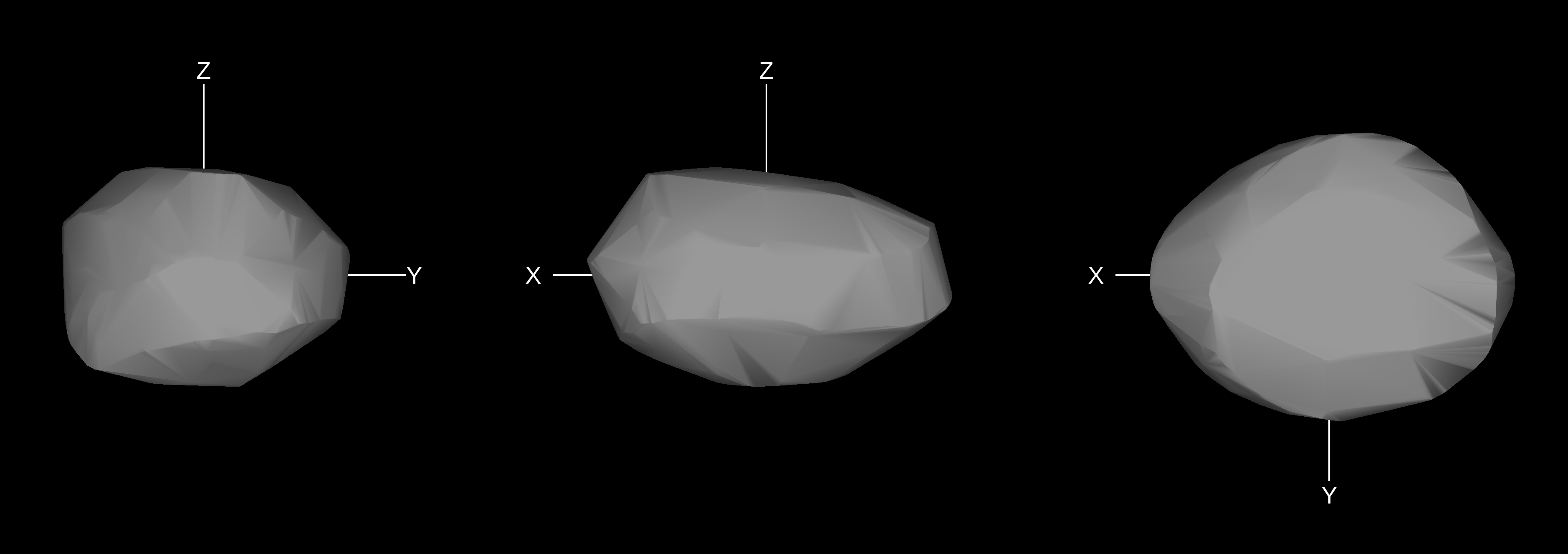}
        \caption{Convex shape model of asteroid (2100) Ra-Shalom shown from the equatorial level ({\it left} and {\it center\/}, $90^\circ$ apart) and pole-on ({\it right\/}).}
        \label{fig:model_2100}
    \end{figure}
  
    \subsection{(2100) Ra-Shalom} \label{obs:rasha}

    Previous photometry data of Ra-Shalom were analyzed by \cite{Dur.ea:12} and later by \cite{Dur.ea:18a}.
    In both cases, the data set was found to be too short to reveal the YORP effect, and the constant period model was found adequate to match the available observations. The best \cite{Dur.ea:18a} could find was that the $3\sigma$ uncertainty interval of the YORP parameter, namely $-1.0 \times 10^{-8} < \upsilon < 1.5 \times 10^{-8}\,\radd$, was slightly asymmetric with respect to the origin. This finding indicated a possible positive $\upsilon$, which we aimed to detect here by adding new observations. The new photometric observations from 2019 and 2022 are listed in Table~\ref{tab:aspect_2100}. Part of these observations was conducted with the $28''$ telescope at the Wise Observatory \citep{Bro.ea:15}. We also included data published by \cite{War.Ste:20a, War.Ste:23a} in our analysis. 

    We also used light-curve observations from a citizen science project spearheaded by the SETI Institute, which connects telescope owners of the eVscope and eQuinox models developed by the France-based company, Unistellar. These digital, automated, and compact telescopes, boasting an 11.3~cm mirror, were tailored for people with very little background in astronomy. Consequently, the telescope is seamlessly operated using a smartphone through a dedicated mobile app.

    An observing campaign focusing on Ra-Shalom commenced in August 2022. A myriad of citizen astronomers collated and submitted their data, covering a nearly two-month span. We selected a subset of data with adequate photometric quality, which is detailed in Table~\ref{tab:aspect_eVscope}. Nonetheless, these light curves were not incorporated into the final modeling because their photometric quality was inferior to that of our primary data set. This discrepancy can be attributed to the small aperture of the telescope and to the fact that the data typically encapsulated limited segments of the complete rotation phase (approximately one to two hours relative to the almost 20-hour rotation period of Ra-Shalom). We still used this as a chance to corroborate the data reliability from the citizen project by juxtaposing it against our model predictions. This endeavor underscores the potential of the Unistellar burgeoning telescope network (comprising nearly 2000 potential contributors of various scientific modalities backed by Unistellar and its partner, the SETI Institute) for the photometry of pertinent targets, especially 
    near-Earth asteroids (NEAs). For Ra-Shalom, a comprehensive data set from experienced professional and amateur astronomers was at our disposal, rendering the Unistellar observations superfluous for the model reconstruction. This burgeoning citizen astronomer network harbors significant potential for the future, however, particularly in capturing photometric data for any sufficiently luminous NEA \citep[refer to the instance of (7335) 1989 JA,][]{Lambert2023}.

    \begin{figure*}
        \centering\includegraphics[width=\textwidth]{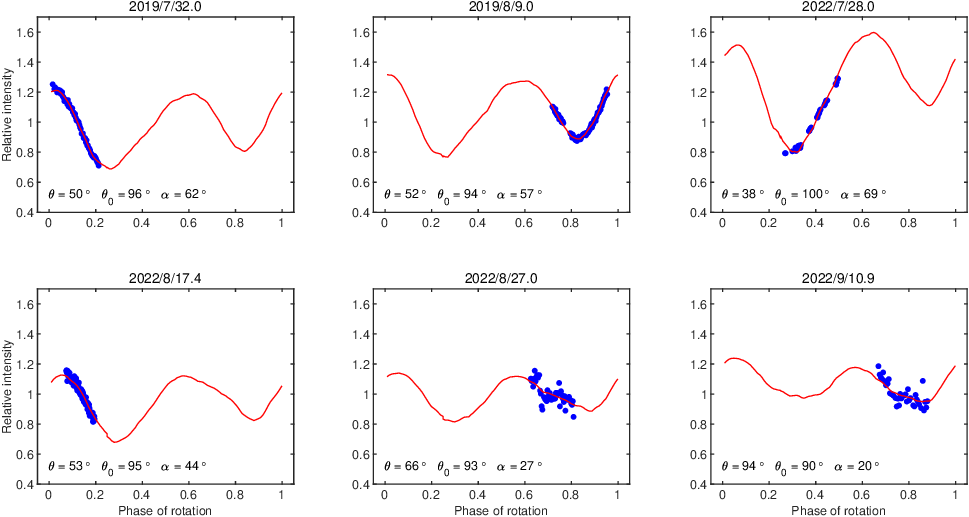}
        \caption{Example light curves of (2100) Ra-Shalom. The blue points show observed data, and the red curve shows our best model with YORP and precession. The last two light curves are examples of Unistellar eVscope data that were not used for the modeling.}
        \label{fig:lc_fit_2100}
    \end{figure*}
    
    The first photometric observations of Ra-Shalom were taken in 1978. The interval covered by available observations therefore spans 44 years, with 190 light curves in total. Being a near-Earth asteroid with the semimajor axis $a = 0.832$\,au, an eccentricity $e = 0.437$, and a rotation period of about 19.82\,h, the solar gravitational torque causes a potentially non-negligible precession of the Ra-Shalom rotation axis. In order to justify this extension of the model, along the lines developed in \cite{Dur.ea:22}, we first theoretically estimated the corresponding perturbation of the rotation pole direction. The model without YORP has a pole direction $(286.4^\circ, -56.4^\circ)$, a rotation period $P = 19.820\,040$\,h, and the dynamical ellipticity of the shape model is $\Delta = 0.232$ (assuming a uniform density distribution). The formal uncertainty in the pole direction is only a few degrees. The precession constant $\alpha$ \citep[see Eq.~(1) in][ for its definition]{Dur.ea:22} may be theoretically estimated with these data. We obtained $\alpha = 2430\,''\,\text{yr}^{-1}$, which corresponds to a shift of $27^\circ$ on the precession cone, or $16^\circ$ in $\lambda$ and $-4^\circ$ in $\beta$ over 44 years. As a result, the pole precession is large enough and must be considered when interpreting the available light curves.

    Following the approach of \citet{Dur.ea:22}, we thus added the precession constant $\alpha$, along with the YORP strength $\upsilon$, as a second free parameter in the model.
    We scanned the values of $\upsilon$ and $\alpha$ on a grid to determine whether the fit level depends on these two parameters. Although the parameter $\alpha$ has to be positive, we formally tested values between $-5000$ and $5000\,''\,\text{yr}^{-1}$. The YORP parameter $\upsilon$ was let free to range the interval $-5$ and $5 \times 10^{-9}\,\radd$. The result is shown in Fig.~\ref{fig:precession_YORP_2100}. The best solution with the lowest root mean square (RMS) residuals is for $\alpha = 3000\,''\,\text{yr}^{-1}$ and $\upsilon = 2.9 \times 10^{-9}\,\radd$. We determined the uncertainty contours in the same way as in \cite{Vok.ea:11} or \cite{Pol:14}, namely by defining an appropriate level of $\chi^2$ with respect to the minimum $\chi^2$. The $3\sigma$ interval defined in this way still covers the $\upsilon = 0$ value, but the $1\sigma$ interval is about $\pm 2 \times 10^{-9}\,\radd$, which is lower than the value of $\upsilon$ itself. Moreover, these $\chi^2$-based error intervals are larger than the uncertainties estimated by bootstrapping \citep[see Appendix A in][]{Dur.ea:22}. The bootstrap $1\sigma$ error of the YORP parameter for $\alpha = 3000''\,\text{yr}^{-1}$ is $\delta\upsilon = 8 \times 10^{-10}\,\radd$. Thus, our conservative conclusion is that the YORP is detected at the $2\sigma$ level and the precession due to solar gravitation torque at the $1\sigma$ level.
    
    The best model has a pole $\lambda_0 = (257 \pm 4)^\circ$, $\beta_0 = (-52 \pm 2)^\circ$, $P_0 = (19.820\,072 \pm 0.000\,008$)\,h (both for JD$_0$ 2443763.0), and $\Delta = 0.259$. With this updated $\Delta$ value, the theoretical precession constant $\alpha$ is now $2720\,''\,\text{yr}^{-1}$. This agrees excellently with the formally best value obtained from observations, given the uncertainty and simplification of the convex shape model. The final shape model is shown in Fig.~\ref{fig:model_2100}, and its synthetic light curves are shown in Fig.~\ref{fig:lc_fit_2100}.
   
        \begin{figure}[t]
        \includegraphics[width=\columnwidth]{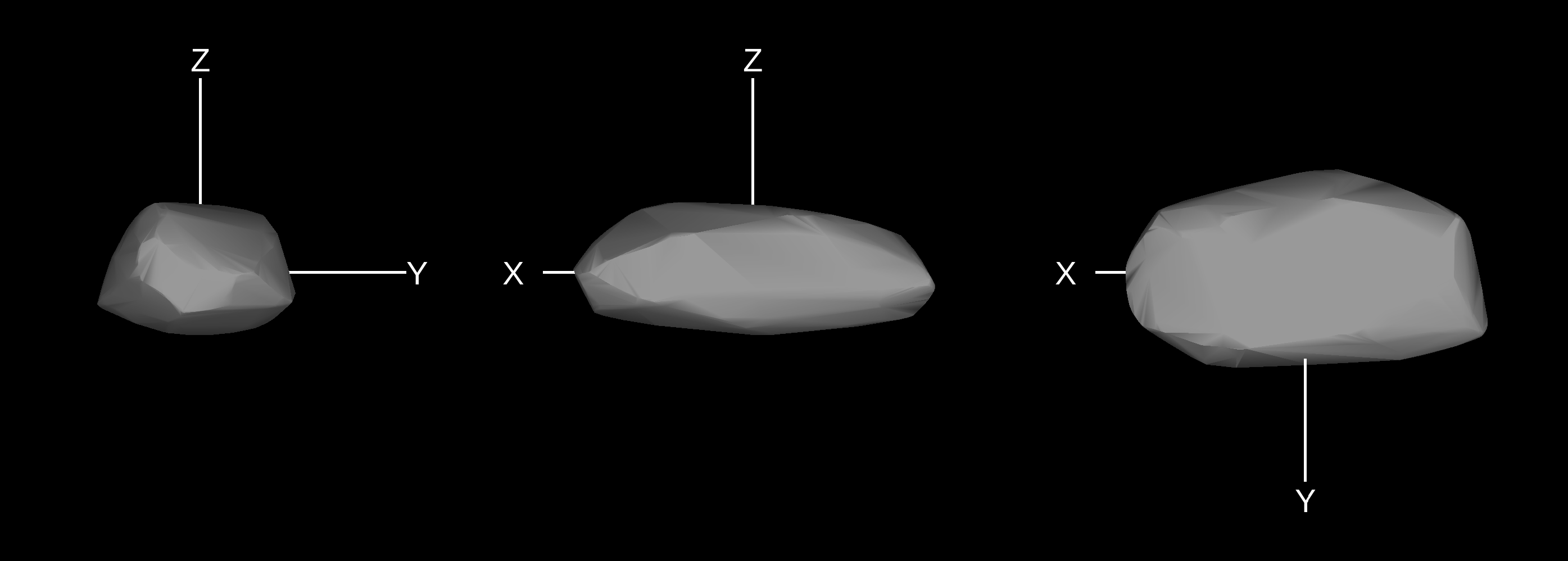}
        \caption{Convex shape model of asteroid (85989) 1999~JD6 shown from the equatorial level ({\it left} and {\it center\/}, $90^\circ$ apart) and pole-on ({\it right\/}).}
        \label{fig:model_85989}
    \end{figure}

    \begin{figure}[t]
        \includegraphics[trim=0.5cm 4cm 1cm 4cm, clip, width=\columnwidth]{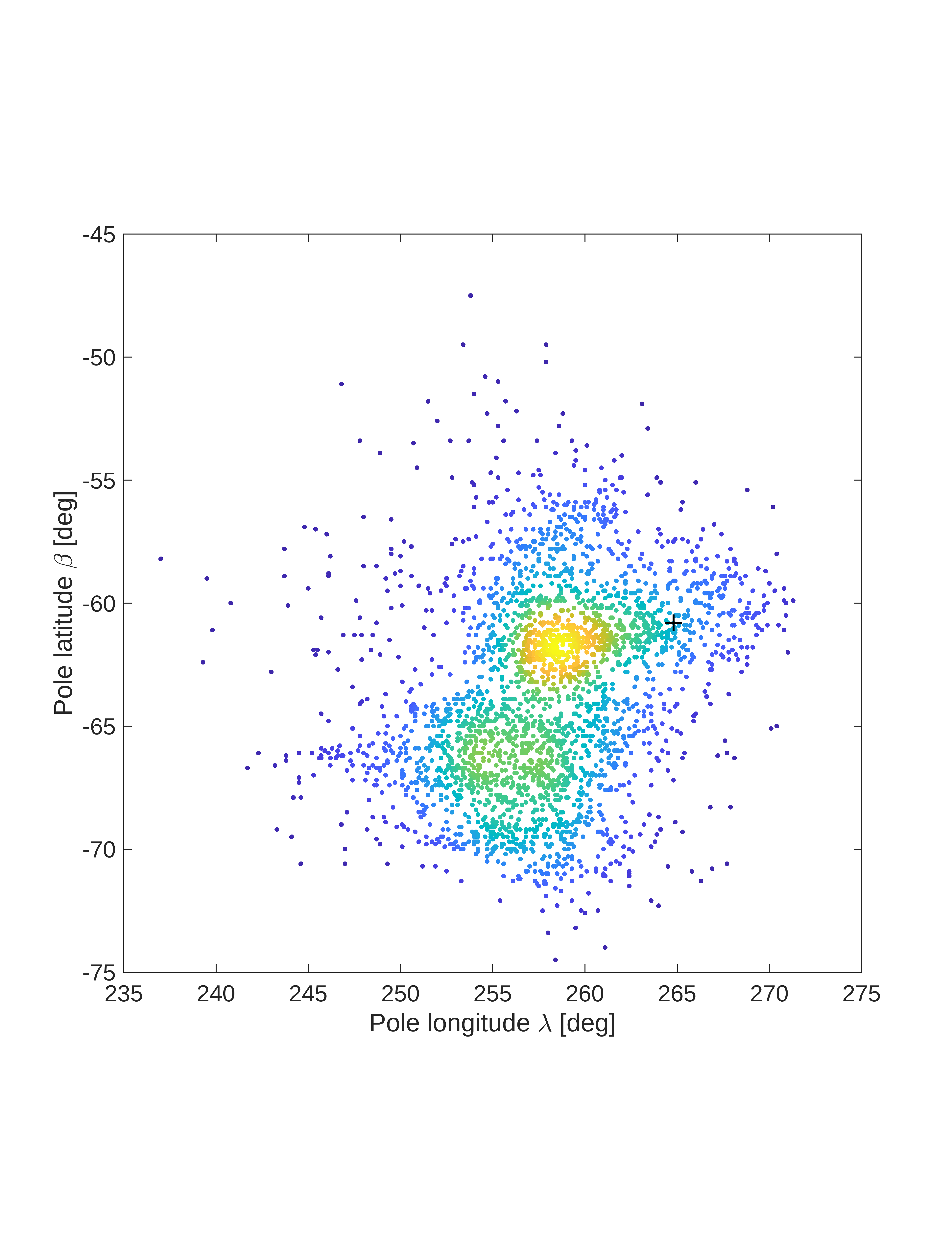}
        \includegraphics[trim=0.5cm 4cm 1cm 4cm, clip, width=\columnwidth]{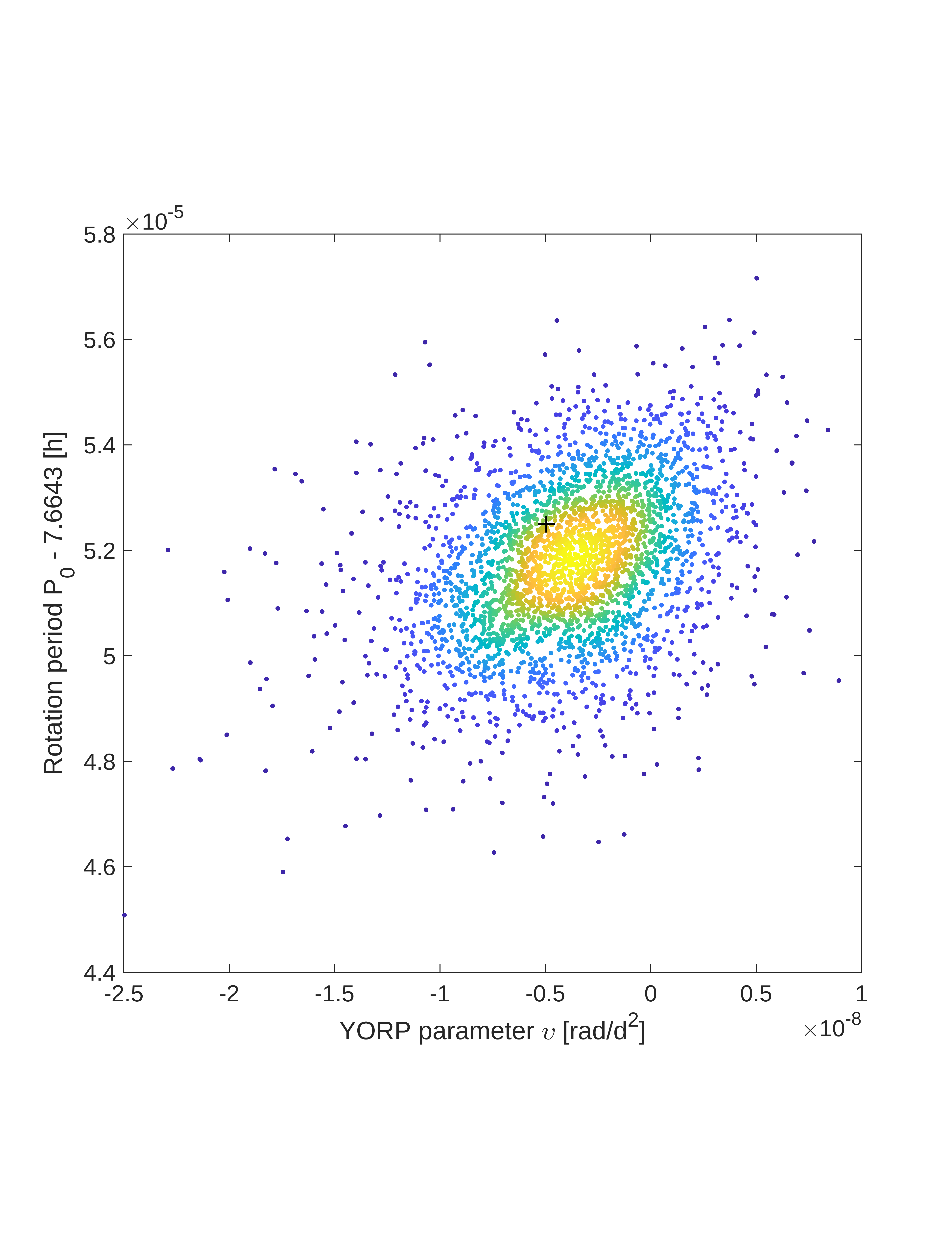}
        \caption{Bootstrap distribution of the pole direction in the ecliptic longitude $\lambda$ and latitude $\beta$ ({\it top\/}) and the YORP parameter $\upsilon$ and the rotation period $P_0$ ({\it bottom\/}) for asteroid (85989) 1999~JD6. The color-coding corresponds to the density of points. The cross marks the solution based on the original (not bootstrapped) data set.}
        \label{fig:BS_85989}
    \end{figure}

    \begin{figure}
        \includegraphics[width=\columnwidth]{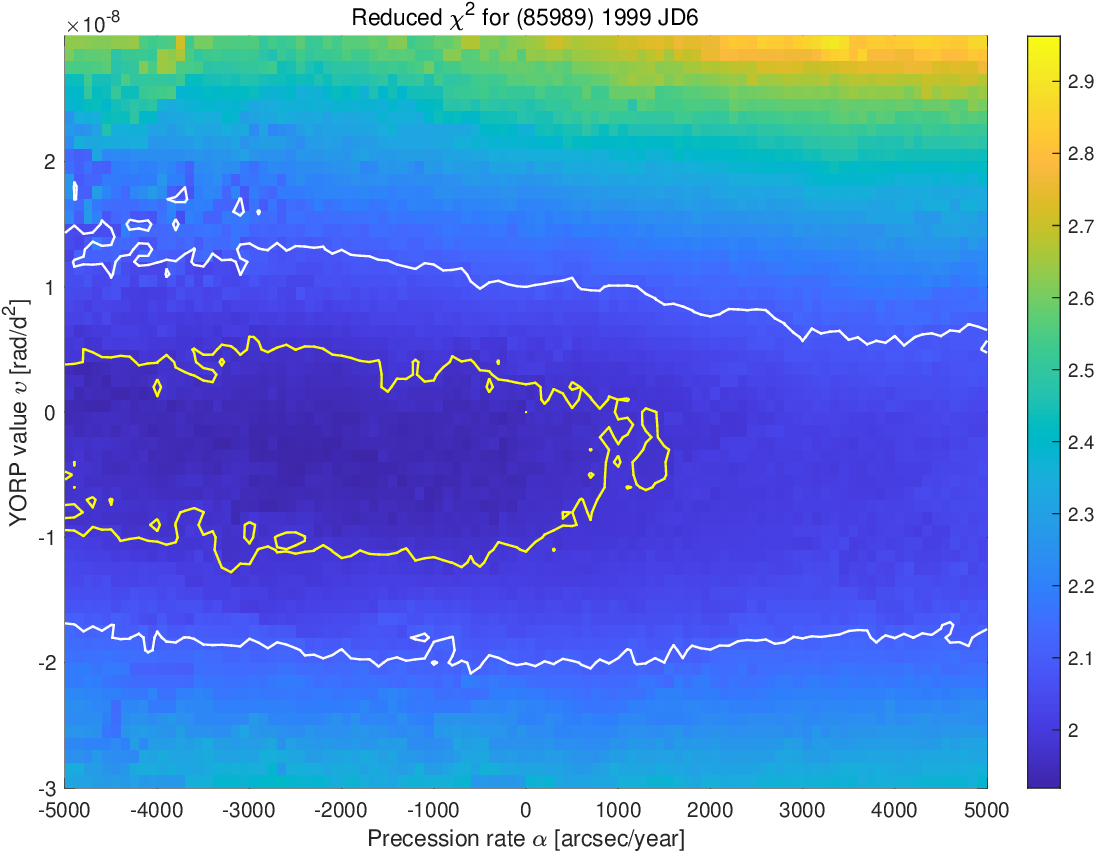}
        \caption{Color map of the reduced $\chi^2$ values of the light-curve fit for asteroid (85989) 1999~JD6 for different values of the precession rate $\alpha$ and the YORP strength $\upsilon$. The inner yellow contour is the estimated $1\sigma$ uncertainty limit, and the outer white contour is a $3\sigma$ boundary.}
        \label{fig:precession_YORP_85989}
    \end{figure}
    
    \subsection{(85989) 1999 JD6} \label{obs:jd6}

    \cite{Marsh.ea:15} and later \cite{marphd} combined Goldstone and Arecibo radar observations obtained during the close approach in 2015 with optical light curves and reconstructed a physical model of this asteroid. The shape model of 1999~JD6 is a contact binary, with two connected lobes and dimensions $3.0 \times 1.2 \times 1.0\,$km. The fortuitous geometry of the radar observations allowed \cite{marphd} to determine the pole direction very accurately, namely $(\lambda, \beta) = (220.3^\circ, -73.43^\circ)$, with an uncertainty of $0.25^\circ$. The sidereal rotation period was $7.664\,346\,4 \pm 0.000\,005\,6$~h, and these authors provided an upper limit on the YORP value $1.6 \times 10^{-6}\,\degd$, which corresponds to $|\upsilon| < 2.8 \times 10^{-8}\,\radd$.

    Recently, \cite{tian2022} analyzed the optical light curves spanning a 20-year time interval (between 2000 and 2020). From this data set, they reported the YORP detection for this asteroid. In particular, their model has a rotation period $7.667\,749 \pm 0.000\,009$\,h (for JD 2451728.0) and a rotation pole direction $\lambda = (232 \pm 2)^\circ$, $\beta = (- 59 \pm 1)^\circ$. In addition, their analysis required a rotation acceleration $\upsilon = (2.4 \pm 0.3) \times 10^{-8}\,\radd$ ($1\sigma$ error), which \cite{tian2022} interpreted as the YORP effect. 

    To reconstruct our independent shape and spin model, we used a similar light-curve data set as \cite{tian2022}, but extended it by additional observations taken in May and June 2023. Altogether, our data extend over ten apparitions from 1999 to 2023 (see Table~\ref{tab:aspect_85989}). To our surprise, we obtained a different result than the previous two studies. Our best model is shown in Fig~\ref{fig:model_85989}, has a pole direction $(260^\circ, -60^\circ)$ and a period $P = 7.664\,354$\,h. When allowing for a nonzero $\upsilon$, the best value is $\upsilon = -5.0 \times 10^{-9}\,\radd$. The bootstrap estimates of the errors in the pole direction, period, and the YORP parameter are shown in Fig.~\ref{fig:BS_85989}. The $1\sigma$ uncertainties are $4.8^\circ$ in $\lambda$, $4.0^\circ$ in $\beta$, $1.5 \times 10^{-6}$\,h in $P$, and $4.1 \times 10^{-9}\,\radd$ in $\upsilon$. Therefore, even if the formally best fit and bootstrap samples suggest a negative YORP acceleration of about $-5.0 \times 10^{-9}\,\radd$, its uncertainty is at about the same level, and zero YORP is still consistent with the current data set.

    \begin{figure*}
        \centering\includegraphics[width=\textwidth]{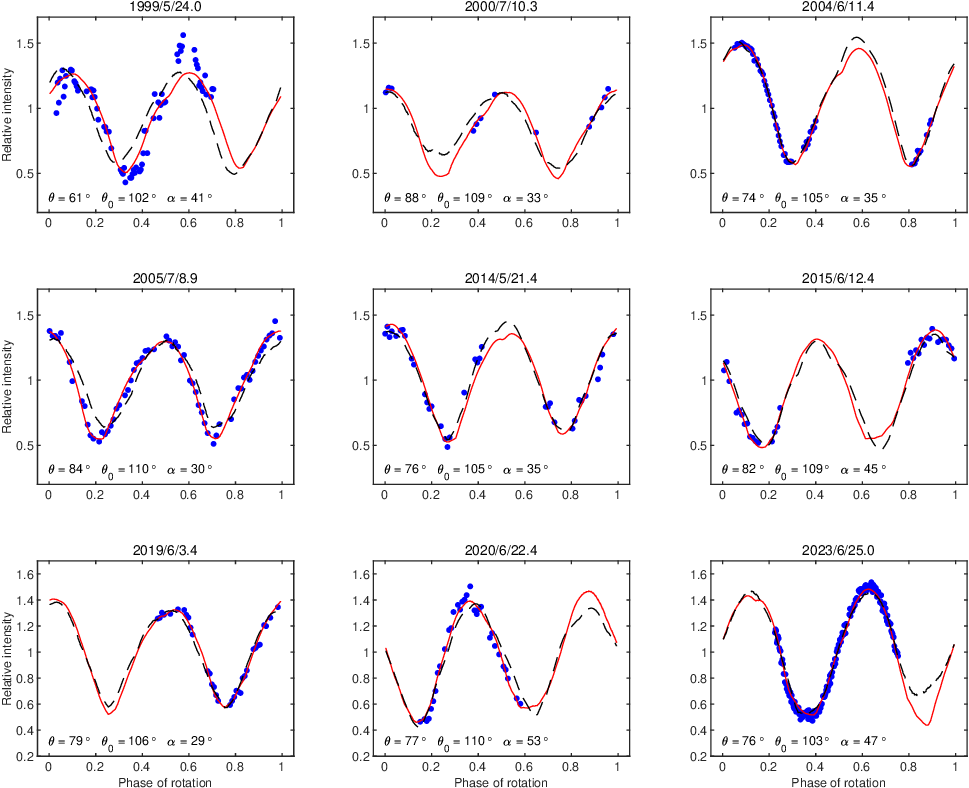}
        \caption{Example light curves of (85989) 1999~JD6. The blue points show observed data, the red curve shows our best model without YORP, and the dashed curve shows the best model with the spin and YORP parameters fixed at values given by \cite{tian2022} (clearly not matching the available observations).}
        \label{fig:lc_fit_85989}
    \end{figure*}
    
    As to the difference with respect to the radar study, we note that our spin-axis direction does not agree with the precise value determined by \cite{marphd}. Fixing the pole direction at their value still gives an acceptable fit, but the $\chi^2$ value increases by about 5\%. When we treat all data sets as relative light curves, the pole direction shifts to $(252\dgr, -67\dgr)$, which is closer to the radar pole, but still outside the error intervals. The reason for this discrepancy is not known to us. We speculate that it might be related to our model being convex, while the real shape of 1999~JD6 is bilobed, as was clearly revealed by radar delay-Doppler images.
       
    The convex shape model shown in Fig.~\ref{fig:model_85989} has a dynamical flattening of $\Delta = 0.40$. With a semimajor axis 0.883\,au, an excentricity 0.633, and an inclination $17.06^\circ$, the theoretically estimated solar precession constant is $\alpha = 2130\,''\,\text{yr}^{-1}$. Over 24 years of observations, this corresponds to a shift of almost $13^\circ$ on the precession cone, which is $7.3^\circ$ in $\lambda$ and $-2.6^\circ$ in $\beta$. These values are comparable to the pole uncertainties, so that the solar precession effect should also be included in the modeling because it has a non-negligible effect, and we wondered whether this omission might not contribute to the difference with respect to the results of \cite{marphd}.

    We proceeded similarly as in the case of asteroid Ra-Shalom: We probed values of $\alpha$ between $-5000$ and $5000\,''\,\text{yr}^{-1}$, and the YORP parameter value $\upsilon$ was tested in between $-3$ and $3 \times 10^{-8}\,\radd$. The result is shown in Fig.~\ref{fig:precession_YORP_85989}. The formally best solutions with the lowest RMS residuals are obtained for $\alpha < 0$, which is unrealistic. However, the estimated $3\sigma$ uncertainty interval is so wide that it includes all tested values of $\alpha$. This means that our data set is not sensitive to $\alpha$, and we currently cannot constrain it from observations. For any value of the precession constant, the constant period, namely $\upsilon = 0$, is still consistent with the data. However, as in the case of the bootstrap samples, the uncertainty interval is not symmetric around zero here either, and negative values of $\upsilon$ are preferred.
       
    \subsection{(138852) 2000 WN10} \label{obs:wn10}

    The data set we used for the shape reconstruction consisted of light curves from 2008 and 2009 obtained by \cite{Ski.ea:12}, from 2015 by \cite{War:16e} downloaded from the Asteroid Lightcurve Database\footnote{\url{https://alcdef.org}} \citep{War.ea:09}, and our new observations listed in Table~\ref{tab:aspect_138852}.

\begin{figure}[t]
        \includegraphics[width=\columnwidth]{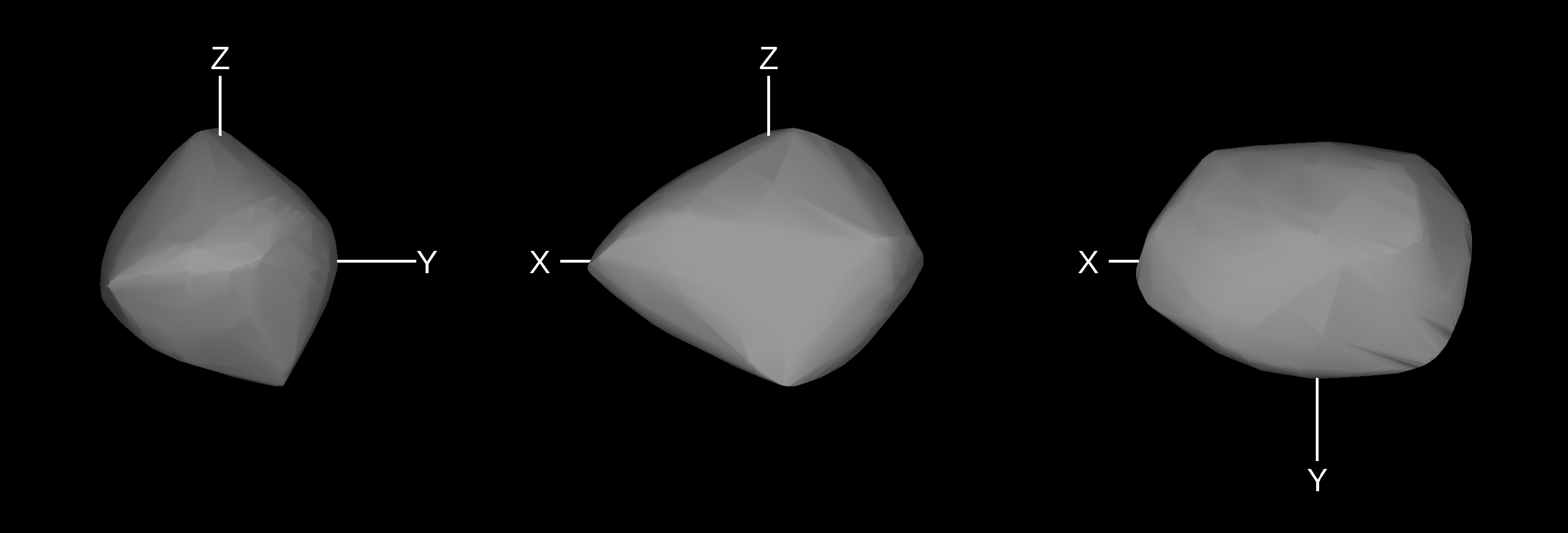}
        \caption{Convex shape model of asteroid (138852) 2000~WN10 shown from the equatorial level ({\it left} and {\it center\/}, $90^\circ$ apart) and pole-on ({\it right\/}).}
        \label{fig:model_138852}
    \end{figure}

\begin{figure}[t]
        \includegraphics[trim=0.5cm 4cm 1cm 4cm, clip, width=\columnwidth]{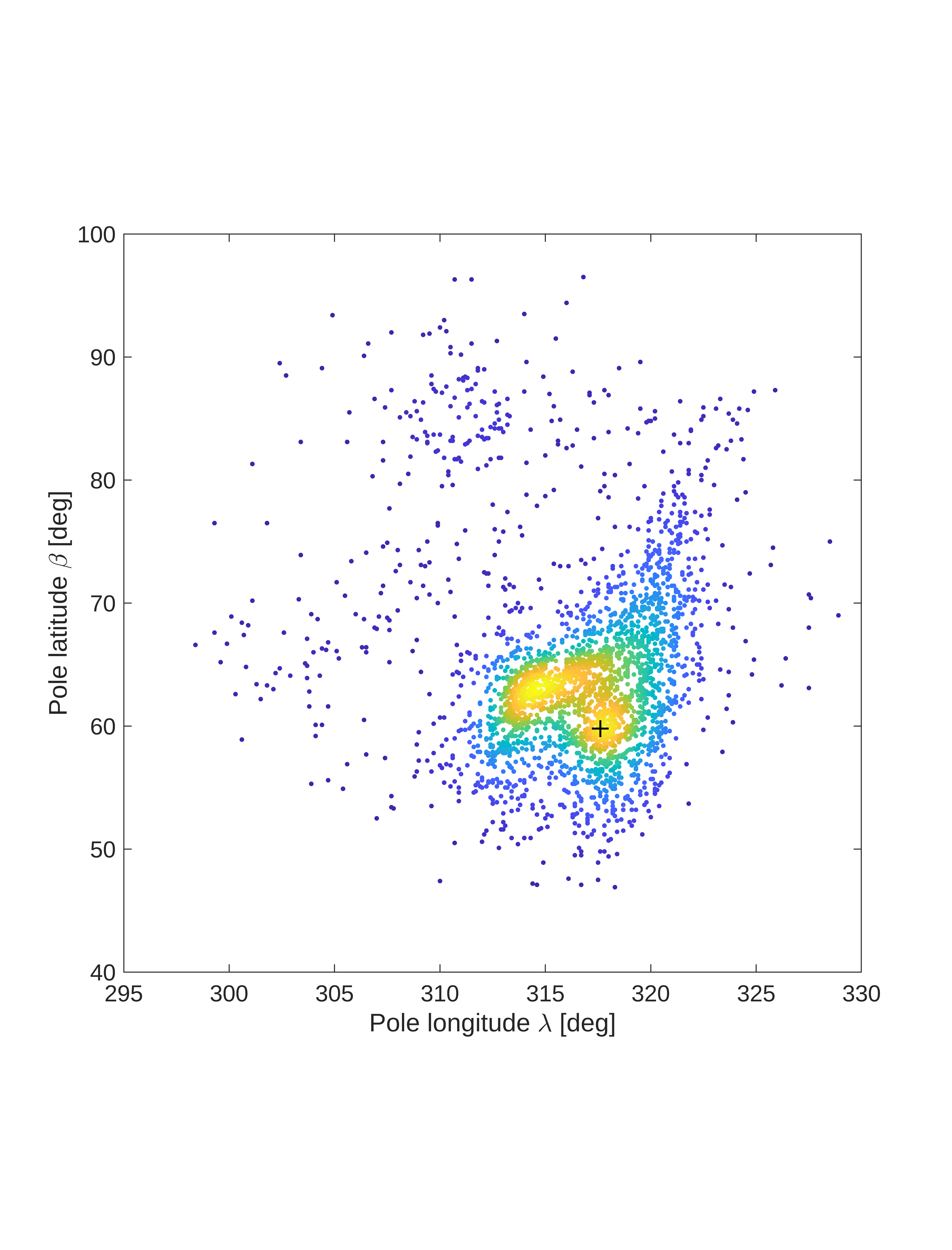}
        \includegraphics[trim=0.5cm 4cm 1cm 4cm, clip, width=\columnwidth]{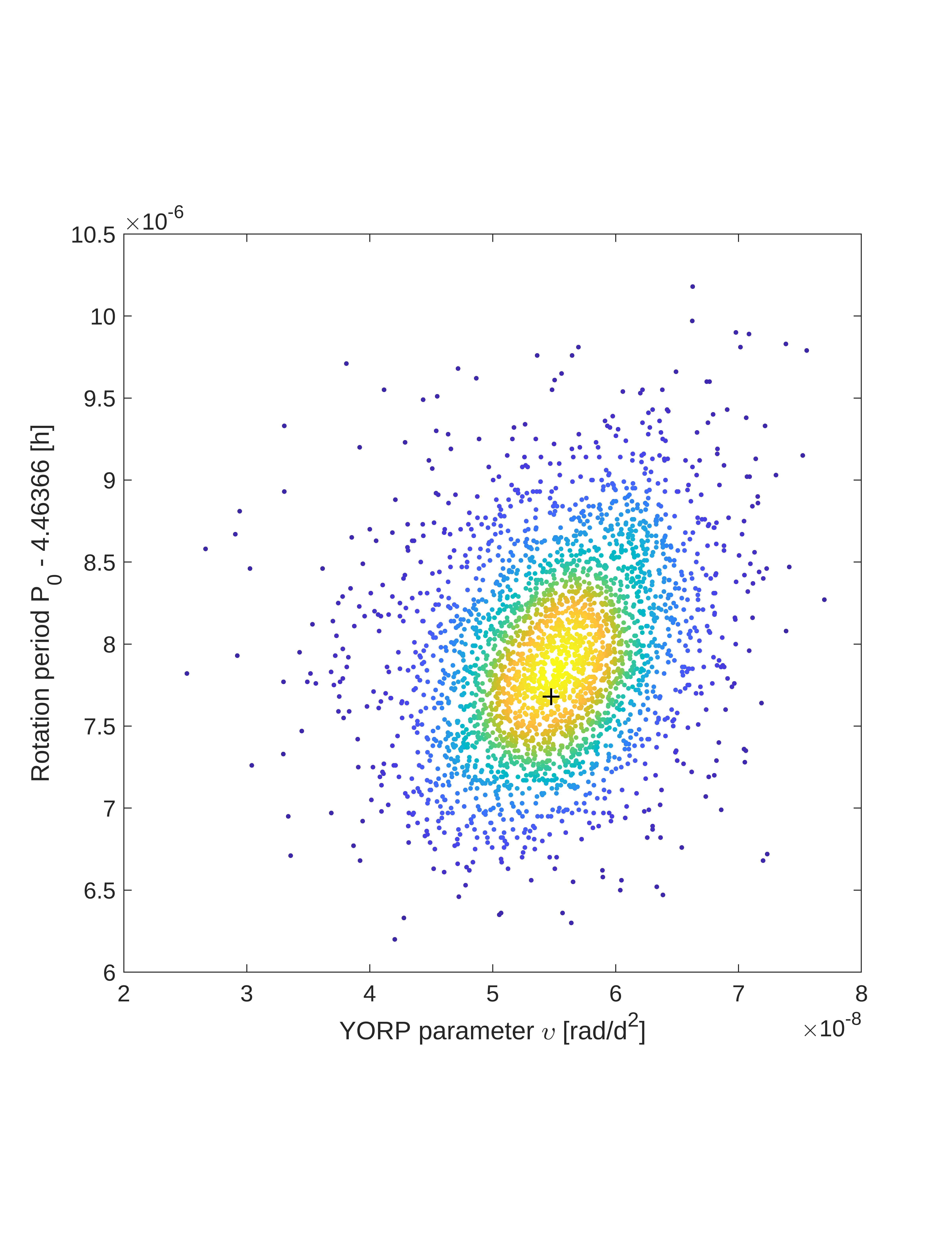}
        \caption{Bootstrap distribution of the pole direction in the ecliptic longitude $\lambda$ and latitude $\beta$ ({\it top\/}), the YORP parameter $\upsilon$, and the rotation period $P_0$ ({\it bottom\/}) for asteroid (138852) 2000~WN10 (the latter at epoch JD$_0$ 2457022.5). The color-coding corresponds to the density of points, and the cross marks the solution based on the original (not bootstrapped) data set. Pole ecliptic latitudes $\beta$ higher than $90^\circ$ correspond to pole directions $\beta - 90^\circ$, $\lambda - 180^\circ$.}
        \label{fig:BS_138852}
    \end{figure}

    The light-curve inversion led to a strong YORP detection.
    The best spin solution has the following parameters: $\lambda = (318 \pm 4)^\circ$,  $\beta = (60 \pm 8)^\circ$, $P = 4.463\,667\,7 \pm 0.000\,000\,6$\,h, (for JD 2457023.5), and $\upsilon = (5.5 \pm 0.7) \times 10^{-8}\,\radd$. The corresponding best-fit shape model of (138852) 2000 WN10 is shown in Fig.~\ref{fig:model_138852}.
    The bootstrap distribution of the poles and rotation parameters is shown in Fig~\ref{fig:BS_138852}.

    In addition to the best-fit solution described above, there was another local minimum in $\chi^2$ for the sidereal rotation $P = 4.465\,939\,1$\,h. However, this solution was significantly worse, as it provided the same fit with nonzero YORP as the best period above without YORP. The pole direction was not unique, likely because of the limited geometry; there were also possible solutions with the pole within $\pm 10^\circ$ around the direction $\beta = 90^\circ$. Some of them are also shown in the upper plot in Fig.~\ref{fig:BS_138852}. However, for this spin-axis direction, a YORP value between  5 and $6 \times 10^{-8}\,\radd$ also provides a significantly better fit than a constant period model.

    \begin{figure*}
        \centering\includegraphics[width=\textwidth]{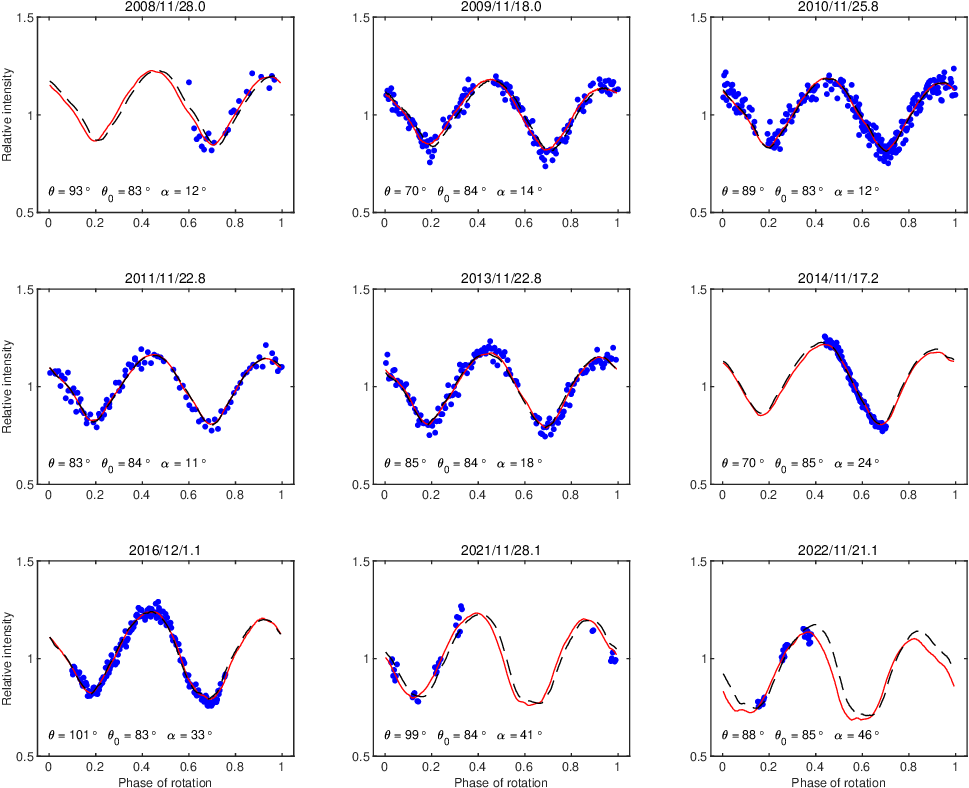}
        \caption{Example light curves of (138852) 2000~WN10. The blue points show the observed data, the red curve shows our best model with YORP, and the dashed curve shows the best model without YORP.}
        \label{fig:lc_fit_138852}
    \end{figure*}
    
    \subsection{ (161989) Cacus} \label{obs:cacus}

    The previous model based on observations from 1978--2016 was published by \citet{Dur.ea:18a}. Its spin parameters were $\lambda = (254 \pm 5)^\circ$, $\beta = (-62 \pm 2)^\circ$, and $\upsilon = (1.9 \pm 0.3) \times 10^{-8}\,\radd$. We updated the model by adding new light curves from 2022 that were observed with TRAPPIST and by \cite{Pan.Paj:23} (see Table~\ref{tab:aspect_161989}).
    
\begin{figure}[t]
        \includegraphics[width=\columnwidth]{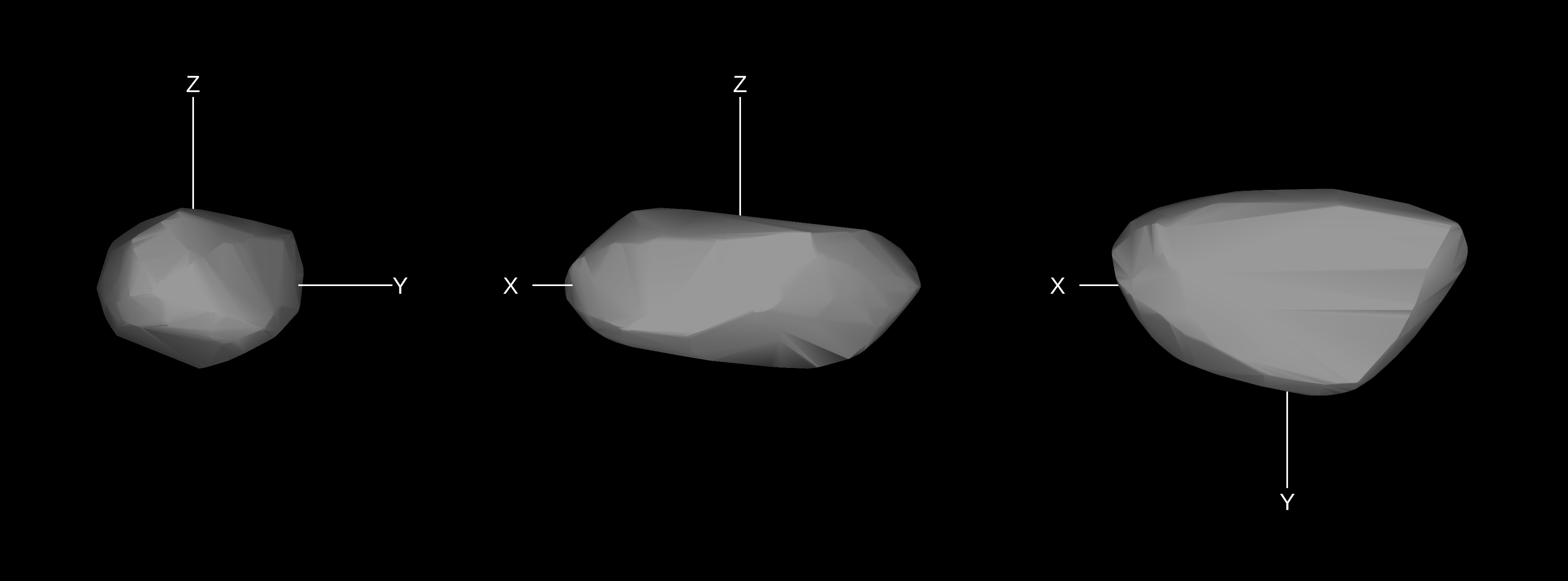}
        \caption{Convex shape model of asteroid (161989)~Cacus shown from the equatorial level ({\it left} and {\it center\/}, $90^\circ$ apart) and pole-on ({\it right\/}).}
        \label{fig:model_161989}
    \end{figure}

    The new model (Fig.~\ref{fig:model_161989}) is essentially the same as the old one, with the same spin parameters, but smaller uncertainties: The pole direction $(252 \pm 3^\circ, -63 \pm 2^\circ)$, period $P = 3.755\,052\,7$ (for JD$_0$ 2451544.5), and the YORP value $\upsilon = 1.86 \times 10^{-8}$\,rad\,d$^{-2}$. 

\begin{figure}[t]
        \includegraphics[trim=0.5cm 4cm 1cm 4cm, clip, width=\columnwidth]{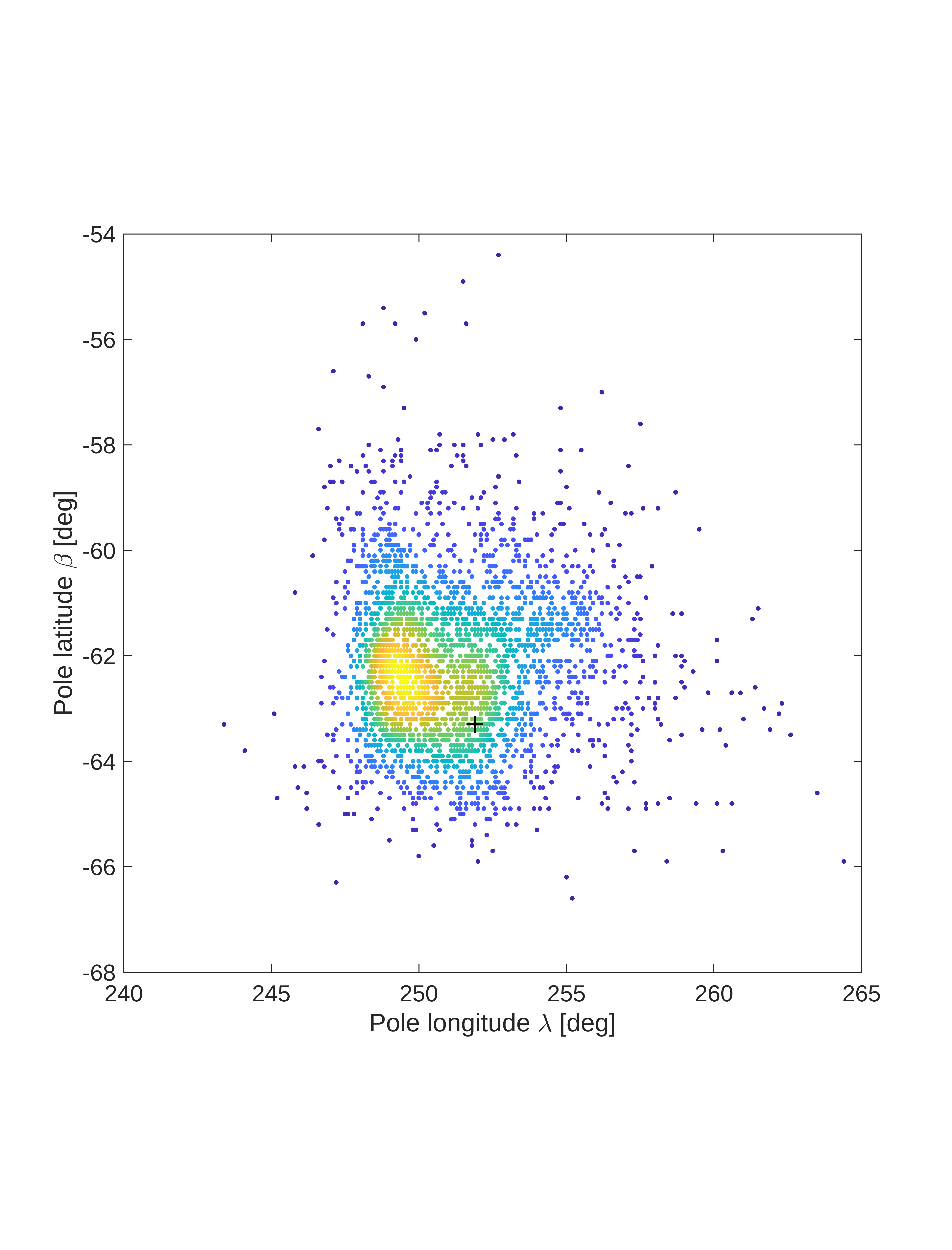}
        \includegraphics[trim=0.5cm 4cm 1cm 4cm, clip, width=\columnwidth]{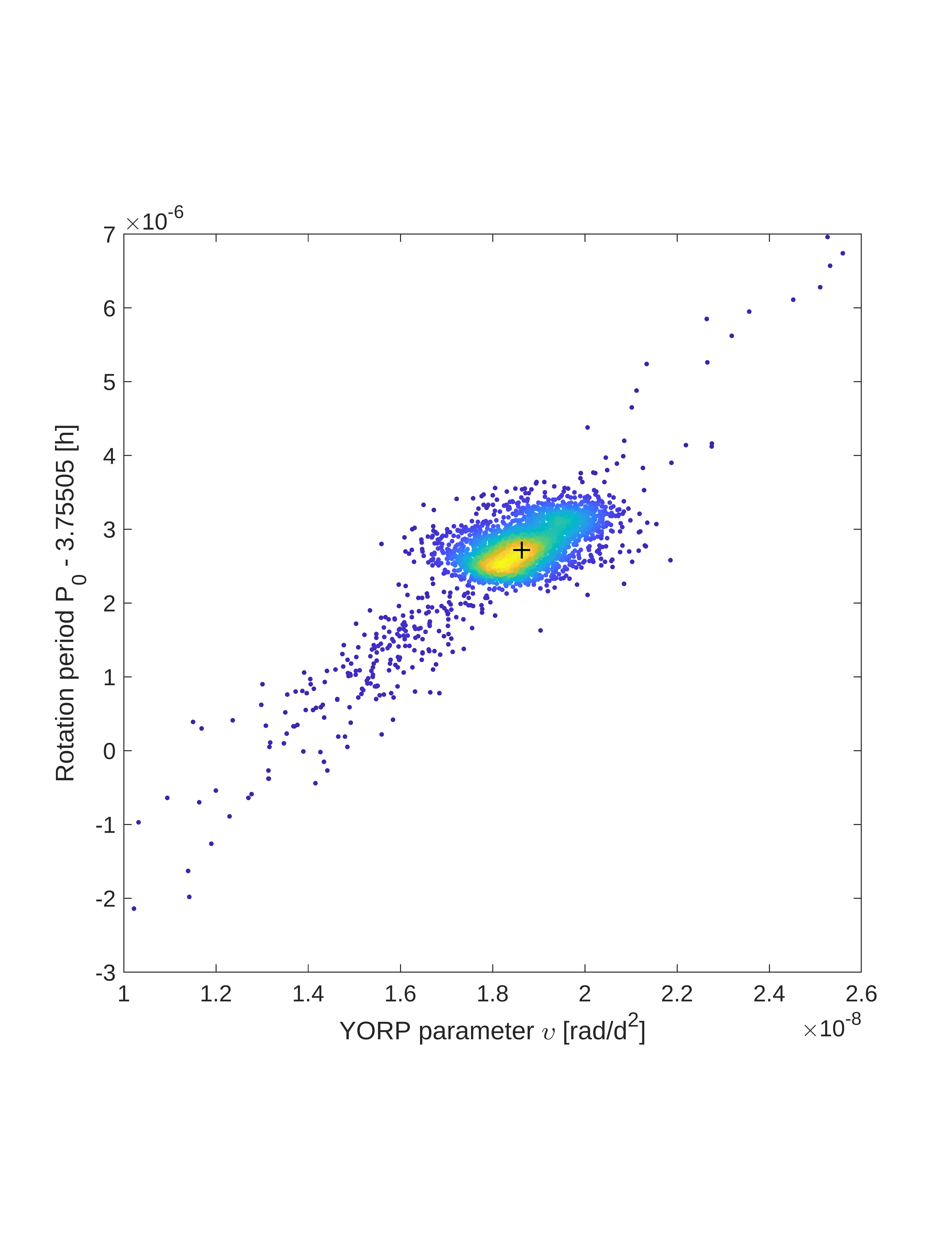}
        \caption{Bootstrap distribution of the pole direction in the ecliptic longitude $\lambda$ and latitude $\beta$ ({\it top\/}), the YORP parameter $\upsilon$, and the rotation period $P_0$ ({\it bottom\/}) for asteroid (161989)~Cacus (the $P_0$ value is given at epoch JD$_0$ 2451544.5). The color-coding corresponds to the density of points, and the cross marks the solution based on the original (not bootstrapped) data set.}
        \label{fig:BS_161989}
    \end{figure}

    However, in contrast to the old model by \cite{Dur.ea:18a}, the YORP detection now does not depend on the isolated data from 1978. Even without the two light curves by \citet{dege1978} and \citet{schu1979}, the best-fit YORP value is $\upsilon = 1.6 \times 10^{-8}\,\radd$. The bootstrap results are shown in Fig.~\ref{fig:BS_161989}. The plot of the period versus YORP shows a core of points and diagonal tails. This pattern corresponds to the unusual time distribution of the photometric observations. Two light curves were taken in 1978, and the remaining part of the data set (another 33 light curves) consists of observations taken between 2003 and 2022. The outlying diagonal points correspond to bootstrap samples that by chance do not contain light curves from 1978. There were 381 such cases out of 3000. For them, the zeropoint lies even before the first observation, so that there is a strong correlation between the period and YORP. On the other hand, the bulk of the points in the center contains observations from 1978, so that the correlation is much weaker. The formal standard deviations for the whole bootstrap sample are $0.000\,000\,6$~h and $1.3 \times 10^{-9}\,\radd$. For a limited subsample including 1978 data, it is $0.000\,000\,3$~h and $8.5 \times 10^{-10}\,\radd$.

% SEC ??? %%%%%%%%%%%%%%%%%%%%%%%%%%%%%%%%%%%%%%%%%%%%%%%%%%%%%%%%%%%%%%%%%%%%%%%%%%%%%
\section{Justification of the rotation rate change using a theoretical
         model}\label{sec:theory}
The secular change in the sidereal rotation rate, expressed
by the parameter $\upsilon$, has been empirically derived in Sec.~\ref{sec:models},
without a particular theoretical interpretation. This approach, namely
including $\upsilon$ into the light-curve inversion method as a free, solved-for
parameter, has been found to be very efficient in most previous studies. However,
it ideally requires a subsequent step of connecting its value to a particular
physical phenomenon. Spontaneous particle ejections \citep{sche2020} or micrometeoroid
impacts \citep{wieg2015} may be examples of these candidate processes. We took
a different standpoint here by adopting a link of the observationally determined $\upsilon$
values to the YORP effect \citep{vetal2015}.

To set up this connection, we must evaluate $\upsilon_{\rm model}$ using
a numerical model. As always, the latter depends on many physical parameters
${\bf p}$, thus $\upsilon_{\rm model}({\bf p})$. We may take the liberty of adjusting 
${\bf p}$ to certain optimum values ${\bf p}_\star$ to match $\upsilon_{\rm model}({\bf p}_\star)\simeq
\upsilon$, but the point is to  ensure that ${\bf p}_\star$ has reasonable 
(expected) values. Unfortunately, the YORP is still slightly more complex:
(i) Some of the parameters are well-constrained observationally or may be estimated
using an educated guess (e.g., the rotation state, the large-scale convex shape, the size,
the surface thermal inertia, and the bulk density, as an example of the latter group), but (ii) others are
not easily accessible from ground-based observations (e.g., shape nonconvexities and
small-scale surface irregularities). The predicted $\upsilon_{\rm model}$ depends
on (ii), however, which may represent a significant perturbation in specific cases
\citep[e.g.,][]{s2009,yorpito2009}. The specific physical phenomena that
were considered to quantitatively test the YORP-dependence on these hidden parameters
had to do with the lateral heat conduction in small-scale surface irregularities
\citep[e.g.,][]{gk2012,Sev.ea:15,gl2022}, or with the anisotropy of the thermal emission (''thermal beaming'')
and mutual irradiation of the surface facets \citep[e.g.,][]{rg2012,rg2013}.
We did not evaluate the contribution of (ii) here, and leave this part
to future studies of specific targets. We only modeled (i). To do this we used our well-tested
numerical approach that we presented in \citet{cv2004,cv2005}. In general terms, the
model solves the 1D heat diffusion problem independently for each of the surface facets,
with nonlinear Robin boundary conditions at the surface and deep interior below
the facet. The time domain of the solution spans one revolution about the Sun, and
the rotation period is slightly adjusted (typically within its uncertainty limits)
to have an integer number of rotation cycles in one revolution period (in this way, the
periodicity of the solution is imposed). The discretization steps in the space and time
domains are chosen to satisfy the von Neumann criterion, and the iterations are repeated until
a subdegree tolerance for all surface facets is satisfied. After reaching convergence,
we evaluated the total thermal torque and acceleration along the heliocentric
revolution using the appropriate sum over all surface facets. More details are provided in the references mentioned above.

For the parameters ${\bf p}$, we used the nominal (best-fit) rotation period, pole orientation, and shape model determined from the light-curve inversion in Sec.~\ref{sec:models}. Noting the strong dependence of the YORP effect on the asteroid shape, we chose an additional ten variant
solutions of the spin state and shape, all providing a statistically acceptable fit to the light curves. We ran the model for all of them and report (i) the median and (ii) the range of the $\upsilon_{\rm model}$ values.
The other physical parameters, such as the size and bulk density, were either taken from previously
published studies, constrained by the observed Yarkovsky effect, or assumed.
Their values are individually discussed in the next sections. We note that $\upsilon_{\rm model}$ does not depend on the surface thermal inertia in the simple model of the YORP effect we use here (see \citet{cv2004} for numerical evidence and \citet{nv2007,bm2008} for analytical proof). Therefore, we do not need to specify this parameter
when we report the $\upsilon_{\rm model}$. 

Even though our model caveat consists of the absence of the observationally hidden
parameters, we further justify it by predicting the thermal acceleration in
heliocentric motion (the Yarkovsky effect). 
All five NEAs analyzed in this paper have a reliably measured
secular drift of the semimajor axis by now. Similarly to the case of the steady
increase in the sidereal rotation rate, we assume that the origin of this 
perturbation is due to the thermal accelerations in the heliocentric motion 
(the Yarkovsky effect). An important justification then arises from the
common prediction of the YORP and Yarkovsky strength based on our model. Importantly, the Yarkovsky
(orbital) part is somewhat less dependent on the details of the asteroid shape and surface roughness (our runs for the best-fit model and the sample of ten variant models usually provides a Yarkovsky effect prediction within $5-15$\% of the median value). As a result, it can provide a reasonable constraint on the bulk density.
The value of the secular drift of the semimajor axis $\dadt$
derived within the orbit determination process has been calculated in several independent
studies published over the past decade or more \citep[see, e.g.,][listed here
sequentially in time]{ches2008,nugent2012,far2013,vetal2015,ches2016,tardioli2017,
vigna2018,green2020}. As soon as the available astrometric information was
accurate and abundant, the $\dadt$ solution across these different sources
converged to a single value. We used the most recently updated solution provided
by the JPL Horizons website.%
\footnote{\url{https://ssd.jpl.nasa.gov/}}
This is important because the statistically significant value of $\dadt$ for
(161989) Cacus only became available after the radar astrometry was taken in
September 2022.

% ----------------------------------------------------
\subsection{(1862) Apollo} \label{the:apollo}
While discovered already in April~1932 \citep[see][becoming later a namesake of its
orbital class among the near-Earth asteroids]{wr1932}, this body
shares the fate of several other NEAs by being lost for many decades and rediscovered
only in 1973. A fortuitous pair of very close encounters with Earth in November 1980 and
April 1982 offered the opportunity to take a sequence of multiwavelength observations:
(i) visible photometry \citep{harris1987}, (ii) infrared \citep{leb1981},
and (iii) even radar observations \citep{gold1981,ostro2002}. This wealth of data
allowed an early determination of many physical parameters (although some needed slight 
corrections later on), making Apollo exclusive even in its category. The parameters included
the rotation state and pole orientation \citep[e.g.,][]{harris1987}, a size between
$1.2$ and $1.6$~km, a geometric albedo of $0.2$ to $0.26$
\citep[e.g.,][]{harris1998,ostro2002}, and a Q-type spectral classification
\citep{tholen1984,bb2002}. Apollo was recognized to be one of the first targets for a detection
of the Yarkovsky effect \citep{vok2005} (see even previous evidence of
nongravitational perturbations of its heliocentric motion by
\citet{ziol1983} and \citet{yeo1991}, who suspected unseen cometary activity).
The actual detection of the Yarkovsky effect was reported in 2008
\citep[see][]{ches2008}, even preceded by the detection of the YORP
effect \citep[see][]{apollo2007,Dur.ea:08}. Apollo has been found to have an unusually
small satellite in its class \citep{ostro2005}, which was uniquely detected in the radar
observations taken in October and November 2005. Because it is so very small, not much
is known about it, but it cannot influence our model determination of either
the Yarkovsky or the YORP effects. Apollo was also taken as an exemplary case to show the possible role of planetary close encounters in explaining asteroidal Q-type
spectra, which are similar to ordinary chondrite analogs \citep[e.g.,][]{nes2010}. In summary, not much is known about (1862)~Apollo, but this body continues to be
an inspiration for interesting concepts of asteroidal science.
% FIG %%%%%%%%%%%%%%%%%%%%%%%%%%%%%%%%%%%%%%%%%%%%%%%%%%%%%%%%%%%%%%%%%%%%%%%%%%%%%%%%
\begin{figure}[t]
 \begin{center} 
 \includegraphics[width=\columnwidth]{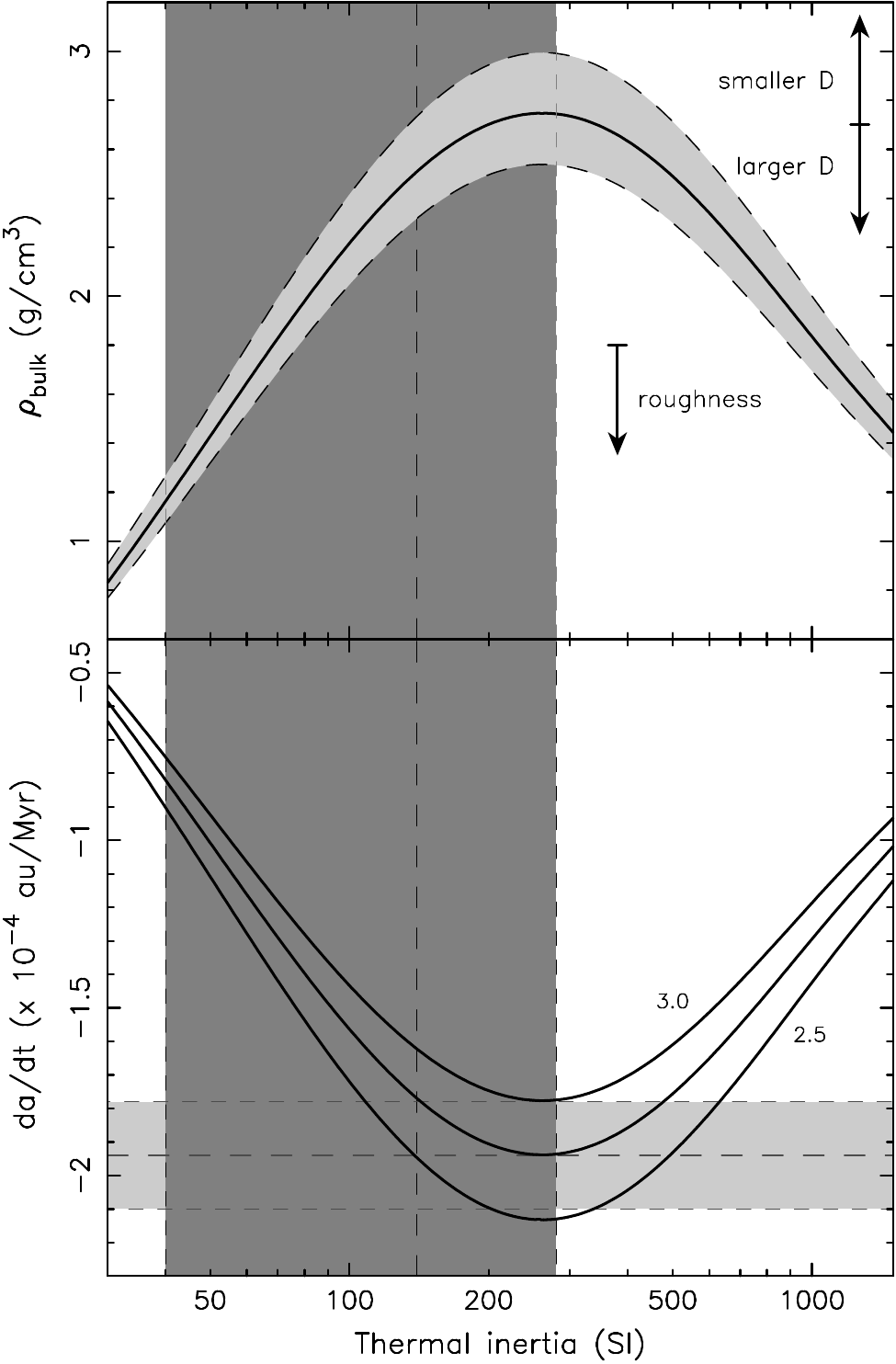}
 \end{center}
 \caption{\label{fig_1862}
  Predicted values of the Yarkovsky effect and density for asteroid (1862)~Apollo.
  {\it Bottom panel:\/} Predicted semimajor axis drift $\dadt$ (ordinate) 
  due to the Yarkovsky effect from our model for three different values of the bulk
  density ($2.5$, $2.75$, and $3$\,g\,cm$^{-3}$; see the labels and the middle curve for
  $2.75$\,g\,cm$^{-3}$) are shown by solid curves. The abscissa is the surface
  thermal inertia in SI units (\sig). We assumed the rotation state and shape model
  from the light-curve inversion in Sec.~\ref{obs:apollo}, and an effective size of $1.55$~km. The
  grayscale horizontal region shows the value $-(1.94\pm 0.16)\times 10^{-4}$\,au\,My$^{-1}$
  from the orbit determination. The grayscale vertical region shows the range of
  the best-fit surface thermal inertia value $140^{+140}_{-100}$~\sig \citep[see][]{rg2012}.
  {\it Top panel:\/} Model-predicted bulk density to match the observed value of the semimajor axis drift shown by the
  grayscale region. The solid line in the middle shows the exact correspondence surrounded by
  a map of the sigma interval of the Yarkovsky drift. A nominal effective size of
  $1.55$~km is used. If this value were higher or lower, the density solution would shift
  in the direction indicated by the arrows (preserving the $\rho_{\rm b}\,D$ value).
  The analysis of ten different shape models of (1862) Apollo reveals a variation of $\pm 12$\% about the median value used in the figure.
  Effects of small-scale surface roughness, if important, would shift the solution toward a
  lower value by typically $10-30$\% \citep[see][]{rg2012}.}
 \end{figure}
%%%%%%%%%%%%%%%%%%%%%%%%%%%%%%%%%%%%%%%%%%%%%%%%%%%%%%%%%%%%%%%%%%%%%%%%%%%%%%%%%%%%%%

One such step has been undertaken by \citet{rozitis2013}, who combined information
about the detected Yarkovsky and YORP effects and interpreted them using a
single thermophysical model of this asteroid \citep[a goal that has been imagined
by][see their Sec.~5, but not achieved before]{cv2004}. We adopted their solution
for an Apollo size $D=1.55\pm 0.07$~km, a geometric albedo $p_V=0.20\pm 0.02$,
a surface thermal inertia $\Gamma = 140^{+140}_{-100}$~\sig, and a bulk density
$\rho_{\rm b} = 2.85^{+0.48}_{-0.68}$\,g\,cm$^{-3}$. In a way, our work is a mere
repetition of the \citet{rozitis2013} calculation, except that we now have updated
and more accurate values of both YORP and Yarkovsky effects. As for the latter,
the current astrometric data set provides $\dadt = -(1.94\pm 0.16)\times 10^{-4}$
au~My$^{-1}$, while the value of \citet{rozitis2013} was higher by about $10$\%. Similarly,
our new value for the YORP effect (Sec.~\ref{obs:apollo}) is about $10$\% smaller and much more
accurate than the value $(5.5\pm 1.2)\times 10^{-8}$\,rad\,d$^{-2}$ from \citet{Dur.ea:08}
considered by \citet{rozitis2013}.

Figure~\ref{fig_1862} shows the comparison between the predicted Yarkovsky drift $\dadt$
from our model and the observed value (the shape variants produce scatter of roughly $\pm 10$\%  about the median value used in Fig.~\ref{fig_1862} in addition to to the bulk density dependence). Assuming a bulk density of $2.75$\,g\,cm$^{-3}$, which we nominally used in the simulation, the predicted semimajor axis drift matches the
observed value very well. Small adjustments of size, surface roughness, and/or thermal inertia
may further help to match any value in the uncertainty interval of $\dadt$. The  same model predicts a median $\upsilon_{\rm model}\simeq 5.0\times 10^{-8}$\,rad\,d$^{-2}$ for the $2.75$\,g\,cm$^{-3}$ bulk density, and a range between $1.80\times 10^{-8}$\,rad\,d$^{-2}$ and $7.25\times 10^{-8}$\,rad\,d$^{-2}$. This is fairly consistent with the
observed value $\upsilon\simeq 4.94\times 10^{-8}$\,rad\,d$^{-2}$ (Sec.~\ref{obs:apollo}). In addition to the global shape dependence, further
differences may readily be explained by the thermal
beaming of a rough surface \citep{rg2012}, and/or simply a slightly larger body size.

% ----------------------------------------------------
\subsection{(2100) Ra-Shalom} \label{the:rasha}
Ra-Shalom has received a wealth of infrared observations, both
from the ground and from space \citep[e.g.,][]{harrisetal1998,delbo2003,tri2010,usui2011}.
In spite of the persisting uncertainty, the results converge to a size
of $2-2.8$~km, a geometric albedo $0.1-0.18$, and an unusually high value
of the surface thermal inertia (possibly $\Gamma \simeq 1000$~\sig).
We adopted the results of \citet{rasharadar2008}, who combined
various data sets, including radar observations, to obtain the most
complete picture of this asteroid. Their preferred size was $D=2.3\pm 0.2$~km
and a geometric albedo $p_V=0.13\pm 0.03$. \citet{rasharadar2008} also
analyzed a composite visible-to-infrared spectrum of Ra-Shalom and found
that it was similar to spectra of CV3 meteorites. Based on this similarity,
they argued that a K-type classification would be the best match for this
asteroid (refining the previous C- or Xc-type classification%
\footnote{Admittedly though, there is an uncertainty in the Ra-Shalom classification,
 since \citet{bin2019} found its spectrum neutral to blueish and classified
 it as a B-type object. The slow rotation makes it unclear whether
 this observation speaks for the whole body or a particular surface feature.
 More data are clearly needed in this respect.}).
Combining the measured grain density of CV3 meteorites and the estimated
$\sim 30$\% porosity of C-class asteroids, \citet{rasharadar2008}
suggested a bulk density of $\rho_{\rm b} = 2.4\pm 0.6$\,g\,cm$^{-3}$.
Finally, these authors also interpreted their radar measurements and
thermal observations as providing evidence for a coarse or rocky surface
with only a thin or unimportant regolith layer.
% FIG %%%%%%%%%%%%%%%%%%%%%%%%%%%%%%%%%%%%%%%%%%%%%%%%%%%%%%%%%%%%%%%%%%%%%%%%%%%%%%%%
\begin{figure}[t]
 \begin{center} 
 \includegraphics[width=\columnwidth]{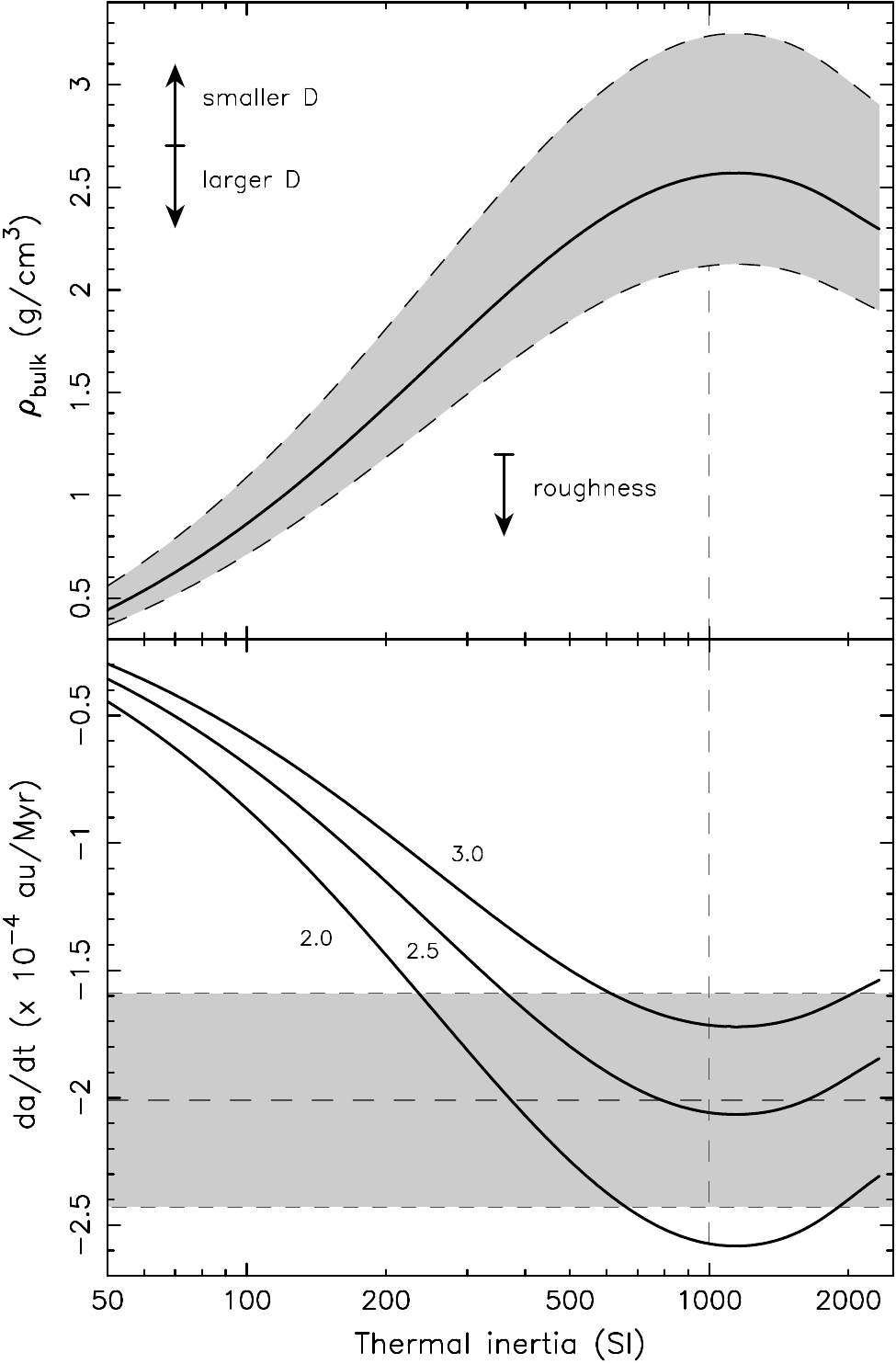}
 \end{center}
 \caption{\label{fig_2100}
 Predicted values of the Yarkovsky effect and density for asteroid (2100)~Ra-Shalom.
  {\it Bottom panel:\/} Predicted semimajor axis drift $\dadt$ (ordinate) 
  due to the Yarkovsky
  effect from our model for three different values of the bulk density ($2$, $2.5$, and
  $3$\,g\,cm$^{-3}$; see the labels) are shown by solid curves. The abscissa is the surface
  thermal inertia in SI units (\sig). We assumed the rotation state and shape model
  from the light-curve inversion in Sec.~\ref{obs:rasha} and an effective size of $2.3$~km. The
  grayscale horizontal region shows the value $-(2.01\pm 0.43)\times 10^{-4}$\,au\,My$^{-1}$
  from the orbit determination. The vertical dashed line indicates the suggested
  surface high-thermal inertia of $\simeq 1000$~\sig.
  {\it Top panel:\/} Model-predicted bulk density to match the observed value of the semimajor axis drift shown by the
  grayscale region. The solid line in the middle shows the exact correspondence surrounded by
  a map of the sigma interval of the Yarkovsky drift. A nominal effective size of
  $2.3$~km is used. If this value were higher or lower, the density solution would shift
  in the direction indicated by the arrows (preserving the $\rho_{\rm b}\,D$ value).
  The analysis of ten different shape models of (2100) Ra-Shalom reveals a variation of $\pm 10$\% about the median value used in the figure.
  Effects of small-scale surface roughness, if important, would shift the solution toward a lower value by typically $10-30$\% \citep[see][]{rg2012}.}
 \end{figure}
%%%%%%%%%%%%%%%%%%%%%%%%%%%%%%%%%%%%%%%%%%%%%%%%%%%%%%%%%%%%%%%%%%%%%%%%%%%%%%%%%%%%%%
% FIG %%%%%%%%%%%%%%%%%%%%%%%%%%%%%%%%%%%%%%%%%%%%%%%%%%%%%%%%%%%%%%%%%%%%%%%%%%%%%%%%
\begin{figure*}[t]
 \begin{center} 
 \includegraphics[width=\textwidth]{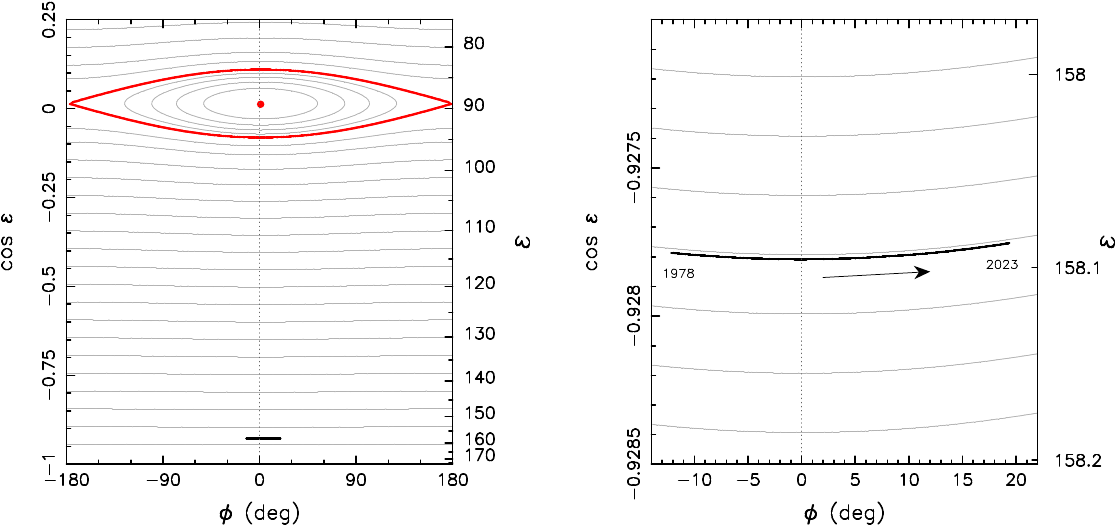}
 \end{center}
 \caption{\label{fig_2100s}
  Rotation pole evolution of (2100) Ra-Shalom due to the solar torque and its
  comparison with the simple Colombo top model. The coordinates on the axes
  are defined in the orbital plane, namely (i) the obliquity $\epsilon$ (or
  $\cos\epsilon$) on the right and left ordinates, and (ii) the longitude $\phi$
  reckoned from the direction $90^\circ$ away from the orbital node \citep[located at
  $\phi=-90^\circ$; e.g.,][]{bnv2005}. The numerically propagated evolution of the
  Ra-Shalom pole over the $44$~yr time span is shown by the solid black line. The gray
  lines are solutions of the simple Colombo top model. For most of the obliquity
  values, they are basically simple straight lines of approximately constant
  $\epsilon$. The exception is the Cassini resonant zone at $\epsilon\simeq 90^\circ$
  (with the separatrix shown by the red curves, and the stationary Cassini state~2
  shown by the red dot). Unlike the case of asteroid
  (433)~Eros, whose spin axis librates in the resonant zone \citep{vok2005Eros},
  the evolution of the Ra-Shalom pole is simpler and only consists of
  nearly regular precession. The zoom in the right panel shows that the change
  in obliquity is minimum, but conforms to the Colombo top solution.}
 \end{figure*}
%%%%%%%%%%%%%%%%%%%%%%%%%%%%%%%%%%%%%%%%%%%%%%%%%%%%%%%%%%%%%%%%%%%%%%%%%%%%%%%%%%%%%%

The large size and favorable orbit of Ra-Shalom, belonging to what 
\citet{milani1989} classified as the Toro orbital group residing near
or inside the mean motion resonances with the Earth, 
offered numerous opportunities to observe this asteroid with radar.
It belongs to the record-holders in the number of radar observations
at different apparitions (in this case, seven observations between 1981 and
2022). Their accuracy and long time-base resulted in a high-quality
orbit determination that allowed a firm detection of the Yarkovsky
effect \citep[see already][]{vok2005}. Interestingly, the related drift
of the semimajor axis $\dadt = -(2.01\pm 0.43)\times 10^{-4}$\,au\,My$^{-1}$
is well matched by our model, adopting a bulk density in between
$2$ and $3$\,g\,cm$^{-3}$ and the suggested high surface thermal inertia
(Fig.~\ref{fig_2100}). These two parameters are correlated: If the thermal inertia were lower, the required bulk density
would also be lower. Additionally, the unaccounted-for effects of the
surface roughness in our model may slightly decrease the bulk density 
solution, as indicated by the arrow in the top panel of Fig.~\ref{fig_2100}.

\citet{Dur.ea:12b} and \citet{Dur.ea:18a} repeatedly sought the YORP effect in
the available photometric data of that date, but were only able to place
an upper limit on the value. In this paper, we finally determined the
value from a series of light-curve data extended to the last year. In contrast to
the conclusion in \citet{Dur.ea:18a}, who speculated that Ra-Shalom might
be the first case for which YORP will be found to decelerate the rotation rate,
we found a weakly accelerating signal of $\upsilon\simeq 2.9\times
10^{-9}$\,rad\,d$^{-2}$ (Sec.~\ref{obs:rasha}). This is even lower than the value found for
asteroid (1685) Toro, even though Toro (i) is larger and (ii) has a larger
semimajor axis. This comparison confirms the unusual weakness of the YORP effect
for Ra-Shalom.

Assuming a bulk density of $2.5$\,g\,cm$^{-3}$, fine-tuned to match the Yarkovsky
effect, our simulation predicts a median $\upsilon_{\rm model}\simeq -2.1\times 10^{-8}$
rad~d$^{-2}$. Not only is the absolute value of $\upsilon_{\rm model}$ higher
than the detected $\upsilon$, but its sign is opposite
\citep[similarly to what has been found in][]{Dur.ea:18a}. Clearly, the nominal convex
shape model, which just has a low resolution at large scales, cannot capture important features of the YORP effect. The full range of predicted 
$\upsilon_{\rm model}$ values from the sample of $11$ equivalent models is
$-0.15\times 10^{-8}$\,rad\,d$^{-2}$ to $-6.1\times 10^{-8}$\,rad\,d$^{-2}$. This is still negative, but some of the extreme values are now closer to the detected $\upsilon$.

A possibly important clue to the solution is the evidence of a significant
surface roughness \citep{rasharadar2008}. This could promote a significant
contribution of the fine-scale surface effects, such as mutual facet irradiation
and/or transverse heat conduction, of which at least the latter preferentially
contributes by an acceleration component in YORP \citep[e.g.,][]{gk2012}.
The situation of (2100) Ra-Shalom may therefore be reminiscent of (25143) Itokawa,
for which the early analyses, making use of the large-scale resolution shape models of this
asteroid, consistently predicted a negative $\upsilon$ value \citep[e.g.,][]{ito2004,
schee2007,yorpito2009}. However, after enough observations were finally available,
the detected $\upsilon$ value proved to be positive \citep{ito2014}, which implied
a spin-up state of Itokawa. \citet{ito2014}, following an earlier theoretical concept
formulated by \citet{sg2008} \citep[see also][]{yorpito2009}, suggested that the perturbing
effect was due to a difference in the bulk density of the two main shape features, notably the
quasi-ellipsoidal lobes in contact. However, the difference required to explain the overall
effect appeared rather large. \citet{Sev.ea:15} proposed an alternative, and perhaps more
natural, explanation based on the novel concept of lateral heat conduction on small-scale
surface features discovered by \citet{gk2012}. We hypothesize that the same model may
explain the disparity in the observed and modeled $\upsilon$ values for Ra-Shalom.
However, we leave a detailed study for the future.

% precession
We now briefly return to the observationally revealed precession of the Ra-Shalom rotation
pole (Sec.~\ref{obs:rasha}). We found that the best-fit shape model corresponds to the dynamical flattening value $\Delta\simeq 0.26$, which provides a theoretical value for the precession constant $\alpha\simeq
2720$~\arcsec~yr$^{-1}$ (which is fully compatible with the best-fit value of this parameter; Fig.~\ref{fig:precession_YORP_2100}). From the simplest possible point of view, in
which the heliocentric orbit would be fixed in space, the pole would precess with
an angular speed of $-\alpha\cos\epsilon\simeq 2522$~\arcsec~yr$^{-1}$, where
$\epsilon\simeq 158^\circ$ is the rotation pole obliquity, about the direction of
the orbital angular momentum \citep[therefore normal to the orbital plane; e.g.,][]{bfv2003}.
The situation becomes more complicated when the orbital plane evolves in time due to
planetary perturbations. In the case of Ra-Shalom, the essence of the generalization may
be described using a model that is still very simple, in which the orbital inclination remains
approximately constant ($I\simeq 15.7^\circ$), but the longitude of node precesses in
inertial space about the sum of the planetary orbital angular momenta with a frequency
$s\simeq -22.98$~\arcsec~yr$^{-1}$. The dynamics of the rotation pole within this setup is
described by the Colombo top model \citep[e.g.,][]{col1966,hapo2020}, whose implementation
for asteroids can be found, for instance, in \citet{vok2006}. We refer
to these references for more details and note here that the Colombo top model has two
fundamental parameters: (i) the orbit inclination $I$, and (ii) the ratio of frequencies
$\kappa=\alpha/(2\sigma)\simeq -59$. The complexity of the pole evolution is determined by
the relation of $\kappa$ to a critical parameter $\kappa_\star=-\frac{1}{2}\left(\sin^{2/3} I
+\cos^{2/3} I\right)^{3/2}\simeq -0.823$. The most complicated situation occurs when
$\kappa$ is just slightly smaller than $\kappa_\star$. In our situation, $\kappa$ is
much smaller than $\kappa_\star$, and this confines the complexity to a narrow obliquity
slab near $90^\circ$, which restores the expected regularity at high and low values of
$\epsilon$. The latter corresponds to the case of Ra-Shalom.

To verify the conclusion, we numerically integrated the evolution of the Ra-Shalom
rotation pole over the $44$~yr time span between the first and last available light-curve
observations (September 1978 to September 2022).%
\footnote{We used the symplectic numerical integrator described in \citet{bnv2005}
 implemented in the widely used orbital dynamics package {\tt swift}
 (e.g., \url{http://www.boulder.swri.edu/~hal/swift.html}).}
The results are summarized in Fig.~\ref{fig_2100s}. The $44$~yr track of the pole evolution
is shown by the solid black segment, and the gray lines show solutions of the Colombo top
model (the proximity of the true evolution of the Ra-Shalom pole from our numerical
simulation to the Colombo top solution starting at nearby initial conditions shows
that the latter is a fairly good but not complete approximation). The existence of the
resonant zone at about $90^\circ$ obliquity perturbs the Ra-Shalom obliquity only very
slightly, producing a negligible variation during the time span of interest. The
principal dynamical effect consists of a $\simeq 31^\circ$ drift in longitudinal angle
$\phi$ associated with the obliquity $\epsilon$ (right panel in Fig.~\ref{fig_2100s}).
The true angular distance $\delta$ between the Ra-Shalom pole direction in 1978 and in 2022
in space is smaller, however. We easily find that $\cos\delta=1-2\sin^2\epsilon
\sin^2 \phi/2$, and therefore, $\delta\simeq 11.5^\circ$. Because the inclination of the Ra-Shalom
orbit is rather small, most of this effect is projected into a change in ecliptic
longitude of the pole, while a small part, about $4^\circ$, is the change in
ecliptic latitude.

% FIG %%%%%%%%%%%%%%%%%%%%%%%%%%%%%%%%%%%%%%%%%%%%%%%%%%%%%%%%%%%%%%%%%%%%%%%%%%%%%%%%
\begin{figure}[t]
 \begin{center} 
 \includegraphics[width=\columnwidth]{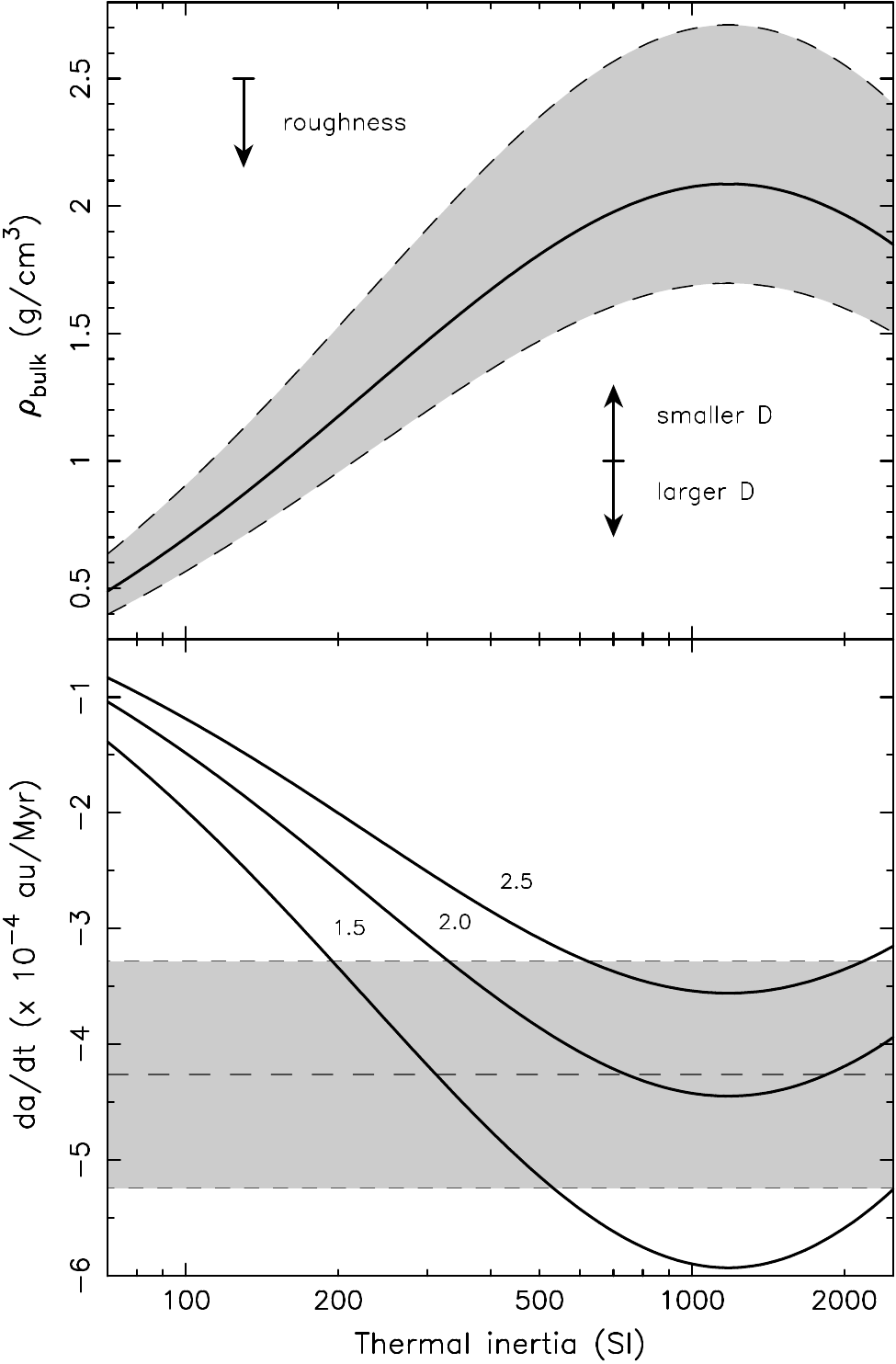}
 \end{center}
 \caption{\label{fig_jd6}
 Predicted values of the Yarkovsky effect and the density for asteroid (85989) 1999~JD6.
  {\it Bottom panel:\/} Predicted semimajor axis drift $\dadt$ (ordinate) 
  due to the Yarkovsky effect from our model for three different values of the bulk
  density ($1.5$, $2$, and $2.5$\,g\,cm$^{-3}$; see the labels) shown by the solid curves; the
  abscissa is the surface thermal inertia in SI units (\sig). We assumed the rotation
  state and shape model from the light-curve inversion in Sec.~\ref{obs:jd6} and an effective
  size of $1.45$~km. The grayscale horizontal region shows the value $-(4.26\pm 0.98)\times
  10^{-4}$\,au\,My$^{-1}$ from the orbit determination. {\it Top panel:\/} Model-predicted bulk density 
  to match the observed value of the semimajor axis drift shown by the
  grayscale region. The solid line in the middle shows the exact correspondence surrounded by
  a map of the sigma interval of the Yarkovsky drift. The nominal effective size of
  $1$~km is used. If this value were lower or higher, the density solution would shift
  in the direction indicated by the arrows (preserving the $\rho_{\rm b}\,D$ value).
  The analysis of ten different shape models of (85989) 1999~JD6 reveals a variation of $\pm 10$\% about the median value used in the figure.}
 \end{figure}
%%%%%%%%%%%%%%%%%%%%%%%%%%%%%%%%%%%%%%%%%%%%%%%%%%%%%%%%%%%%%%%%%%%%%%%%%%%%%%%%%%%%%%

% ----------------------------------------------------
\subsection{(85989) 1999~JD6} \label{the:jd6}
The most complex study of this asteroid was presented by \citet{marphd},
who combined several sets of multiwavelength data from visible photometry and infrared observations to very detailed radar sensing during its 2015 apparition.
The results revealed the highly elongated shape of a contact binary with a 
volume equivalent size of $D=1.45\pm 0.14$~km and a surface thermal inertia
$\Gamma\simeq 280$ \sig \citep[see also][]{camp2009}. The radar data were complemented with
light-curve observations between 1999 and 2015, which allowed determining the sidereal rotation
period to $7.664\,346\,4 \pm 0.000\,005\,6$\,hr. The abundant radar data in 2015 and their
fortuitous viewing geometry discussed in \citet{marphd} resulted in an unusually
accurate determination of the asteroid rotation pole, namely $(\lambda,\beta)
=(220.3^\circ,-73.43^\circ)$, with only a fraction of a degree of uncertainty. This implies an
obliquity of nearly $180^\circ$. \citet{marphd} also searched for the YORP signal in
the data available until 2015, but he set only an upper limit $|\upsilon| < 2.6\times 10^{-8}$
rad~d$^{-2}$. The recent analysis of the Yarkovsky effect with Gaia DR3 precise astrometric
measurements included by \citet{dzia2023} provided $\dadt = -(4.26\pm 0.98)\times
10^{-4}$\,au\,My$^{-1}$ for this asteroid. The taxonomic classification of (85989) 1999~JD6
is not clearly constrained, with suggestions of K class \citet{thomas2014}, X class
\citet{carry2016}, and L class \citet{bin2019}, which means that only some of the primitive
classes are excluded. From BVR photometry, \cite{Pol.Bro:08} concluded that RaShalom belongs to the K or S class.

Recently, \citet{tian2022} analyzed a larger set of photometric observations taken 
between 2000 and 2020 and claimed to have detected the YORP effect with $\upsilon =
(2.4\pm 0.3)\times 10^{-8}$\,rad\,d$^{-2}$. However, the results we presented in Sec.~\ref{obs:jd6} do not
confirm their finding. Instead, we obtained $\upsilon = -(5\pm 4)\times 10^{-9}$\,rad\,d$^{-2}$,
with a zero value still statistically acceptable. This is a tighter constraint than was found by
\citet{marphd}, but it does not yet prove the YORP signal. The reason for the
difference between our results and that in \citet{tian2022} eludes us. The best we could have done
is to double-check our solution, which is compatible for the YORP nondetection with that of \citet{marphd}.

We also used our rotation state and convex shape model to estimate the expected value of
the semimajor axis secular drift (the Yarkovsky effect) and the sidereal rotation rate
secular change (the YORP effect). We assumed a $2$\,g\,cm$^{-3}$  bulk density and
ran simulations for a wide range of surface thermal inertia. The results are shown
in Fig.~\ref{fig_jd6}. The predicted semimajor axis drift is generally consistent
with the value determined by \citet{dzia2023}, but if the low thermal inertia of the
surface holds, the bulk density also needs to be low ($\simeq 1.3-1.7$\,g\,cm$^{-3}$).
The intrinsic bulk density might be higher because by using the convex-hull
model, we overestimated the volume that is occupied by matter for this contact binary object.

As expected, the formal median value $\upsilon_{\rm model}\simeq 2.7\times 10^{-7}$\,rad\,d$^{-2}$ obtained with our nominal shape model and a bulk density $2$\,g\,cm$^{-3}$ is significantly different from the upper limit $|\upsilon|\leq
10^{-8}$\,rad\,d$^{-2}$ (with even a preference for the negative values).
However, the full range of $\upsilon_{\rm model}$ from our $11$ shape/spin variant solutions extends from $-3.5\times 10^{-7}$\,rad\,d$^{-2}$ to
$3.7\times 10^{-7}$\,rad\,d$^{-2}$, indicating that the YORP effect cannot be accurately predicted from the simple convex models and available photometric observations.

% ----------------------------------------------------
\subsection{(138852) 2000~WN10} \label{sec_wn10}
The population of the Earth coorbitals is an interesting transient subclass
of NEAs with a characteristic dynamical lifetime of some tens of thousands
of years \citep[e.g.,][]{mm2002}. Coorbitals, or objects that are orbitally very
close to them, may offer a unique possibility among NEAs to be observable
every year for a decade or more. Because these annual Earth encounters may
be close, even very small NEAs may be observed. In spite of the slight drawback that they
typically repeat a similar observing geometry, coorbitals are particularly suitable in 
situations when a series of observations seeks to determine the cumulative effect of 
a weak perturbation. This is the case of the Yarkovsky and YORP effects. For this
reason, \citet{vok2005} considered the detection of the Yarkovsky effect for a handful of
small coorbital asteroids, including (54509) YORP \citep[for which the Yarkovsky
effect was swiftly reported by][]{ches2008} and (138852) 2000~WN10 \citep[for
which the Yarkovsky effect was reported by][due its larger size and unfortunate absence
of the radar astrometry]{vetal2015}. Even more importantly, (54509) YORP was one of
the first two objects for which the YORP effect was detected 
\citep{yorp12007,yorp22007}. The YORP detection for (138852) 2000~WN10 proved
to be harder and had to wait until the present paper. This is due to its larger
size and intrinsically weaker YORP strength.%
\footnote{The rotation rate acceleration of (54509) YORP reads $\upsilon\simeq
 350\times 10^{-8}$\,rad\,d$^{-2}$. If we were to plainly scale this value for
 about three times larger and about twice denser (138852) 2000~WN10, we may expect
 the YORP strength of about $\upsilon\simeq 350/2/3^2\times 10^{-8}\simeq 20\times
 10^{-8}$\,rad\,d$^{-2}$ (note that both asteroids have about the same orbit and
 both have extreme values of the obliquity). Yet, the observed $\upsilon$ is four times smaller (Sec.~\ref{obs:wn10}), perhaps due to a more regular shape.}

To compare the observed $\dadt$ and $\upsilon$ (from Sec.~\ref{obs:wn10}) with the
theoretical values, we need to know or assume several physical parameters of the
asteroid. Because 2000~WN10 is small and lacks radar observations, the information
about this asteroid is unfortunately sparse. The light-curve inversion provides complete information about the spin state and a rough (convex) shape model.
Based on broadband photometry, \citet{ieva2018} found (138852) 2000~WN10 to be compatible with
the S-type group of asteroids. \citet{perna2018} conducted a spectroscopic survey of small 
NEAs and included (138852) 2000~WN10 in their program. The visible spectrum corresponds to the
Sq taxonomic type \citep[similarly][found it to be a Q-type object]{bin2019}.
Adopting a mean geometric albedo of $0.24$ for this group,
they argued for a rather small size of $\simeq 250$~m. However, without
infrared observations, the albedo and size values are simply an educated guess.
If we were to assume the mean albedo value of the NEA population $0.15$, the size would
recalibrate to $\simeq 320$~m. At this moment, we adopt $D\simeq 300$~m for our
thermal model. In practice, we rescale unconstrained dimensions of the shape
model from Sec.~\ref{obs:wn10} such that its volume is equivalent to a sphere with a diameter of
$300$~m. Finally, we need to adopt some value of (i) the surface thermal
inertia $\Gamma$ and (ii) the bulk density $\rho_{\rm b}$. In this case, we
ran a series of simulations for different $\Gamma$, sufficiently sampling
the interval of values from thermophysical modeling of small NEAs
\citep[see, e.g.,][]{delb2015}. For the bulk density, we used $2$
g~cm$^{-3}$. The results for different $\rho_{\rm b}$ values were readily obtained
with the $\dadt\propto \rho_{\rm b}^{-1}$ relation.
% FIG %%%%%%%%%%%%%%%%%%%%%%%%%%%%%%%%%%%%%%%%%%%%%%%%%%%%%%%%%%%%%%%%%%%%%%%%%%%%%%%%
\begin{figure}[t]
 \begin{center} 
 \includegraphics[width=\columnwidth]{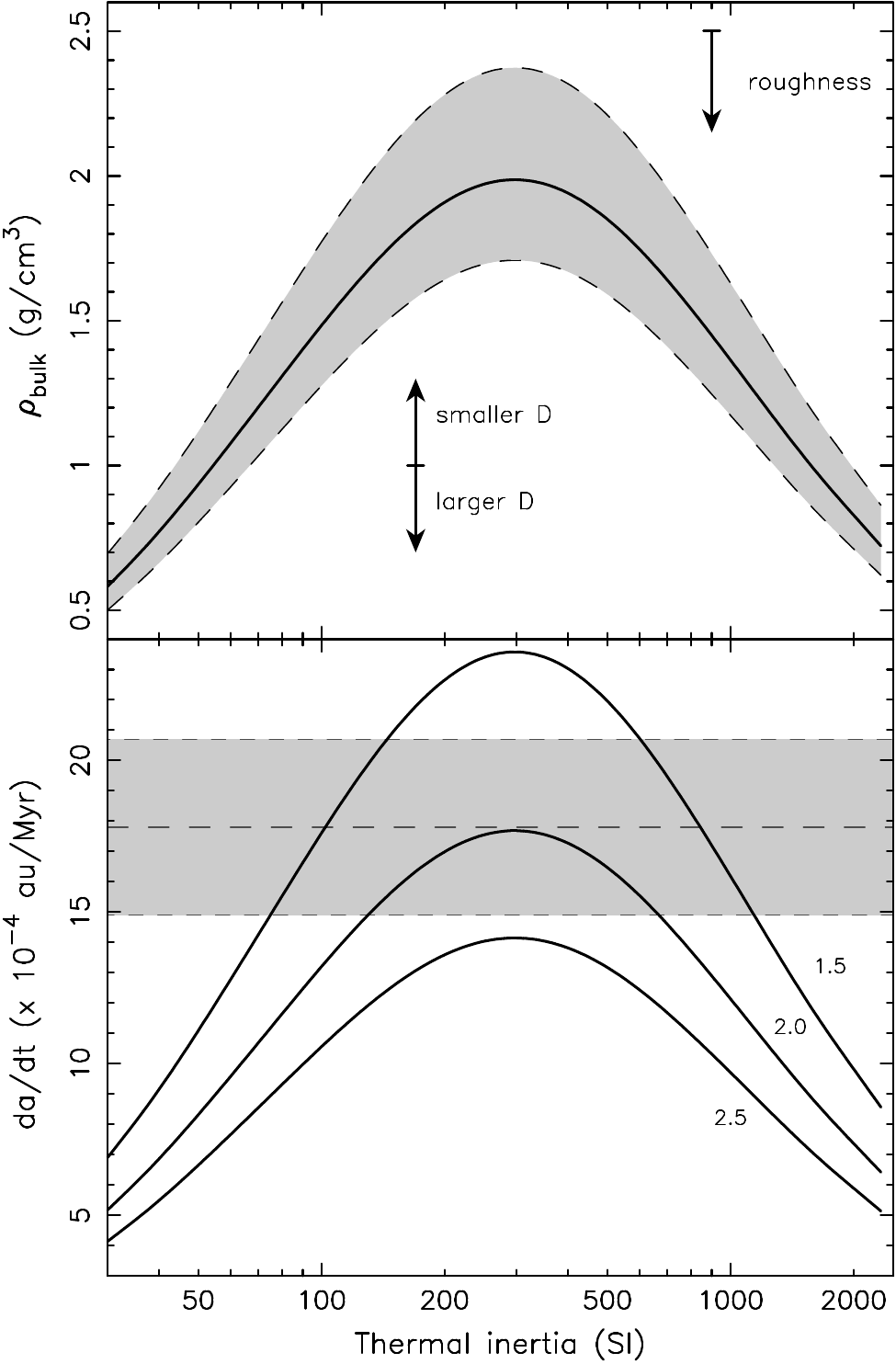}
 \end{center}
 \caption{\label{fig_WN10}
 Predicted values of the Yarkovsky effect and density for asteroid (138852) 2000~WN10.
  {\it Bottom panel:\/} Predicted semimajor axis drift $\dadt$ (ordinate) 
  due to the Yarkovsky
  effect from our model for three different values of the bulk density ($1.5$, $2$, and
  $2.5$\,g\,cm$^{-3}$; see the labels) are shown by the solid curves; the abscissa is the surface
  thermal inertia in SI units (\sig). We assumed the rotation state and shape model
  from the light-curve inversion in Sec.~\ref{obs:wn10} and an effective size of $300$~m. The
  grayscale horizontal region shows the value $(17.8\pm 2.9)\times 10^{-4}$\,au\,My$^{-1}$
  from the orbit determination. 
  {\it Top panel:\/} Model-predicted bulk density 
  to match the observed value of the semimajor axis drift shown by the
  grayscale region. The solid line in the middle shows the exact correspondence surrounded by
  a map of the sigma interval of the Yarkovsky drift. The nominal effective size of
  $300$~m is used. If this value were higher or lower, the density solution would shift
  in the direction indicated by the arrows (preserving the $\rho_{\rm b}\,D$ value).
  The analysis of ten different shape models of (138852) 2000~WN10 reveals a variation of $\pm 3$\% about the median value used in the figure.  
  The effects of small-scale surface roughness, if important, would shift the solution toward a lower
  value by typically $10-30$\% \citep[see][]{rg2012}.}
 \end{figure}
%%%%%%%%%%%%%%%%%%%%%%%%%%%%%%%%%%%%%%%%%%%%%%%%%%%%%%%%%%%%%%%%%%%%%%%%%%%%%%%%%%%%%%

Figure~\ref{fig_WN10} shows our theoretical predictions for the Yarkovsky semimajor
axis drift as compared to the observed value $\dadt = (17.8\pm 2.9)\times 10^{-4}$\,au\,My$^{-1}$ (the variation in the predicted values due to different shape models is small in this case, only $\pm 3$\%).
The plausible range $1.7$ to $2.4$\,g\,cm$^{-3}$ for the bulk density is fully acceptable,
but is shifted to slightly lower values than those of the small Q-type asteroid
(6489) Golevka, as also determined by the Yarkovsky effect \citep[see][]{gol2003}. At the same
time, the required thermal inertia in between $150$ and $600$~\sig\ favorably compares
to what is expected from limited cases of small S-type NEAs with good visible and
infrared observations \citep{delb2015}. There is clearly room for further adjustments
due to small changes in the effective size $D$ and the effects of small-scale roughness
(the arrows in the upper panel schematically illustrate their influence).
%%%%%%%%%%%%%%%%%%%%%%%%%%%%%%%%%%%%%%%%%%%%%%%%%%%%%%%%%%%%%%%%%%%%%%%%%%%%%%%%
\begin{figure}[t]
 \begin{center} 
 \includegraphics[width=\columnwidth]{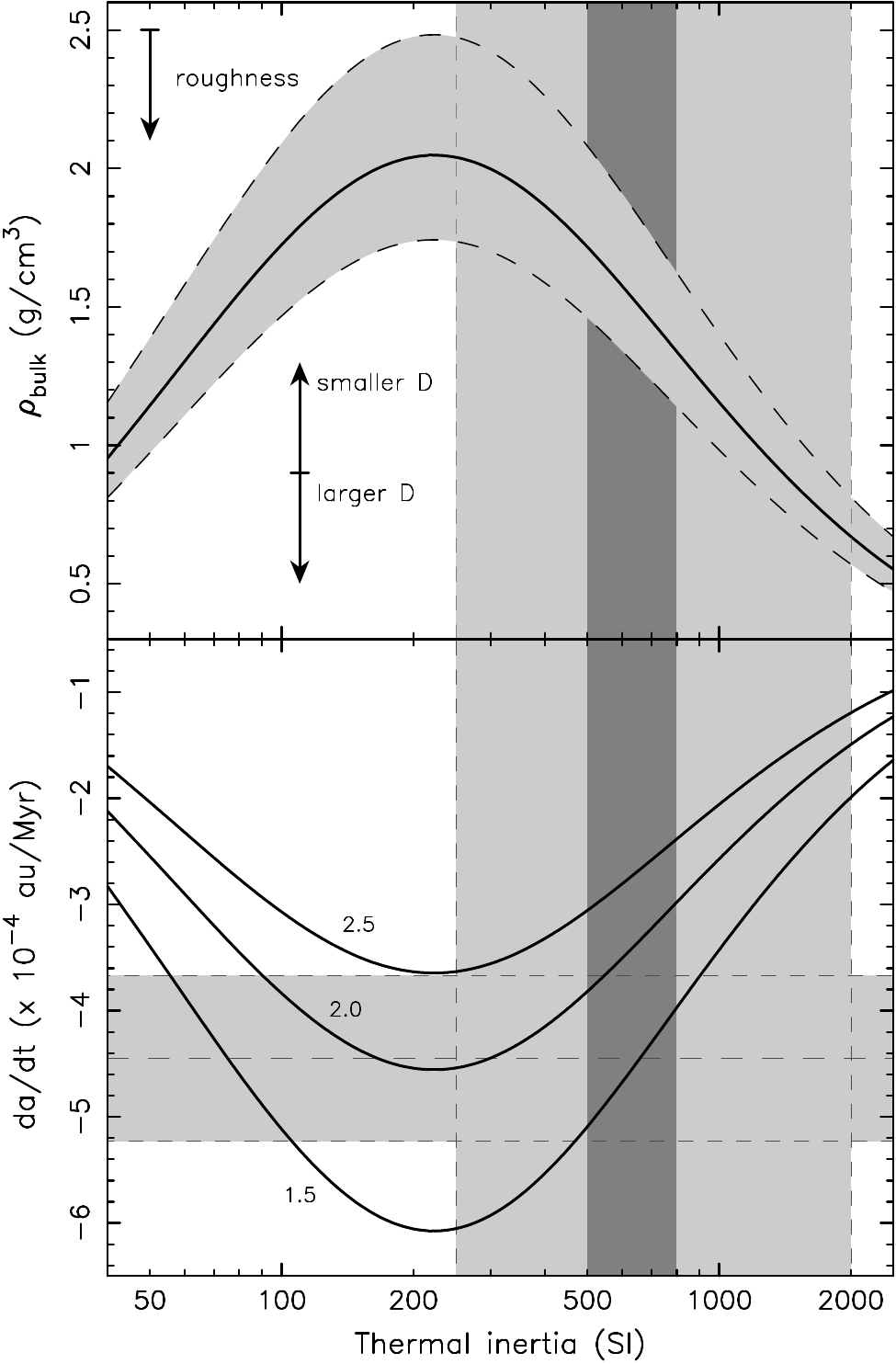}
 \end{center}
 \caption{\label{fig_cacus}
 Predicted values of the Yarkovsky effect and density for asteroid (161989)~Cacus.
 {\it Bottom panel:\/} Predicted semimajor axis drift $\dadt$ (ordinate) 
  due to the Yarkovsky effect from our model for three different values of the bulk
  density ($1.5$, $2$ and $2.5$\,g\,cm$^{-3}$; see the labels) are shown by the solid curves; the
  abscissa is the surface thermal inertia in SI units (\sig). We assumed the rotation
  state and shape model from the light-curve inversion in Sec.~\ref{obs:cacus} and an effective
  size of $1$~km. The grayscale horizontal region shows the value $-(4.45\pm 0.78)\times
  10^{-4}$\,au\,My$^{-1}$ from the orbit determination. The grayscale vertical region shows
  the range of the acceptable surface thermal inertia value $200$ to $2000$~\sig
  and the best-fitting core $500$ to $800$~\sig \citep[see][]{Dur.ea:18a}.
  {\it Top panel:\/} Model-predicted bulk density 
  to match the observed value of the semimajor axis drift shown by the
  grayscale region. The solid line in the middle shows the exact correspondence surrounded by
  a map of the sigma interval of the Yarkovsky drift. The nominal effective size of
  $1$~km is used. If this value were higher or lower, the density solution would shift
  in the direction indicated by the arrows (preserving the $\rho_{\rm b}\,D$ value).
  The analysis of ten different shape models of (161989) Cacus reveals a variation of $\pm 3$\% about the median value used in the figure.  
  The effects of small-scale surface roughness, if important, would shift the solution toward a lower
  value by typically $10-30$\% \citep[see][]{rg2012}.}
 \end{figure}
%%%%%%%%%%%%%%%%%%%%%%%%%%%%%%%%%%%%%%%%%%%%%%%%%%%%%%%%%%%%%%%%%%%%%%%%%%%%%%%%%%%%%%

The same simulations that provided satisfactory results for the Yarkovsky effect were
also used to compute the YORP effect. The light-curve data analysis resulted in a low and positive value $\upsilon\simeq
5.5\times 10^{-8}$\,rad\,d$^{-2}$, indicating that the rotation rate of 2000~WN10 increases slowly.
However, using the obtained shape model, the bulk density of $2$\,g\,cm$^{-3}$, and the effective size
of $300$~m, our simplified nominal model provides $\upsilon_{\rm model}\simeq -8.4\times 10^{-8}$\,rad\,d$^{-2}$.
This is a similar mismatch as for Ra-Shalom, presented in Sec.~\ref{the:rasha}.
Here, however, we find a much stronger variation of the predicted  $\upsilon_{\rm model}$ values computed for $11$ possible solutions from light-curve fitting: $-15\times 10^{-7}$\,rad\,d$^{-2}$ to 
$26\times 10^{-7}$\,rad\,d$^{-2}$. Clearly, the sample of resulting convex shape models cannot predict the YORP effect accurately. This
likely originates from two effects: (i) the very low
obliquity of $\simeq 10^\circ$ means that the polar flattening of the shape model from the light-curve inversion is only poorly resolved,
and (ii) the coorbital nature of the 2000~WN10 orbit inherently limits the viewing geometry for
an Earth-bound observed for subsequent seasons \citep[see discussion in][]{yorp12007}. 
Because of the low obliquity, the solar aspect angle of our observations has been always limited to between 82 and $88^\circ$. The shape-modeling procedure clearly interpolates the surface even in the zone that was poorly sampled by the observations. This produces a huge missing piece of information, however, that would be needed to determine $\upsilon_{\rm model}$. At this moment, we must satisfy ourselves by observing that $\upsilon$ and $\upsilon_{\rm model}$ are generally on the same order of magnitude (or at least within the expected limits).

% ----------------------------------------------------
\subsection{(161989) Cacus} \label{the:cacus}
This asteroid was discovered during its close encounter with the Earth in
February 1978, but it was subsequently lost until its rediscovery in January
2003 (the close encounter in September 1997 was missed because the position of Cacus was at a
low declination on the southern sky). Nevertheless, the 1978 apparition is important
by providing an astrometric tie point, but mainly by providing the early
photometric observations \citep{dege1978,schu1979}. Cacus has been regularly 
observed since 2003, and both astrometry and photometry were accumulated.
The radar observations
taken from the Goldstone DSS-14 dish in August and September 2022 have significantly increased the orbit accuracy. A low-accuracy
detection of the YORP effect was reported by \citet{Dur.ea:18a} and was improved
in this paper by adding photometric observations taken in 2022. The post-radar orbit
also allows us to constrain the Yarkovsky effect fairly well, that is, the semimajor axis
secular drift of $\dadt = -(4.45\pm 0.78)\times 10^{-4}$\,au\,My$^{-1}$. Cacus
was also fortuitously observed by the WISE spacecraft in its cryogenic phase
\citep{wise2011}, which allowed \citet{Dur.ea:18a} to determine its effective size
$D=1.0\pm 0.2$~km and the most likely range of the surface thermal inertia 
$\Gamma = 500-800$~\sig. Finally, \citet{thomas2014} and \citet{bin2019} reported
spectroscopic observations, classifying Cacus consistently as a Q- or S-class
object. Overall, in a quite short interval of time, Cacus has been elevated to the rank
of NEAs with fairly complete information.

We used the best-fit spin state and shape model found in Sec.~\ref{obs:cacus}, together with the
assumed effective size of $1$~km, to compute the Yarkovsky effect for a plausible range of
surface thermal inertia and bulk density values. The results are shown in Fig.~\ref{fig_cacus}.
The observed semimajor axis drift $\dadt = -(4.45\pm 0.78)\times 10^{-4}$\,au\,My$^{-1}$
is well compatible with a bulk density between $1.7$ and $2.4$\,g\,cm$^{-3}$ (Fig.~\ref{fig_cacus}).
Because the taxonomic class is similar to (1862) Apollo and (138852) 2000~WN10, the inferred bulk
density for Cacus appears to be acceptable (a slightly smaller size and/or lower thermal
inertia may push the value higher and close to the value inferred for Apollo or Golevka).

As above, we were less successful when we tried to match the observed acceleration of the
rotation rate $\upsilon\simeq 1.94\times 10^{-8}$\,rad\,d$^{-2}$ (Sec.~\ref{obs:cacus}). Using the nominal model
as above for the Yarkovsky effect (and specifically, $2$\,g\,cm$^{-3}$ bulk density), we obtained a median $\upsilon_{\rm model}\simeq -2\times 10^{-8}$\,rad\,d$^{-2}$ and a full range between $-20\times 10^{-8}$\,rad\,d$^{-2}$ and
$18\times 10^{-8}$\,rad\,d$^{-2}$ for a sample of ten variant shape or spin solutions.
Similarly to the case of 2000~WN10, we suspect that the overall uncertainty in the shape model plays a fundamental role. The obliquity of Cacus is basically
$180^\circ$, implying that the spin axis stretch of the model, as well as other details of the shape model along this direction, are very loosely constrained. 
This means that the result strongly depends on model details, even at medium
and large scales. At this moment, the data simply do not provide us with an accurate enough
model to firmly predict the $\upsilon_{\rm model}$ value.

% ---------------------------------------------------
\section{Discussion and conclusions} \label{sec:discussion}

  \begin{table*}[h]
    \caption{\label{tab:YORP_detections} 
      YORP detections as of August 2023. The first segment of 12 objects corresponds to highly reliable cases. The two additional objects at the end of the table list candidate cases for which additional observations in the future will likely result in a YORP detection.}
      \centering
      \begin{tabular}{r l c c d c c c c r}
        \hline \hline
        \multicolumn{2}{c}{Asteroid} & \rule{0pt}{2ex} $\upsilon$ & $\upsilon/\delta\upsilon$ & \multicolumn{1}{c}{$C_Y$}     & $P$   & $D$   & $\epsilon$    & Reference \\
                &    & [$10^{-8}\,\radd$]      &   &         & [h]   & [km]  & [deg]         & \\ [1pt]
        \hline
        \multicolumn{9}{c}{\it \rule{0pt}{3ex} -- YORP detections --} \\  
\rule{0pt}{3ex}        
        1620 & Geographos   & $1.14 \pm 0.03$   & 38.0  & 0.013     &  $\phantom{15}5.2233360$        & $\phantom{2}2.56$  & 154   & (1)   \\ 
        1685 & Toro         & $0.33 \pm 0.03$   & 11.0  &  0.0080   &  10.197826        & 3.5   & 160   & (1)   \\
        1862 & Apollo       & $4.94 \pm 0.09$   & 54.9  &  0.025    &  $\phantom{15}3.0654226$        & $\phantom{2}1.55$ & 163  & This work \\
        2100 & Ra-Shalom    & $0.29 \pm 0.2\phantom{9}$    & $\phantom{5}1.5$   &  0.0011   &  19.820072        & 2.3   & 158   & This work \\
        3103 & Eger         & $1.1\phantom{9} \pm 0.5\phantom{9}$     & $\phantom{5}2.2$   &  0.0077   & $\phantom{5}5.710156$     & $\phantom{2}1.78$  & 176 & (2) \\  
        10115 & 1992 SK     & $8.3\phantom{9} \pm 0.6\phantom{9}$     & 13.8  &  0.014    &  $\phantom{5}7.320232$  & 1.0   & 161   & (1)   \\
        25143 & Itokawa     & $3.54 \pm 0.38$   & $\phantom{5}9.3$   &  0.00072  & 12.132371         &  $\phantom{1}0.32$ & 178 & (3) \\
        54509 & YORP        & $350 \pm 35\phantom{0}$  & 10.0 & 0.0053  &  $\phantom{555}0.20290020$  & $\phantom{15}0.114$ & 173 & (4,5) \\
        68346 & 2001~KZ66   & $8.43 \pm 0.69$   & 12.2 & 0.013 & $\phantom{1}4.985997$ & $\phantom{15}0.797$ & 159 & (6) \\   
        101955 & Bennu      & $6.34 \pm 0.91$   & $\phantom{5}7.0$   & 0.0010    & $\phantom{1}4.296057$  & $\phantom{15}0.490$ & 178 & (7) \\ 
        138852 & 2000 WN10  & $5.5\phantom{9} \pm 0.7\phantom{9}$     & $\phantom{5}7.9$   &  0.00056  & $\phantom{51}4.4636677$  & 0.3   & $\phantom{1}10$   & This work \\
        161989 & Cacus      & $1.86 \pm 0.09$   & 20.7  &  0.0027   & $\phantom{55}3.7550527$         &  1.0  & 179  & This work\\ [6pt]
             \multicolumn{9}{c}{\it -- Possible, but weak YORP signal --} \\  \rule{0pt}{3ex}
        85989 & 1999~JD6    & $-0.5 \pm 0.4$    & $\phantom{55}1.25$  & 0.00083   & $\phantom{1}7.664354$          & 1.53  & 160  & This work\\ 
        85990 & 1999~JV6    & $\phantom{-}3.1 \pm 2.4$     & $\phantom{5}1.3$   & 0.00069  & $\phantom{1}6.536787$ & $\phantom{5}0.442$  &   174 & (8) \\ [2pt]        
        \hline
      \end{tabular}
      \tablefoot{The first two columns provide the number and designation of the asteroid. The third column gives the change in the secular rotation-rate $\omega$, $\upsilon=d\omega/dt$, which is empirically determined from the observations, and the fourth column gives the statistical significance of $\upsilon$. The fifth column lists the parameter $C_Y$ introduced in Eq.~(1) of \citet{rg2013MNRASdet} \citep[see also][]{retal2009}. Columns 5 to 8 give the rotation period $P$ (at the epoch of detection), the size $D$, and the obliquity $\epsilon$. The last column provides the source.}
      \tablebib{(1)~\cite{Dur.ea:22}, (2)~\cite{Dur.ea:18a},      
      (3)~\cite{ito2014}, (4)~\cite{yorp12007}, (5)~\cite{yorp22007},
      (6)~\citet{zeg2021}, (7)~\cite{her2019}, (8)~\citet{roz2019}}
    \end{table*}
    
It has become tradition over the past years to conclude a
paper that reported a new detection of the YORP effect by recalling the
previous detections and noting that in all cases, the rotation rate 
increases (i.e., $\upsilon$ is positive for all asteroids). We continue this tradition. For brevity, we recall Table~3 of
\citet{tian2022}, who listed the relevant last
YORP detections. Our work represents two modifications:
(i) an extension by two new detections for (2100) Ra-Shalom and
(138852) 2000~WN10, but (ii) also one retraction, at least as far as
the situation stands now, namely a nondetection of YORP for (85989)
1999~JD6. Taken together, the YORP effect has been detected for 12 asteroids, and the rotation rate has indeed been found to increase in all cases.\footnote{Ra-Shalom spins up 
 contrary to expectations from \citet{Dur.ea:18a}. Another YORP-detection candidate for which
 previous analysis predicted the rotation rate to slow down
 is (1917) Cuyo \citep[see][]{rozek2019}. In this case, it would also have the first YORP detection for an Amor-class NEA (i.e., perihelion
 higher than 1\,au).} 
 Based on our new results, we provide a list of YORP detections in Table~\ref{tab:YORP_detections}. In addition to the spin parameters and the size, the table also lists the parameter $C_Y$, which is a nondimensional coefficient introduced by \citet{rg2013MNRASdet} \citep[see also][]{retal2009} to describe the YORP strength for a particular asteroid. The $C_Y$ parameter is the $\upsilon$ parameter normalized for the semimajor axis $a$, the eccentricity $e$, the size $D$, and the density $\rho$ (we used the same density of $2500$\,kg\,cm$^{-3}$ according to the formula
 \begin{equation}
 C_Y = \upsilon\, \frac{a^2 \sqrt{1-e^2} \,\rho \,D^2}{G_1}\,,
 \end{equation}
 where $G_1 \approx 6.4 \times 10^{16}$\,kg\,m\,s$^{-2}$ is a modified constant of solar radiation ($2/\pi$ times the solar radiation pressure per unit area at a unit distance $1 \times 10^{17}$\,kg\,m\,s$^{-2}$). The values of $C_Y$ allowed us to directly compare the YORP strength after removing the effects of the size and heliocentric distance. Apollo has the highest value of $C_Y = 0.025$, which is also likely the reason for the good agreement of its theoretical value of $\upsilon_\mathrm{model}$ with the observed value and why it is stable with respect to bootstrap shape variants.  For the other three asteroids we modeled in this work, Ra-Shalom, 2000~WN10, and Cacus, $C_Y$ is smaller by at least one order of magnitude. This means that the detected YORP is weak and thus is more sensitive to the shape details. The $\upsilon_\mathrm{model}$ values are therefore spread much more widely for these three cases.
 
While not an exception from the rule, we
may mention for the sake of interest that (138852) 2000~WN10 is the first case
in which the YORP effect has been determined for a prograde-rotating asteroid.
This is expected, however, because the theory does not
expect any difference between the prograde- and retrograde-rotating
cases. Retrograde rotators predominate in the NEA population and represent a fraction of
about two-thirds of the total. This is confirmed by direct observations
of their rotation pole \citep[e.g.,][]{spina2004,kry2007}, but it is also fairly
well reflected in the statistics of the Yarkovsky detections
\citep[e.g.,][]{far2013,tardioli2017,green2020}. In this respect, it seems a small portion that the YORP effect is detected for only one prograde-rotator in a sample of 12 objects (if YORP acts equally on prograde- and retrograde-rotating bodies).
Using a binomial distribution, we find that the probability that this occurs by chance is $\simeq 5$\%. As far as we know, no apriori selection bias exists to
preferentially detect the YORP effect for retrograde-rotating NEAs. As a result,
the situation may still indicate that the sample of NEAs
for which the YORP effect was detected is still too small to draw bold conclusions.

Following the same line of argument, the binomial distribution may indicate
the likelihood that none of the YORP detections revealed negative $\upsilon$ if the model were not to give a preference to its sign. Simple
algebra provides a probability of only $0.05$\% that 12 cases have the same sign of $\upsilon$ by chance alone. This would all become more
realistic if the model were to give a preference to spinning up, rather
than spinning down, by YORP.  \citet{gk2012} discovered this
hidden element in the theory by accounting for the lateral heat conduction
in the small-scale surface features of the asteroid \citep[an interesting attempt
for a combined model of the YORP effect may be found in][]{gs2019}. If the still
uncertain fraction to accelerate the rotation rate is about $70-80$\%, 
the chance of missing a deceleration case out of 12 trials would increase
to $\simeq (1.5-7)$\%. Here again, we conclude that the sample
of asteroids in which the YORP effect has been detected is small. This motivates
further efforts to search for asteroids for which the YORP effect may be
detected. Of particular interest may be a future follow-up of (85989) 1999~JD6.
If the tendency from our solution in Sec.~\ref{obs:jd6} is confirmed, it may become the first
asteroid for which the detected YORP effect decelerates the rotation rate.

% ......
Our new detections add new insight into the properties of the sample of objects for which the YORP effect has been determined in several respects. For instance,
$\upsilon = (2.9 \pm 2.0)\times 10^{-9}\,\radd$ for (2100)~Ra-Shalom is the lowest value
detected so far. This confirms that it might be possible to extend YORP detections to large asteroids,
eventually even beyond the near-Earth asteroid group if enough data are available in the
future. Moreover, Ra-Shalom has the longest rotation period for which the YORP effect has been detected so far. At the same time, the Ra-Shalom YORP detection came together with the detection of the spin-axis precession. In this respect, we note that the detection of the YORP effect for an asteroid in a tumbling state is yet another novelty. 
An indication of such detection can be found in the analysis of the rotation state of the very small asteroid 2012~TC4 \citep[see][]{lee2021}. 
The accurately monitored rotation state of (99942)~Apophis \citep[e.g.][]{Pra.ea:14,lee2022,dur.apo.2022} might indicate that this is a very suitable candidate, especially if high-quality data were obtained from both space-borne and ground-based observations before its very close approach to the Earth in April~2029. 

The accuracy of the observationally determined YORP strength for (1862) Apollo,
expressed by a value $\upsilon/\delta\upsilon\simeq 55$ for the signal-to-noise ratio, is
superior to all currently known cases \citep[followed with not much worse solution
for (1620) Geographos; see][]{Dur.ea:22}. Future observations, maybe within the next
decade or two,
will certainly continue to improve the solution. Additionally, for asteroids as large
as (1862)~Apollo, these observations do not require large instruments: $0.5-1$~m
scale telescopes may be suitable. As an example, if a good-quality light-curve observation of (1862) Apollo is obtained during its close approach in October 2030, the signal-to-noise ratio of the  YORP detection would increase to $\simeq 70-75$.
The question arises, however, whether this is even
needed and what the scientific justification would be.

A straightforward answer is provided by continuing a high-accuracy theoretical
modeling work along the line of \citet{rozitis2013}. If the theoretical prediction
of the Yarkovsky and YORP strength could still be improved,
spectacular constraints might be derived on parameters such as the bulk density
\citep[see][ where only the Yarkovsky side was considered]{far2021}. In cases like
(2100)~Ra-Shalom (Sec.~\ref{obs:rasha}), (1620)~Geographos, and/or (1685)~Toro \citep{Dur.ea:22},
further observations are clearly motivated to better decorrelate the YORP
effect (thus also to improve its accuracy) from the effect of the regular
precession of the spin axis. The observationally determined value of the precession constant
may also contribute significantly to the consistency of the whole approach and might
even help to answer interesting scientific questions (e.g.,
improve the shape model or suggest a possible inhomogeneity of the density
distribution). One additional aspect of the YORP detection
methods that have been used so far may be clarified in view of the prospects for further
significant improvements in the detection accuracy.
% FIG %%%%%%%%%%%%%%%%%%%%%%%%%%%%%%%%%%%%%%%%%%%%%%%%%%%%%%%%%%%%%%%%%%%%%%%%%%%%%%%%
\begin{figure}[t]
 \begin{center} 
 \includegraphics[width=\columnwidth]{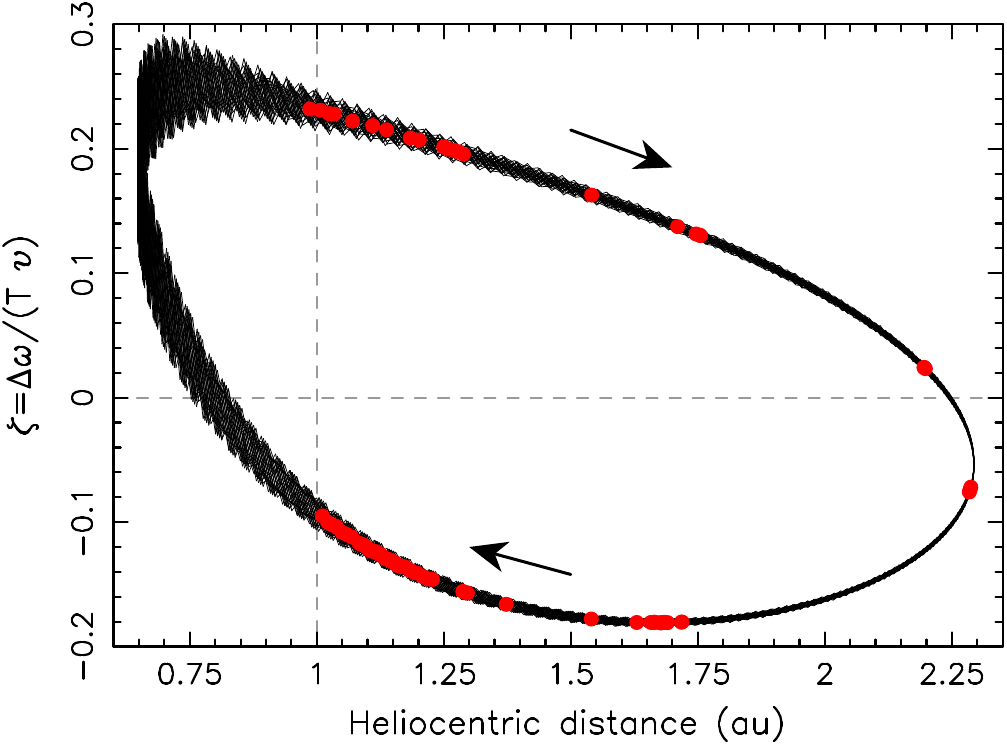}
 \end{center}
 \caption{\label{fig_domega}
  Nondimensional factor $\zeta$ (on the ordinate) related to the periodic part
  of the rotation frequency change $\Delta\omega$ (see the text), expressed as a
  function of the heliocentric distance (at the abscissa). The sense of motion
  is clarified by the arrows, which imply that the upper branch of the curve
  corresponds to the motion from perihelion to aphelion, and vice versa for the
  lower branch of the curve. This prediction corresponds to the specific case
  of (1862)~Apollo; the secular part with the $\upsilon$ value determined in
  Sec.~\ref{obs:apollo} has been subtracted from the rotation-rate $\omega(t)$ evolution and used
  for the normalization of the $\zeta$ function. The high eccentricity ($\simeq
  0.56$) of the Apollo orbit implies that the amplitude of the effect is not too small.
  The red symbols indicate the epochs along the orbits for which light-curve observations
  are available. Some of the observations were taken during close encounters on
  extreme values of phase angle exceeding $90^\circ$ at heliocentric distances smaller than
  $1$\,au (indicated by the vertical dashed line). However, Apollo is large enough to
  also be easily observable near the aphelion of its orbit at a heliocentric distance of
  $2.29$\,au (see also Table~\ref{tab:aspect_1862}).}
 \end{figure}
%%%%%%%%%%%%%%%%%%%%%%%%%%%%%%%%%%%%%%%%%%%%%%%%%%%%%%%%%%%%%%%%%%%%%%%%%%%%%%%%%%%%%%

The baseline model with the constant sidereal rotation frequency $\omega(t)=\omega_0$
has always been generalized in the simplest way, namely using the linear model $\omega(t)=
\omega_0+\upsilon t$ so far (implying that $\omega_0$ only holds at some arbitrary epoch
$t=0$). However, when the asteroid orbit is highly eccentric, all this is
slightly more general because the YORP torque depends at any moment along
the orbits on the instantaneous heliocentric distance, on the orientation of the
body frame with respect to the heliocentric position vector, and on other factors
\citep[e.g.,][]{rub2000,cv2004}. The
empirical factor $\upsilon$ in the formula above is only the result of averaging the
YORP effect over one revolution about the Sun. As a result, the correct model for
the rotation frequency evolution in time should read $\omega(t)=\omega_0+\upsilon t+
\Delta \omega(t)$, where $\Delta \omega(t)$ is a periodic function with the
baseline periodicity $T$ of the asteroid revolution about the Sun. We may rearrange
the formula to provide the variation in the rotation rate, so that it reads $\omega(t)-\omega_0 =
\upsilon \left[t+ T\,\zeta\left(t\right)\right]$, where we introduced a nondimensional
factor $\zeta(t)= \Delta \omega(t)/(T\,\upsilon)$. Figure~\ref{fig_domega} shows
$\zeta$ as a function of heliocentric distance rather than time, which is possible
because the heliocentric motion on a fixed ellipse is periodic (here we clearly may
neglect planetary perturbations). We note that the variation has a non-negligible
amplitude of $\simeq 0.3$. However, the principal point is that it does not accumulate
in time. At maximum, its positive value lasts half of the orbital period $T/2$, and
the variation in the rotation frequency is $\simeq \upsilon\,\zeta_{\rm max}\,T/2$,
producing an accumulated phase shift of $\simeq \upsilon\,\zeta_{\rm max}\,T^2/4$. For (1862)~Apollo, this amounts to $\simeq 0.1^\circ$. By repeating the same calculation for (101955) Bennu, we found that the periodic part of the YORP signal only has an amplitude of $\simeq 0.006^\circ$ (because Bennu has a lower eccentricity and shorter orbital period). This is likely too small even for the very accurate observations of the OSIRIS-REx spacecraft mission.

We may conclude that in the case
of large asteroids, such as Apollo, with currently reachable YORP detection, we may
only rely on the secular change in the rotation rate $\upsilon$, neglecting
variations in the instantaneous radiation torques along the orbit. The situation would
be reversed in a hypothetical experiment, when an extremely strong torque would
affect the rotation of a very small body and the available observations would cover
a timescale comparable to or even shorter than the revolution period $T$ about the
Sun. The future may perhaps offer this interesting possibility.

% REFERENCES %%%%%%%%%%%%%%
  \begin{acknowledgements}
    This work has been supported by the Grant Agency of the Czech Republic, grants 20-04431S and 23-04946S. The work at Modra was supported by the Slovak Grant Agency for Science VEGA, Grant 1/0530/22. YK thanks for supporting the European Federation of Academies of Sciences and Humanities (grants ALLEA EFDS-FL1-18). AR and CS acknowledge support from the UK Science and Technology Facilities Council. This project has received funding from the European Union’s Horizon 2020 research and innovation programme under grant agreement No 870403 (NEOROCKS). DP is thankful for the Wise Observatory staff.
  \end{acknowledgements}

%  \bibliographystyle{aa}
%  \bibliography{bibliography_all, lit}

\begin{thebibliography}{119}
\expandafter\ifx\csname natexlab\endcsname\relax\def\natexlab#1{#1}\fi

\bibitem[{{Archinal} {et~al.}(2018){Archinal}, {Acton}, {A'Hearn}, {Conrad},
  {Consolmagno}, {Duxbury}, {Hestroffer}, {Hilton}, {Kirk}, {Klioner},
  {McCarthy}, {Meech}, {Oberst}, {Ping}, {Seidelmann}, {Tholen}, {Thomas}, \&
  {Williams}}]{archie2018}
{Archinal}, B.~A., {Acton}, C.~H., {A'Hearn}, M.~F., {et~al.} 2018, Celestial
  Mechanics and Dynamical Astronomy, 130, 22

\bibitem[{{Bertotti} {et~al.}(2003){Bertotti}, {Farinella}, \&
  {Vokrouhlick{\'y}}}]{bfv2003}
{Bertotti}, B., {Farinella}, P., \& {Vokrouhlick{\'y}}, D. 2003, {Physics of
  the Solar System - Dynamics and Evolution, Space Physics, and Spacetime
  Structure.} (Kluwer Academic Press, Dordrecht)

\bibitem[{{Binzel} {et~al.}(2019){Binzel}, {DeMeo}, {Turtelboom}, {Bus},
  {Tokunaga}, {Burbine}, {Lantz}, {Polishook}, {Carry}, {Morbidelli}, {Birlan},
  {Vernazza}, {Burt}, {Moskovitz}, {Slivan}, {Thomas}, {Rivkin}, {Hicks},
  {Dunn}, {Reddy}, {Sanchez}, {Granvik}, \& {Kohout}}]{bin2019}
{Binzel}, R.~P., {DeMeo}, F.~E., {Turtelboom}, E.~V., {et~al.} 2019, \icarus,
  324, 41

\bibitem[{{Breiter} {et~al.}(2009){Breiter}, {Bartczak}, {Czekaj}, {Oczujda},
  \& {Vokrouhlick{\'y}}}]{yorpito2009}
{Breiter}, S., {Bartczak}, P., {Czekaj}, M., {Oczujda}, B., \&
  {Vokrouhlick{\'y}}, D. 2009, \aap, 507, 1073

\bibitem[{{Breiter} \& {Michalska}(2008)}]{bm2008}
{Breiter}, S. \& {Michalska}, H. 2008, \mnras, 388, 927

\bibitem[{{Breiter} {et~al.}(2005){Breiter}, {Nesvorn{\'y}}, \&
  {Vokrouhlick{\'y}}}]{bnv2005}
{Breiter}, S., {Nesvorn{\'y}}, D., \& {Vokrouhlick{\'y}}, D. 2005, \aj, 130,
  1267

\bibitem[{{Brosch} {et~al.}(2015){Brosch}, {Kaspi}, {Niv}, \&
  {Manulis}}]{Bro.ea:15}
{Brosch}, N., {Kaspi}, S., {Niv}, S., \& {Manulis}, I. 2015, \apss, 359, 9

\bibitem[{{Bus} \& {Binzel}(2002)}]{bb2002}
{Bus}, S.~J. \& {Binzel}, R.~P. 2002, \icarus, 158, 146

\bibitem[{{Campins} {et~al.}(2009){Campins}, {Kelley}, {Fern{\'a}ndez},
  {Licandro}, \& {Hargrove}}]{camp2009}
{Campins}, H., {Kelley}, M.~S., {Fern{\'a}ndez}, Y., {Licandro}, J., \&
  {Hargrove}, K. 2009, Earth Moon and Planets, 105, 159

\bibitem[{{\SortNoop{Capek2004}{\v{C}}apek} \&
  {Vokrouhlick{\'y}}(2004)}]{cv2004}
{\SortNoop{Capek2004}{\v{C}}apek}, D. \& {Vokrouhlick{\'y}}, D. 2004, \icarus,
  172, 526

\bibitem[{{\SortNoop{Capek2005}{\v{C}}apek} \&
  {Vokrouhlick{\'y}}(2005)}]{cv2005}
{\SortNoop{Capek2005}{\v{C}}apek}, D. \& {Vokrouhlick{\'y}}, D. 2005, in IAU
  Colloq. 197: Dynamics of Populations of Planetary Systems, ed.
  Z.~{Kne{\v{z}}evi{\'c}} \& A.~{Milani}, 171--178

\bibitem[{{Carry} {et~al.}(2016){Carry}, {Solano}, {Eggl}, \&
  {DeMeo}}]{carry2016}
{Carry}, B., {Solano}, E., {Eggl}, S., \& {DeMeo}, F.~E. 2016, \icarus, 268,
  340

\bibitem[{{Chandrasekhar}(1995)}]{chandra}
{Chandrasekhar}, S. 1995, {Newton'Principia for the Common Reader} (Oxford
  University Press, Oxford)

\bibitem[{{Chesley} {et~al.}(2016){Chesley}, {Farnocchia}, {Pravec}, \&
  {Vokrouhlick{\'y}}}]{ches2016}
{Chesley}, S.~R., {Farnocchia}, D., {Pravec}, P., \& {Vokrouhlick{\'y}}, D.
  2016, in Asteroids: New Observations, New Models, ed. S.~R. {Chesley},
  A.~{Morbidelli}, R.~{Jedicke}, \& D.~{Farnocchia}, Vol. 318, 250--258

\bibitem[{{Chesley} {et~al.}(2003){Chesley}, {Ostro}, {Vokrouhlick{\'y}},
  {{\v{C}}apek}, {Giorgini}, {Nolan}, {Margot}, {Hine}, {Benner}, \&
  {Chamberlin}}]{gol2003}
{Chesley}, S.~R., {Ostro}, S.~J., {Vokrouhlick{\'y}}, D., {et~al.} 2003,
  Science, 302, 1739

\bibitem[{{Chesley} {et~al.}(2008){Chesley}, {Vokrouhlick{\'y}}, {Ostro},
  {Benner}, {Margot}, {Matson}, {Nolan}, \& {Shepard}}]{ches2008}
{Chesley}, S.~R., {Vokrouhlick{\'y}}, D., {Ostro}, S.~J., {et~al.} 2008, in LPI
  Contributions, Vol. 1405, Asteroids, Comets, Meteors 2008, ed. {LPI Editorial
  Board}, 8330

\bibitem[{{Colombo}(1966)}]{col1966}
{Colombo}, G. 1966, \aj, 71, 891

\bibitem[{{D'Alembert}(1749)}]{alem1749}
{D'Alembert}, J. 1749, {Recherches sur la pr{\'{e}}cession des equinoxes, et
  sur la nutation de l'axe de la terre, dans le systeme Newtonien} (Michel
  Antoine David, Paris)

\bibitem[{{Degewij} {et~al.}(1978){Degewij}, {Lebofsky}, \&
  {Lebofsky}}]{dege1978}
{Degewij}, J., {Lebofsky}, L., \& {Lebofsky}, M. 1978, \iaucirc, 3193, 1

\bibitem[{{Del Vigna} {et~al.}(2018){Del Vigna}, {Faggioli}, {Milani}, {Spoto},
  {Farnocchia}, \& {Carry}}]{vigna2018}
{Del Vigna}, A., {Faggioli}, L., {Milani}, A., {et~al.} 2018, \aap, 617, A61

\bibitem[{{Delb{\'o}} {et~al.}(2003){Delb{\'o}}, {Harris}, {Binzel}, {Pravec},
  \& {Davies}}]{delbo2003}
{Delb{\'o}}, M., {Harris}, A.~W., {Binzel}, R.~P., {Pravec}, P., \& {Davies},
  J.~K. 2003, \icarus, 166, 116

\bibitem[{{Delb{\'o}} {et~al.}(2015){Delb{\'o}}, {Mueller}, {Emery}, {Rozitis},
  \& {Capria}}]{delb2015}
{Delb{\'o}}, M., {Mueller}, M., {Emery}, J.~P., {Rozitis}, B., \& {Capria},
  M.~T. 2015, in Asteroids IV, ed. P.~{Michel}, F.~E. {DeMeo}, \& W.~F.
  {Bottke}, 107--128

\bibitem[{{\v{D}urech} {et~al.}(2012{\natexlab{a}}){\v{D}urech}, {Delbo}, \&
  {Carry}}]{Dur.ea:12}
{\v{D}urech}, J., {Delbo}, M., \& {Carry}, B. 2012{\natexlab{a}}, LPI
  Contributions, 1667, 6118

\bibitem[{{\v{D}urech} {et~al.}(2012{\natexlab{b}}){\v{D}urech},
  {Vokrouhlick{\'y}}, {Baransky}, {Breiter}, {Burkhonov}, {Cooney}, {Fuller},
  {Gaftonyuk}, {Gross}, {Inasaridze}, {Kaasalainen}, {Krugly}, {Kvaratshelia},
  {Litvinenko}, {Macomber}, {Marchis}, {Molotov}, {Oey}, {Polishook},
  {Pollock}, {Pravec}, {S{\'a}rneczky}, {Shevchenko}, {Slyusarev}, {Stephens},
  {Szab{\'o}}, {Terrell}, {Vachier}, {Vanderplate}, {Viikinkoski}, \&
  {Warner}}]{Dur.ea:12b}
{\v{D}urech}, J., {Vokrouhlick{\'y}}, D., {Baransky}, A.~R., {et~al.}
  2012{\natexlab{b}}, \aap, 547, A10

\bibitem[{{\v{D}urech} {et~al.}(2008){\v{D}urech}, {Vokrouhlick\'y},
  {Kaasalainen}, {Weissman}, {Lowry}, \& {Beshore}}]{Dur.ea:08}
{\v{D}urech}, J., {Vokrouhlick\'y}, D., {Kaasalainen}, M., {et~al.} 2008, \aap,
  488, 345

\bibitem[{{\v{D}urech} {et~al.}(2018){\v{D}urech}, {Vokrouhlick{\'y}},
  {Pravec}, {Hanu{\v s}}, {Farnocchia}, {Krugly}, {Inasaridze}, {Ayvazian},
  {Fatka}, {Chiorny}, {Gaftonyuk}, {Gal{\'a}d}, {Groom}, {Hornoch}, {Ku{\v
  c}{\'a}kov{\'a}}, {Ku{\v s}nir{\'a}k}, {Lehk{\'y}}, {Kvaratskhelia}, {Masi},
  {Molotov}, {Oey}, {Pollock}, {Shevchenko}, {Vra{\v s}til}, \&
  {Warner}}]{Dur.ea:18a}
{\v{D}urech}, J., {Vokrouhlick{\'y}}, D., {Pravec}, P., {et~al.} 2018, \aap,
  609, A86

\bibitem[{{\v{D}urech} {et~al.}(2022{\natexlab{a}}){\v{D}urech},
  {Vokrouhlicky}, {Pravec}, {Hornoch}, {Kusnirak}, {Fatka}, \&
  {Kucakova}}]{dur.apo.2022}
{\v{D}urech}, J., {Vokrouhlicky}, D., {Pravec}, P., {et~al.}
  2022{\natexlab{a}}, in European Planetary Science Congress, EPSC2022--36

\bibitem[{{\v{D}urech} {et~al.}(2022{\natexlab{b}}){\v{D}urech},
  {Vokrouhlick{\'y}}, {Pravec}, {Krugly}, {Kim}, {Polishook}, {Ayvazian},
  {Bonev}, {Choi}, {Datashvili}, {Donchev}, {Ehgamberdiev}, {Hornoch},
  {Inasaridze}, {Kapanadze}, {Kim}, {Ku{\v{c}}{\'a}kov{\'a}}, {Kusakin},
  {Ku{\v{s}}nir{\'a}k}, {Lee}, {Molotov}, {Moon}, {Mykhailova}, {Nikolenko},
  {Novichonok}, {Oey}, {Omarov}, {Pollock}, {Reva}, {Rumyantsev}, \&
  {Zhornichenko}}]{Dur.ea:22}
{\v{D}urech}, J., {Vokrouhlick{\'y}}, D., {Pravec}, P., {et~al.}
  2022{\natexlab{b}}, \aap, 657, A5

\bibitem[{{Dziadura} {et~al.}(2023){Dziadura}, {Oszkiewicz}, {Spoto}, {Carry},
  {Tanga}, \& {Bartczak}}]{dzia2023}
{Dziadura}, K., {Oszkiewicz}, D., {Spoto}, F., {et~al.} 2023, \aap, in press

\bibitem[{{Farnocchia} {et~al.}(2021){Farnocchia}, {Chesley}, {Takahashi},
  {Rozitis}, {Vokrouhlick{\'y}}, {Rush}, {Mastrodemos}, {Kennedy}, {Park},
  {Bellerose}, {Lubey}, {Velez}, {Davis}, {Emery}, {Leonard}, {Geeraert},
  {Antreasian}, \& {Lauretta}}]{far2021}
{Farnocchia}, D., {Chesley}, S.~R., {Takahashi}, Y., {et~al.} 2021, \icarus,
  369, 114594

\bibitem[{{Farnocchia} {et~al.}(2013){Farnocchia}, {Chesley},
  {Vokrouhlick{\'y}}, {Milani}, {Spoto}, \& {Bottke}}]{far2013}
{Farnocchia}, D., {Chesley}, S.~R., {Vokrouhlick{\'y}}, D., {et~al.} 2013,
  \icarus, 224, 1

\bibitem[{{Goldstein} {et~al.}(1981){Goldstein}, {Jurgens}, \&
  {Yeomans}}]{gold1981}
{Goldstein}, R.~M., {Jurgens}, R.~F., \& {Yeomans}, D.~K. 1981, \icarus, 48, 59

\bibitem[{{Golubov} \& {Krugly}(2012)}]{gk2012}
{Golubov}, O. \& {Krugly}, Y.~N. 2012, \apjl, 752, L11

\bibitem[{{Golubov} \& {Lipatova}(2022)}]{gl2022}
{Golubov}, O. \& {Lipatova}, V. 2022, \aap, 666, A146

\bibitem[{{Golubov} \& {Scheeres}(2019)}]{gs2019}
{Golubov}, O. \& {Scheeres}, D.~J. 2019, \aj, 157, 105

\bibitem[{{Greenberg} {et~al.}(2020){Greenberg}, {Margot}, {Verma}, {Taylor},
  \& {Hodge}}]{green2020}
{Greenberg}, A.~H., {Margot}, J.-L., {Verma}, A.~K., {Taylor}, P.~A., \&
  {Hodge}, S.~E. 2020, \aj, 159, 92

\bibitem[{{Haponiak} {et~al.}(2020){Haponiak}, {Breiter}, \&
  {Vokrouhlick{\'y}}}]{hapo2020}
{Haponiak}, J., {Breiter}, S., \& {Vokrouhlick{\'y}}, D. 2020, Celestial
  Mechanics and Dynamical Astronomy, 132, 24

\bibitem[{{Harris}(1998)}]{harris1998}
{Harris}, A.~W. 1998, \icarus, 131, 291

\bibitem[{{Harris} {et~al.}(1998){Harris}, {Davies}, \&
  {Green}}]{harrisetal1998}
{Harris}, A.~W., {Davies}, J.~K., \& {Green}, S.~F. 1998, \icarus, 135, 441

\bibitem[{{Harris} {et~al.}(1987){Harris}, {Young}, {Goguen}, {Hammel}, {Hahn},
  {Tedesco}, \& {Tholen}}]{harris1987}
{Harris}, A.~W., {Young}, J.~W., {Goguen}, J., {et~al.} 1987, \icarus, 70, 246

\bibitem[{{Hergenrother} {et~al.}(2019){Hergenrother}, {Maleszewski}, {Nolan},
  {Li}, {Drouet D'Aubigny}, {Shelly}, {Howell}, {Kareta}, {Izawa}, {Barucci},
  {Bierhaus}, {Campins}, {Chesley}, {Clark}, {Christensen}, {Dellagiustina},
  {Fornasier}, {Golish}, {Hartzell}, {Rizk}, {Scheeres}, {Smith}, {Zou},
  {Lauretta}, \& {OSIRIS-REx Team}}]{her2019}
{Hergenrother}, C.~W., {Maleszewski}, C.~K., {Nolan}, M.~C., {et~al.} 2019,
  Nature Communications, 10, 1291

\bibitem[{{Ieva} {et~al.}(2018){Ieva}, {Dotto}, {Mazzotta Epifani}, {Perna},
  {Rossi}, {Barucci}, {Di Paola}, {Speziali}, {Micheli}, {Perozzi}, {Lazzarin},
  \& {Bertini}}]{ieva2018}
{Ieva}, S., {Dotto}, E., {Mazzotta Epifani}, E., {et~al.} 2018, \aap, 615, A127

\bibitem[{{Kaasalainen} \& {Torppa}(2001)}]{Kaa.Tor:01}
{Kaasalainen}, M. \& {Torppa}, J. 2001, Icarus, 153, 24

\bibitem[{{Kaasalainen} {et~al.}(2001){Kaasalainen}, {Torppa}, \&
  {Muinonen}}]{Kaa.ea:01}
{Kaasalainen}, M., {Torppa}, J., \& {Muinonen}, K. 2001, Icarus, 153, 37

\bibitem[{{Kaasalainen} {et~al.}(2007){Kaasalainen}, {{\v{D}}urech}, {Warner},
  {Krugly}, \& {Gaftonyuk}}]{apollo2007}
{Kaasalainen}, M., {{\v{D}}urech}, J., {Warner}, B.~D., {Krugly}, Y.~N., \&
  {Gaftonyuk}, N.~M. 2007, \nat, 446, 420

\bibitem[{{Konopliv} {et~al.}(2014){Konopliv}, {Asmar}, {Park}, {Bills},
  {Centinello}, {Chamberlin}, {Ermakov}, {Gaskell}, {Rambaux}, {Raymond},
  {Russell}, {Smith}, {Tricarico}, \& {Zuber}}]{kono2014}
{Konopliv}, A.~S., {Asmar}, S.~W., {Park}, R.~S., {et~al.} 2014, \icarus, 240,
  103

\bibitem[{{Kryszczy{\'n}ska} {et~al.}(2007){Kryszczy{\'n}ska}, {La Spina},
  {Paolicchi}, {Harris}, {Breiter}, \& {Pravec}}]{kry2007}
{Kryszczy{\'n}ska}, A., {La Spina}, A., {Paolicchi}, P., {et~al.} 2007,
  \icarus, 192, 223

\bibitem[{{Kwiatkowski} {et~al.}(2021){Kwiatkowski}, {Kole{\'n}czuk},
  {Kryszczy{\'n}ska}, {Oszkiewicz}, {Kami{\'n}ski}, {Kami{\'n}ska},
  {Troianskyi}, {Skiff}, {Moskowitz}, {Kashuba}, {Kim}, {Kim}, {Mottola},
  {Santana-Ros}, {Kluwak}, {Buzzi}, {Bacci}, {Birtwhistle}, {Miles}, \&
  {Chatelain}}]{Kwi.ea:21}
{Kwiatkowski}, T., {Kole{\'n}czuk}, P., {Kryszczy{\'n}ska}, A., {et~al.} 2021,
  \aap, 656, A126

\bibitem[{{La Spina} {et~al.}(2004){La Spina}, {Paolicchi}, {Kryszczy{\'n}ska},
  \& {Pravec}}]{spina2004}
{La Spina}, A., {Paolicchi}, P., {Kryszczy{\'n}ska}, A., \& {Pravec}, P. 2004,
  \nat, 428, 400

\bibitem[{{Lambert} {et~al.}(2023){Lambert}, {Marchis}, {Hanu{\v{s}}},
  {Archer}, {Billiani}, {Bradley}, {Breeze-Lamb}, {Camilleri}, {Davy}, {Deitz},
  {Donnelly}, {Fairfax}, {Fukui}, {Gamurot}, {Goto}, {Guillet}, {Kardel},
  {Knight}, {Langvad}, {Loose}, {Meneghelli}, {Mitchell}, {Nikiforov},
  {Parker}, {Pickering}, {Primm}, {Randolph}, {Ribas}, {Richardot}, {Rivett},
  {Shimizu}, {Simard}, {Smallen}, {Teng}, {van Dam}, {Verveen}, \&
  {Widi}}]{Lambert2023}
{Lambert}, R., {Marchis}, F., {Hanu{\v{s}}}, J., {et~al.} 2023, Minor Planet
  Bulletin, 50, 16

\bibitem[{{Lebofsky} {et~al.}(1981){Lebofsky}, {Veeder}, {Rieke}, {Lebofsky},
  {Matson}, {Kowal}, {Wynn-Williams}, \& {Becklin}}]{leb1981}
{Lebofsky}, L.~A., {Veeder}, G.~J., {Rieke}, G.~H., {et~al.} 1981, \icarus, 48,
  335

\bibitem[{{Lee} {et~al.}(2022){Lee}, {Kim}, {Marciniak}, {Kim}, {Moon}, {Choi},
  {Zo{\l}a}, {Chatelain}, {Lister}, {Gomez}, {Greenstreet}, {P{\'a}l},
  {Szak{\'a}ts}, {Erasmus}, {Lees}, {Janse van Rensburg}, {Og{\l}oza},
  {Dr{\'o}{\.z}d{\.z}}, {{\.Z}ejmo}, {Kami{\'n}ski}, {Kami{\'n}ska}, {Duffard},
  {Roh}, {Yim}, {Kim}, {Mottola}, {Yoshida}, {Reichart}, {Sonbas}, {Caton},
  {Kaplan}, {Erece}, \& {Yang}}]{lee2022}
{Lee}, H.~J., {Kim}, M.~J., {Marciniak}, A., {et~al.} 2022, \aap, 661, L3

\bibitem[{{Lee} {et~al.}(2021){Lee}, {{\v{D}}urech}, {Vokrouhlick{\'y}},
  {Pravec}, {Moon}, {Ryan}, {Kim}, {Kim}, {Choi}, {Bacci}, {Pollock}, \&
  {Apitzsch}}]{lee2021}
{Lee}, H.-J., {{\v{D}}urech}, J., {Vokrouhlick{\'y}}, D., {et~al.} 2021, \aj,
  161, 112

\bibitem[{{Lowry} {et~al.}(2007){Lowry}, {Fitzsimmons}, {Pravec},
  {Vokrouhlick{\'y}}, {Boehnhardt}, {Taylor}, {Margot}, {Gal{\'a}d}, {Irwin},
  {Irwin}, \& {Kusnir{\'a}k}}]{yorp22007}
{Lowry}, S.~C., {Fitzsimmons}, A., {Pravec}, P., {et~al.} 2007, Science, 316,
  272

\bibitem[{{Lowry} {et~al.}(2014){Lowry}, {Weissman}, {Duddy}, {Rozitis},
  {Fitzsimmons}, {Green}, {Hicks}, {Snodgrass}, {Wolters}, {Chesley},
  {Pittichov{\'a}}, \& {van Oers}}]{ito2014}
{Lowry}, S.~C., {Weissman}, P.~R., {Duddy}, S.~R., {et~al.} 2014, \aap, 562,
  A48

\bibitem[{{Mainzer} {et~al.}(2011){Mainzer}, {Grav}, {Bauer}, {Masiero},
  {McMillan}, {Cutri}, {Walker}, {Wright}, {Eisenhardt}, {Tholen}, {Spahr},
  {Jedicke}, {Denneau}, {DeBaun}, {Elsbury}, {Gautier}, {Gomillion}, {Hand},
  {Mo}, {Watkins}, {Wilkins}, {Bryngelson}, {Del Pino Molina}, {Desai},
  {G{\'o}mez Camus}, {Hidalgo}, {Konstantopoulos}, {Larsen}, {Maleszewski},
  {Malkan}, {Mauduit}, {Mullan}, {Olszewski}, {Pforr}, {Saro}, {Scotti}, \&
  {Wasserman}}]{wise2011}
{Mainzer}, A., {Grav}, T., {Bauer}, J., {et~al.} 2011, \apj, 743, 156

\bibitem[{{Marshall}(2017)}]{marphd}
{Marshall}, S.~E. 2017, PhD thesis, Cornell University, New York

\bibitem[{{Marshall} {et~al.}(2015){Marshall}, {Howell}, {Brozovi{\'c}},
  {Taylor}, {Campbell}, {Benner}, {Naidu}, {Giorgini}, {Jao}, {Lee},
  {Richardson}, {Rodriguez-Ford}, {Rivera-Valentin}, {Ghigo}, {Kobelski},
  {Busch}, {Pravec}, {Warner}, {Reddy}, {Hicks}, {Crowell}, {Fernandez},
  {Vervack}, {Nolan}, {Magri}, {Sharkey}, \& {Bozek}}]{Marsh.ea:15}
{Marshall}, S.~E., {Howell}, E.~S., {Brozovi{\'c}}, M., {et~al.} 2015, in
  AAS/Division for Planetary Sciences Meeting Abstracts, Vol.~47, AAS/Division
  for Planetary Sciences Meeting Abstracts \#47, 204.09

\bibitem[{{Milani} {et~al.}(1989){Milani}, {Carpino}, {Hahn}, \&
  {Nobili}}]{milani1989}
{Milani}, A., {Carpino}, M., {Hahn}, G., \& {Nobili}, A.~M. 1989, \icarus, 78,
  212

\bibitem[{{Monteiro} {et~al.}(2020){Monteiro}, {Silva}, {Tamayo}, {Rodrigues},
  \& {Lazzaro}}]{Mon.ea:20}
{Monteiro}, F., {Silva}, J.~S., {Tamayo}, F., {Rodrigues}, T., \& {Lazzaro}, D.
  2020, \mnras, 495, 3990

\bibitem[{{Morais} \& {Morbidelli}(2002)}]{mm2002}
{Morais}, M.~H.~M. \& {Morbidelli}, A. 2002, \icarus, 160, 1

\bibitem[{{Nesvorn{\'y}} {et~al.}(2010){Nesvorn{\'y}}, {Bottke},
  {Vokrouhlick{\'y}}, {Chapman}, \& {Rafkin}}]{nes2010}
{Nesvorn{\'y}}, D., {Bottke}, W.~F., {Vokrouhlick{\'y}}, D., {Chapman}, C.~R.,
  \& {Rafkin}, S. 2010, \icarus, 209, 510

\bibitem[{{Nesvorn{\'y}} \& {Vokrouhlick{\'y}}(2007)}]{nv2007}
{Nesvorn{\'y}}, D. \& {Vokrouhlick{\'y}}, D. 2007, \aj, 134, 1750

\bibitem[{{Nugent} {et~al.}(2012){Nugent}, {Margot}, {Chesley}, \&
  {Vokrouhlick{\'y}}}]{nugent2012}
{Nugent}, C.~R., {Margot}, J.~L., {Chesley}, S.~R., \& {Vokrouhlick{\'y}}, D.
  2012, \aj, 144, 60

\bibitem[{{Ostro} {et~al.}(2005){Ostro}, {Benner}, {Giorgini}, {Nolan}, {Hine},
  {Howell}, {Margot}, {Magri}, \& {Shepard}}]{ostro2005}
{Ostro}, S.~J., {Benner}, L.~A.~M., {Giorgini}, J.~D., {et~al.} 2005, \iaucirc,
  8627, 2

\bibitem[{{Ostro} {et~al.}(2002){Ostro}, {Rosema}, {Campbell}, \&
  {Shapiro}}]{ostro2002}
{Ostro}, S.~J., {Rosema}, K.~D., {Campbell}, D.~B., \& {Shapiro}, I.~I. 2002,
  \icarus, 156, 580

\bibitem[{{Panfichi} \& {Pajuelo}(2023)}]{Pan.Paj:23}
{Panfichi}, A.~M. \& {Pajuelo}, M. 2023, Minor Planet Bulletin, 50, 128

\bibitem[{{Perna} {et~al.}(2018){Perna}, {Barucci}, {Fulchignoni}, {Popescu},
  {Belskaya}, {Fornasier}, {Doressoundiram}, {Lantz}, \& {Merlin}}]{perna2018}
{Perna}, D., {Barucci}, M.~A., {Fulchignoni}, M., {et~al.} 2018, \planss, 157,
  82

\bibitem[{{Polishook}(2014)}]{Pol:14}
{Polishook}, D. 2014, \icarus, 241, 79

\bibitem[{{Polishook} \& {Brosch}(2008)}]{Pol.Bro:08}
{Polishook}, D. \& {Brosch}, N. 2008, \icarus, 194, 111

\bibitem[{{Pravec} {et~al.}(2014){Pravec}, {Scheirich}, {{\v D}urech},
  {Pollock}, {Ku{\v s}nir{\'a}k}, {Hornoch}, {Gal{\'a}d}, {Vokrouhlick{\'y}},
  {Harris}, {Jehin}, {Manfroid}, {Opitom}, {Gillon}, {Colas}, {Oey}, {Vra{\v
  s}til}, {Reichart}, {Ivarsen}, {Haislip}, \& {LaCluyze}}]{Pra.ea:14}
{Pravec}, P., {Scheirich}, P., {{\v D}urech}, J., {et~al.} 2014, \icarus, 233,
  48

\bibitem[{{Rossi} {et~al.}(2009){Rossi}, {Marzari}, \& {Scheeres}}]{retal2009}
{Rossi}, A., {Marzari}, F., \& {Scheeres}, D.~J. 2009, \icarus, 202, 95

\bibitem[{{Ro{\.z}ek} {et~al.}(2019{\natexlab{a}}){Ro{\.z}ek}, {Lowry},
  {Nolan}, {Taylor}, {Benner}, {Fitzsimmons}, {Zegmott}, {Weissman}, {Green},
  {Rozitis}, {Snodgrass}, {Smythe}, {Hicks}, {Howell}, {Virkki},
  {Aponte-Hernandez}, {Rivera-Valent{\'\i}n}, {Rodriguez-Ford},
  {Zambrano-Marin}, {Brozovi{\'c}}, {Naidu}, {Giorgini}, {Snedeker}, {Jao}, \&
  {Ghigo}}]{roz2019}
{Ro{\.z}ek}, A., {Lowry}, S.~C., {Nolan}, M.~C., {et~al.} 2019{\natexlab{a}},
  \aap, 631, A149

\bibitem[{{Ro{\.z}ek} {et~al.}(2019{\natexlab{b}}){Ro{\.z}ek}, {Lowry},
  {Rozitis}, {Green}, {Snodgrass}, {Weissman}, {Fitzsimmons}, {Hicks},
  {Lawrence}, {Duddy}, {Wolters}, {Roberts-Borsani}, {Behrend}, \&
  {Manzini}}]{rozek2019}
{Ro{\.z}ek}, A., {Lowry}, S.~C., {Rozitis}, B., {et~al.} 2019{\natexlab{b}},
  \aap, 627, A172

\bibitem[{{Rozitis} {et~al.}(2013){Rozitis}, {Duddy}, {Green}, \&
  {Lowry}}]{rozitis2013}
{Rozitis}, B., {Duddy}, S.~R., {Green}, S.~F., \& {Lowry}, S.~C. 2013, \aap,
  555, A20

\bibitem[{{Rozitis} \& {Green}(2012)}]{rg2012}
{Rozitis}, B. \& {Green}, S.~F. 2012, \mnras, 423, 367

\bibitem[{{Rozitis} \& {Green}(2013{\natexlab{a}})}]{rg2013}
{Rozitis}, B. \& {Green}, S.~F. 2013{\natexlab{a}}, \mnras, 433, 603

\bibitem[{{Rozitis} \& {Green}(2013{\natexlab{b}})}]{rg2013MNRASdet}
{Rozitis}, B. \& {Green}, S.~F. 2013{\natexlab{b}}, \mnras, 430, 1376

\bibitem[{{Rubincam}(2000)}]{rub2000}
{Rubincam}, D.~P. 2000, \icarus, 148, 2

\bibitem[{{Scheeres} {et~al.}(2007){Scheeres}, {Abe}, {Yoshikawa}, {Nakamura},
  {Gaskell}, \& {Abell}}]{schee2007}
{Scheeres}, D.~J., {Abe}, M., {Yoshikawa}, M., {et~al.} 2007, \icarus, 188, 425

\bibitem[{{Scheeres} \& {Gaskell}(2008)}]{sg2008}
{Scheeres}, D.~J. \& {Gaskell}, R.~W. 2008, \icarus, 198, 125

\bibitem[{{Scheeres} {et~al.}(2020){Scheeres}, {McMahon}, {Brack}, {French},
  {Chesley}, {Farnocchia}, {Vokrouhlick{\'y}}, {Ballouz}, {Emery}, {Rozitis},
  {Nolan}, {Hergenrother}, \& {Lauretta}}]{sche2020}
{Scheeres}, D.~J., {McMahon}, J.~W., {Brack}, D.~N., {et~al.} 2020, Journal of
  Geophysical Research (Planets), 125, e06284

\bibitem[{{Schuster} {et~al.}(1979){Schuster}, {Surdej}, \&
  {Surdej}}]{schu1979}
{Schuster}, H.~E., {Surdej}, A., \& {Surdej}, J. 1979, \aaps, 37, 483

\bibitem[{{\SortNoop{Sevecek1}{\v S}eve{\v c}ek}
  {et~al.}(2015){\SortNoop{Sevecek1}{\v S}eve{\v c}ek}, {Bro{\v z}}, {{\v
  C}apek}, \& {{\v D}urech}}]{Sev.ea:15}
{\SortNoop{Sevecek1}{\v S}eve{\v c}ek}, P., {Bro{\v z}}, M., {{\v C}apek}, D.,
  \& {{\v D}urech}, J. 2015, \mnras, 450, 2104

\bibitem[{{Shepard} {et~al.}(2008){Shepard}, {Clark}, {Nolan}, {Benner},
  {Ostro}, {Giorgini}, {Vilas}, {Jarvis}, {Lederer}, {Lim}, {McConnochie},
  {Bell}, {Margot}, {Rivkin}, {Magri}, {Scheeres}, \&
  {Pravec}}]{rasharadar2008}
{Shepard}, M.~K., {Clark}, B.~E., {Nolan}, M.~C., {et~al.} 2008, \icarus, 193,
  20

\bibitem[{{Skiff} {et~al.}(2012){Skiff}, {Bowell}, {Koehn}, {Sanborn},
  {McLelland}, \& {Warner}}]{Ski.ea:12}
{Skiff}, B.~A., {Bowell}, E., {Koehn}, B.~W., {et~al.} 2012, Minor Planet
  Bulletin, 39, 111

\bibitem[{{Skogl{\"o}v}(1997)}]{sko1997}
{Skogl{\"o}v}, E. 1997, \planss, 45, 439

\bibitem[{{Skogl{\"o}v}(1998)}]{sko1998}
{Skogl{\"o}v}, E. 1998, \planss, 47, 11

\bibitem[{{Skogl{\"o}v} \& {Erikson}(2002)}]{sko2002}
{Skogl{\"o}v}, E. \& {Erikson}, A. 2002, \icarus, 160, 24

\bibitem[{{Skogl{\"o}v} {et~al.}(1996){Skogl{\"o}v}, {Magnusson}, \&
  {Dahlgren}}]{sko1996}
{Skogl{\"o}v}, E., {Magnusson}, P., \& {Dahlgren}, M. 1996, \planss, 44, 1177

\bibitem[{{Statler}(2009)}]{s2009}
{Statler}, T.~S. 2009, \icarus, 202, 502

\bibitem[{{Tardioli} {et~al.}(2017){Tardioli}, {Farnocchia}, {Rozitis},
  {Cotto-Figueroa}, {Chesley}, {Statler}, \& {Vasile}}]{tardioli2017}
{Tardioli}, C., {Farnocchia}, D., {Rozitis}, B., {et~al.} 2017, \aap, 608, A61

\bibitem[{{Taylor} {et~al.}(2007){Taylor}, {Margot}, {Vokrouhlick{\'y}},
  {Scheeres}, {Pravec}, {Lowry}, {Fitzsimmons}, {Nolan}, {Ostro}, {Benner},
  {Giorgini}, \& {Magri}}]{yorp12007}
{Taylor}, P.~A., {Margot}, J.-L., {Vokrouhlick{\'y}}, D., {et~al.} 2007,
  Science, 316, 274

\bibitem[{{Tholen}(1984)}]{tholen1984}
{Tholen}, D.~J. 1984, PhD thesis, University of Arizona

\bibitem[{{Thomas} {et~al.}(2014){Thomas}, {Emery}, {Trilling}, {Delb{\'o}},
  {Hora}, \& {Mueller}}]{thomas2014}
{Thomas}, C.~A., {Emery}, J.~P., {Trilling}, D.~E., {et~al.} 2014, \icarus,
  228, 217

\bibitem[{{Tian} {et~al.}(2022){Tian}, {Zhao}, \& {Li}}]{tian2022}
{Tian}, J., {Zhao}, H.-B., \& {Li}, B. 2022, Research in Astronomy and
  Astrophysics, 22, 125004

\bibitem[{{Trilling} {et~al.}(2010){Trilling}, {Mueller}, {Hora}, {Harris},
  {Bhattacharya}, {Bottke}, {Chesley}, {Delbo}, {Emery}, {Fazio}, {Mainzer},
  {Penprase}, {Smith}, {Spahr}, {Stansberry}, \& {Thomas}}]{tri2010}
{Trilling}, D.~E., {Mueller}, M., {Hora}, J.~L., {et~al.} 2010, \aj, 140, 770

\bibitem[{{Usui} {et~al.}(2011){Usui}, {Kuroda}, {M{\"u}ller}, {Hasegawa},
  {Ishiguro}, {Ootsubo}, {Ishihara}, {Kataza}, {Takita}, {Oyabu}, {Ueno},
  {Matsuhara}, \& {Onaka}}]{usui2011}
{Usui}, F., {Kuroda}, D., {M{\"u}ller}, T.~G., {et~al.} 2011, \pasj, 63, 1117

\bibitem[{{Vokrouhlick\'y} {et~al.}(2015){Vokrouhlick\'y}, {Bottke}, {Chesley},
  {Scheeres}, \& {Statler}}]{vetal2015}
{Vokrouhlick\'y}, D., {Bottke}, W.~F., {Chesley}, S.~R., {Scheeres}, D.~J., \&
  {Statler}, T.~S. 2015, in Asteroids IV, ed. P.~{Michel}, F.~E. {DeMeo}, \&
  W.~F. {Bottke}, 509--531

\bibitem[{{Vokrouhlick{\'y}} {et~al.}(2005{\natexlab{a}}){Vokrouhlick{\'y}},
  {Bottke}, \& {Nesvorn{\'y}}}]{vok2005Eros}
{Vokrouhlick{\'y}}, D., {Bottke}, W.~F., \& {Nesvorn{\'y}}, D.
  2005{\natexlab{a}}, \icarus, 175, 419

\bibitem[{{Vokrouhlick{\'y}} {et~al.}(2006){Vokrouhlick{\'y}}, {Nesvorn{\'y}},
  \& {Bottke}}]{vok2006}
{Vokrouhlick{\'y}}, D., {Nesvorn{\'y}}, D., \& {Bottke}, W.~F. 2006, \icarus,
  184, 1

\bibitem[{{Vokrouhlick{\'y}} {et~al.}(2011){Vokrouhlick{\'y}}, {{\v D}urech},
  {Polishook}, {Krugly}, {Gaftonyuk}, {Burkhonov}, {Ehgamberdiev}, {Karimov},
  {Molotov}, {Pravec}, {Hornoch}, {Ku{\v s}nir{\'a}k}, {Oey}, {Gal{\'a}d}, \&
  {{\v Z}i{\v z}ka}}]{Vok.ea:11}
{Vokrouhlick{\'y}}, D., {{\v D}urech}, J., {Polishook}, D., {et~al.} 2011, \aj,
  142, 159

\bibitem[{{Vokrouhlick{\'y}} {et~al.}(2005{\natexlab{b}}){Vokrouhlick{\'y}},
  {{\v{C}}apek}, {Chesley}, \& {Ostro}}]{vok2005}
{Vokrouhlick{\'y}}, D., {{\v{C}}apek}, D., {Chesley}, S.~R., \& {Ostro}, S.~J.
  2005{\natexlab{b}}, \icarus, 173, 166

\bibitem[{{Vokrouhlick{\'y}} {et~al.}(2004){Vokrouhlick{\'y}}, {{\v{C}}apek},
  {Kaasalainen}, \& {Ostro}}]{ito2004}
{Vokrouhlick{\'y}}, D., {{\v{C}}apek}, D., {Kaasalainen}, M., \& {Ostro}, S.~J.
  2004, \aap, 414, L21

\bibitem[{{Warner}(2014)}]{War:14k}
{Warner}, B.~D. 2014, Minor Planet Bulletin, 41, 213

\bibitem[{{Warner}(2015)}]{War:15r}
{Warner}, B.~D. 2015, Minor Planet Bulletin, 42, 256

\bibitem[{{Warner}(2016)}]{War:16e}
{Warner}, B.~D. 2016, Minor Planet Bulletin, 43, 143

\bibitem[{{Warner}(2018)}]{War:18m}
{Warner}, B.~D. 2018, Minor Planet Bulletin, 45, 366

\bibitem[{{Warner} {et~al.}(2009){Warner}, {Harris}, \& {Pravec}}]{War.ea:09}
{Warner}, B.~D., {Harris}, A.~W., \& {Pravec}, P. 2009, Icarus, 202, 134

\bibitem[{{Warner} \& {Stephens}(2019)}]{War.Ste:19p}
{Warner}, B.~D. \& {Stephens}, R.~D. 2019, Minor Planet Bulletin, 46, 423

\bibitem[{{Warner} \& {Stephens}(2020{\natexlab{a}})}]{War.Ste:20a}
{Warner}, B.~D. \& {Stephens}, R.~D. 2020{\natexlab{a}}, Minor Planet Bulletin,
  47, 23

\bibitem[{{Warner} \& {Stephens}(2020{\natexlab{b}})}]{War.Ste:20g}
{Warner}, B.~D. \& {Stephens}, R.~D. 2020{\natexlab{b}}, Minor Planet Bulletin,
  47, 290

\bibitem[{{Warner} \& {Stephens}(2022)}]{War.Ste:22c}
{Warner}, B.~D. \& {Stephens}, R.~D. 2022, Minor Planet Bulletin, 49, 83

\bibitem[{{Warner} \& {Stephens}(2023)}]{War.Ste:23a}
{Warner}, B.~D. \& {Stephens}, R.~D. 2023, Minor Planet Bulletin, 50, 21

\bibitem[{{Wiegert}(2015)}]{wieg2015}
{Wiegert}, P.~A. 2015, \icarus, 252, 22

\bibitem[{{Wolf} \& {Reinmuth}(1932)}]{wr1932}
{Wolf}, M. \& {Reinmuth}, K. 1932, Astronomische Nachrichten, 245, 401

\bibitem[{{Yeomans}(1991)}]{yeo1991}
{Yeomans}, D.~K. 1991, \aj, 101, 1920

\bibitem[{{Zegmott} {et~al.}(2021){Zegmott}, {Lowry}, {Ro{\.z}ek}, {Rozitis},
  {Nolan}, {Howell}, {Green}, {Snodgrass}, {Fitzsimmons}, \&
  {Weissman}}]{zeg2021}
{Zegmott}, T.~J., {Lowry}, S.~C., {Ro{\.z}ek}, A., {et~al.} 2021, \mnras, 507,
  4914

\bibitem[{{Ziolkowski}(1983)}]{ziol1983}
{Ziolkowski}, K. 1983, in Asteroids, Comets, and Meteors, 171--174

\end{thebibliography}

  \newcommand{\SortNoop}[1]{}

  \onecolumn
  
  \begin{appendix}
  
  \section{New photometric observations}

    The aspect tables below list the asteroid distance from the Sun $r$ and from the Earth $\Delta$, the solar phase angle $\alpha$, the geocentric ecliptic coordinates of the asteroid $(\lambda, \beta)$, and the observatory at which the data were taken (see Table~\ref{tab:telescopes}) or a reference to the original publication.
    
    \begin{table*}[h]
      \caption{\label{tab:telescopes} 
      Observatories and telescopes.}
      \centering
      \begin{tabular}{llrr}
        \hline \hline
        Abbreviation    & Telescope/Observatory & Telescope aperture [cm]      & MPC code        \\
        \hline
        D65         & Ond\v{r}ejov Observatory, Czech Republic              & 65    & 557   \\
        DK154       & Danish Telescope, ESO, La Silla, Chile                & 154   & W74   \\
        Rozhen      & Rozhen Observatory, Bulgaria                          & 200   & 071   \\
        Simeiz      & Simeiz Observatory, Crimea, Ukraine                   & 100   & 094   \\
        Abastumani  & Abastumani Observatory, Georgia                       & 70    & 119   \\
        Maidanak    & Maidanak Observatory, Uzbekistan                      & 150   & 188   \\
        Kharkiv     & Chuguyiv Observatory, Kharkiv, Ukraine                & 70    & 121   \\
        Modra       & Modra Observatory, Slovakia                           & 60    & 118   \\
        PROMPT      & Cerro Tololo Inter-American Observatory, Chile        & 41    &       \\
        SM          & Sugarloaf Mountain Observatory, South Deerfield, MA, USA  & 50  &       \\
        Wi          & Wise Observatory, Israel                              & 71    & 097   \\
        BMO         & Blue Mountains Observatory, Australia                 & 35    & Q68   \\
        TS          & TRAPPIST South, ESO, La Silla, Chile                  & 60    & I40   \\
        \hline
      \end{tabular}
      \tablefoot{The table lists telescopes used in our work.}
    \end{table*}

    \begin{table*}[h]
      \caption{\label{tab:aspect_1862} 
      Aspect data for the new observations of (1862) Apollo.}
      \centering
      \begin{tabular}{cccrrrl}
        \hline \hline
        Date    & $r$   & $\Delta$      & $\alpha\phantom{g}$   & \multicolumn{1}{c}{$\lambda$} & \multicolumn{1}{c}{$\beta$}     & Observatory or        \\
        & [au]  & [au]          & [deg]                 & \multicolumn{1}{c}{[deg]}         & \multicolumn{1}{c}{[deg]}     &       Reference \\
        \hline
        2014 03 20.8  & 1.784    & 0.793  &  4.7     & 182.1     & $ 8.2$ & D65           \\
        2014 03 25.2  & 1.757    & 0.766  &  5.3     & 179.6     & $ 8.1$ & (1)           \\
        2014 03 26.2  & 1.751    & 0.760  &  5.8     & 179.0     & $ 8.0$ & (1)           \\
        2014 03 27.3  & 1.744    & 0.754  &  6.4     & 178.3     & $ 8.0$ & (1)           \\
        2014 03 28.9  & 1.734    & 0.747  &  7.6     & 177.4     & $ 7.9$ & D65           \\
        2014 03 30.1  & 1.726    & 0.742  &  8.4     & 176.7     & $ 7.8$ & PROMPT        \\
        2014 04 03.1  & 1.700    & 0.727  & 11.7     & 174.2     & $ 7.6$ & SM           \\
        2014 12 28.7  & 1.679    & 0.707  &  7.8     &  97.1     & $13.3$ & D65           \\
        2014 12 29.7  & 1.686    & 0.714  &  7.7     &  96.5     & $13.3$ & D65           \\
        2017 01 21.1  & 2.166    & 1.196  &  5.8     & 126.8     & $11.6$ & DK154           \\
        2017 01 22.3  & 2.170    & 1.198  &  5.5     & 126.3     & $11.6$ & DK154           \\
        2019 02 09.2  & 2.292    & 1.318  &  5.2     & 146.0     & $10.6$ & DK154           \\
        2019 02 11.3  & 2.291    & 1.315  &  4.7     & 145.1     & $10.6$ & DK154           \\
        2019 02 13.3  & 2.290    & 1.313  &  4.5     & 144.3     & $10.6$ & DK154           \\
        2021 12 11.1  & 1.151    & 0.326  & 52.0     &  12.3     & $10.9$ & (2)           \\
        2021 12 12.2  & 1.161    & 0.336  & 51.0     &  14.1     & $10.9$ & (2)           \\
        2023 02 17.3  & 1.787    & 1.098  & 29.3     & 210.1     & $ 5.9$ & DK154           \\
        2023 02 23.3  & 1.750    & 1.003  & 28.3     & 211.0     & $ 6.0$ & DK154           \\
        2023 02 25.3  & 1.738    & 0.972  & 27.9     & 211.3     & $ 6.1$ & DK154           \\
        \hline
      \end{tabular}
      \tablebib{(1) \cite{War:14k}; (2) \cite{War.Ste:22c}; PROMPT observation by Joe Pollock}
    \end{table*}

    \begin{longtable}{cccrrrl}
      \caption{\label{tab:aspect_2100} 
      Aspect data for the new observations of (2100) Ra-Shalom.}\\
        \hline \hline
        Date    & $r$   & $\Delta$      & $\alpha\phantom{g}$   & \multicolumn{1}{c}{$\lambda$} & \multicolumn{1}{c}{$\beta$}     & Observatory or        \\
        & [au]  & [au]          & [deg]                 & \multicolumn{1}{c}{[deg]}         & \multicolumn{1}{c}{[deg]}     &       Reference \\
        \hline
        \endfirsthead
       \caption{continued.}\\
       \hline \hline
        Date	& $r$	& $\Delta$	& $\alpha\phantom{g}$	& \multicolumn{1}{c}{$\lambda$}	& \multicolumn{1}{c}{$\beta$}	& Observatory or	\\
        & [au] 	& [au] 		& [deg] 		& \multicolumn{1}{c}{[deg]} 	& \multicolumn{1}{c}{[deg]} 	&	              Reference \\
      \hline
      \endhead
      \hline
     \endfoot
      \hline
        \endlastfoot
        2019 07 25.0  & 1.122    & 0.449  & 64.9     &  29.9     & $18.3$ & Wi \\
        2019 07 26.0  & 1.125    & 0.444  & 64.4     &  30.0     & $18.2$ & Wi \\
        2019 07 26.9  & 1.129    & 0.440  & 64.0     &  30.1     & $18.1$ & Wi \\
        2019 07 27.0  & 1.129    & 0.440  & 63.9     &  30.1     & $18.1$ & D65 \\
        2019 07 32.0  & 1.145    & 0.414  & 61.5     &  30.4     & $17.3$ & Wi \\
        2019 08 04.0  & 1.154    & 0.398  & 59.9     &  30.4     & $16.7$ & Wi \\
        2019 08 05.0  & 1.157    & 0.392  & 59.4     &  30.4     & $16.5$ & D65 \\
        2019 08 08.0  & 1.164    & 0.376  & 57.7     &  30.2     & $15.9$ & Wi \\
        2019 08 09.0  & 1.166    & 0.370  & 57.1     &  30.2     & $15.7$ & Wi \\
        2019 08 09.0  & 1.167    & 0.370  & 57.0     &  30.2     & $15.7$ & D65 \\
        2019 08 16.4  & 1.181    & 0.327  & 52.0     &  29.0     & $13.8$ &     (1) \\
        2019 08 17.4  & 1.182    & 0.321  & 51.2     &  28.7     & $13.5$ &     (1) \\
        2019 08 19.4  & 1.185    & 0.310  & 49.5     &  28.1     & $12.8$ &     (1) \\
        2019 08 20.4  & 1.187    & 0.304  & 48.6     &  27.8     & $12.5$ &     (1) \\
        2019 08 21.4  & 1.188    & 0.298  & 47.8     &  27.4     & $12.1$ &     (1) \\
        2019 08 22.4  & 1.189    & 0.292  & 46.8     &  27.0     & $11.7$ &     (1) \\
        2019 08 23.4  & 1.190    & 0.287  & 45.8     &  26.6     & $11.3$ &     (1) \\
        2019 08 23.9  & 1.190    & 0.284  & 45.3     &  26.3     & $11.1$ & Wi \\
        2019 08 24.1  & 1.191    & 0.283  & 45.2     &  26.2     & $11.1$ & D65 \\
        2019 08 25.0  & 1.191    & 0.278  & 44.2     &  25.8     & $10.7$ & Wi \\
        2019 08 26.0  & 1.192    & 0.272  & 43.1     &  25.2     & $10.2$ & D65 \\
        2019 09 11.6  & 1.192    & 0.196  & 17.4     &   9.0     & $-1.4$ &        BMO \\
        2019 09 22.4  & 1.178    & 0.180  & 12.8     & 350.5     & $-12.7$ &        BMO \\
        2019 09 23.5  & 1.176    & 0.180  & 15.2     & 348.5     & $-13.8$ &        BMO \\
        2019 09 24.5  & 1.175    & 0.181  & 17.2     & 346.8     & $-14.7$ &        BMO \\
        2019 09 25.5  & 1.173    & 0.182  & 19.5     & 344.9     & $-15.7$ &        BMO \\
        2019 09 26.4  & 1.171    & 0.183  & 21.6     & 343.1     & $-16.5$ &        BMO \\
        2019 09 28.6  & 1.166    & 0.187  & 26.5     & 339.2     & $-18.4$ &        BMO \\
        2022 07 18.0  & 1.030    & 0.329  & 78.4     &  33.4     & $34.3$ & D65 \\
        2022 07 19.0  & 1.036    & 0.325  & 77.5     &  32.9     & $34.3$ & D65 \\
        2022 07 20.0  & 1.041    & 0.322  & 76.7     &  32.5     & $34.3$ & D65 \\
        2022 07 22.0  & 1.052    & 0.315  & 74.9     &  31.6     & $34.3$ & D65 \\
        2022 07 24.1  & 1.062    & 0.307  & 73.0     &  30.7     & $34.3$ & D65 \\
        2022 07 27.0  & 1.076    & 0.296  & 70.3     &  29.1     & $34.3$ & D65 \\
        2022 07 28.0  & 1.081    & 0.293  & 69.4     &  28.6     & $34.3$ & D65 \\
        2022 08 03.0  & 1.107    & 0.269  & 63.3     &  24.5     & $34.2$ & D65 \\
        2022 08 04.0  & 1.111    & 0.265  & 62.1     &  23.6     & $34.1$ & D65 \\
        2022 08 04.9  & 1.115    & 0.261  & 61.1     &  22.9     & $34.1$ & D65 \\
        2022 08 08.0  & 1.126    & 0.249  & 57.4     &  20.0     & $33.8$ & D65 \\
        2022 08 16.1  & 1.152    & 0.218  & 46.0     &  10.1     & $32.3$ & D65 \\
        2022 08 16.4  & 1.153    & 0.217  & 45.6     &   9.7     & $32.2$ &     (2) \\
        2022 08 16.9  & 1.154    & 0.215  & 44.7     &   8.8     & $32.0$ & D65 \\
        2022 08 17.4  & 1.156    & 0.214  & 43.9     &   8.1     & $31.8$ &     (2) \\
        2022 08 17.9  & 1.157    & 0.212  & 43.1     &   7.4     & $31.6$ & D65 \\
        2022 08 18.4  & 1.158    & 0.211  & 42.3     &   6.6     & $31.5$ &     (2) \\
        2022 08 19.0  & 1.160    & 0.209  & 41.1     &   5.6     & $31.2$ & D65 \\
        2022 08 19.3  & 1.160    & 0.208  & 40.8     &   5.2     & $31.1$ &     (2) \\
        2022 08 20.4  & 1.163    & 0.204  & 38.7     &   3.3     & $30.6$ &     (2) \\
        2022 08 21.3  & 1.165    & 0.202  & 37.2     &   1.8     & $30.2$ &     (2) \\
        2022 08 22.3  & 1.168    & 0.199  & 35.3     & 360.0     & $29.6$ &     (2) \\
        2022 08 23.3  & 1.170    & 0.197  & 33.5     & 358.2     & $29.0$ &     (2) \\
        2022 08 25.2  & 1.174    & 0.193  & 30.0     & 354.8     & $27.7$ &     (2) \\
        2022 08 26.0  & 1.175    & 0.192  & 28.6     & 353.3     & $27.2$ & D65 \\
        2022 08 29.0  & 1.181    & 0.188  & 23.2     & 347.7     & $24.7$ & D65 \\
        2022 08 29.9  & 1.182    & 0.188  & 21.6     & 345.8     & $23.8$ & D65 \\
        2022 09 06.9  & 1.192    & 0.193  & 16.5     & 331.6     & $15.3$ & D65 \\
    \end{longtable}
    \tablebib{(1)~\cite{War.Ste:20a}; (2)~\cite{War.Ste:23a}}

\newpage

   \begin{table*}
      \caption{\label{tab:aspect_eVscope} 
      Aspect data for Unistellar observations of (2100) Ra-Shalom.}
      \centering
      \begin{tabular}{cccrrrl}
        \hline \hline
        Date    & $r$   & $\Delta$      & $\alpha\phantom{g}$   & \multicolumn{1}{c}{$\lambda$} & \multicolumn{1}{c}{$\beta$}     & Observer        \\
        & [au]  & [au]          & [deg]                 & \multicolumn{1}{c}{[deg]}         & \multicolumn{1}{c}{[deg]}     &        \\
        \hline
        2022 08 15.0  & 1.149    & 0.222  & 47.7     &  11.6     & $32.6$ & M. Billiani \\
        2022 08 18.7  & 1.159    & 0.210  & 41.7     &   6.1     & $31.3$ &    K. Fukui \\
        2022 08 20.0  & 1.162    & 0.206  & 39.5     &   4.0     & $30.8$ &  B. Guillet \\
        2022 08 20.9  & 1.164    & 0.203  & 37.8     &   2.4     & $30.3$ & P. Tikkanen \\
        2022 08 21.0  & 1.165    & 0.203  & 37.8     &   2.4     & $30.3$ & P. Kuossari \\
        2022 08 24.9  & 1.173    & 0.194  & 30.6     & 355.4     & $28.0$ & anonymous \\
        2022 08 25.9  & 1.175    & 0.192  & 28.7     & 353.5     & $27.2$ & P. Kuossari \\
        2022 08 25.9  & 1.175    & 0.192  & 28.6     & 353.4     & $27.2$ &    S. Price \\
        2022 08 25.9  & 1.175    & 0.192  & 28.7     & 353.5     & $27.2$ & anonymous \\
        2022 08 27.0  & 1.177    & 0.191  & 26.7     & 351.4     & $26.4$ &  B. Guillet \\
        2022 08 27.9  & 1.179    & 0.190  & 25.1     & 349.7     & $25.6$ & P. Kuossari \\
        2022 08 27.2  & 1.178    & 0.190  & 26.3     & 350.9     & $26.2$ &   S. Kardel \\
        2022 08 28.5  & 1.180    & 0.189  & 24.0     & 348.5     & $25.1$ &  M. Shimizu \\
        2022 08 28.8  & 1.180    & 0.189  & 23.4     & 347.9     & $24.8$ &  O. Clerget \\
        2022 08 29.0  & 1.181    & 0.188  & 23.1     & 347.6     & $24.6$ & M. Lauvernier \\
        2022 08 28.7  & 1.180    & 0.189  & 23.7     & 348.2     & $24.9$ &    K. Fukui \\
        2022 08 29.2  & 1.181    & 0.188  & 22.8     & 347.2     & $24.5$ &    M. Loose \\
        2022 08 29.6  & 1.182    & 0.188  & 22.1     & 346.4     & $24.1$ &     K. Oura \\
        2022 08 31.7  & 1.185    & 0.187  & 18.9     & 342.4     & $22.0$ & I. Chairman \\
        2022 09 01.7  & 1.186    & 0.187  & 17.8     & 340.6     & $21.0$ & I. Chairman \\
        2022 09 01.2  & 1.185    & 0.187  & 18.4     & 341.6     & $21.6$ &   E. Hickok \\
        2022 09 01.8  & 1.186    & 0.188  & 17.7     & 340.4     & $20.9$ &  O. Clerget \\
        2022 09 01.1  & 1.185    & 0.187  & 18.5     & 341.7     & $21.6$ &     S. Will \\
        2022 09 01.1  & 1.185    & 0.187  & 18.5     & 341.7     & $21.6$ &  J. Randolph \\
        2022 09 01.9  & 1.186    & 0.188  & 17.6     & 340.3     & $20.8$ &   D. Martin \\
        2022 09 02.9  & 1.187    & 0.188  & 16.8     & 338.5     & $19.8$ & P. Kuossari \\
        2022 09 03.0  & 1.188    & 0.188  & 16.6     & 338.2     & $19.6$ &  A. Schmidt \\
        2022 09 02.1  & 1.187    & 0.188  & 17.4     & 339.9     & $20.6$ &   G. Simard \\
        2022 09 03.8  & 1.189    & 0.189  & 16.2     & 336.8     & $18.7$ & P. Kuossari \\
        2022 09 03.9  & 1.189    & 0.189  & 16.2     & 336.7     & $18.7$ & A. Katterfeld \\
        2022 09 03.9  & 1.189    & 0.189  & 16.2     & 336.7     & $18.7$ &  O. Clerget \\
        2022 09 03.6  & 1.188    & 0.189  & 16.3     & 337.2     & $19.0$ &      W. Yue \\
        2022 09 08.9  & 1.193    & 0.197  & 18.1     & 328.6     & $13.1$ & P. Tikkanen \\
        2022 09 10.9  & 1.194    & 0.203  & 20.5     & 325.7     & $10.9$ &   Y. Lorand \\
        2022 09 11.8  & 1.195    & 0.206  & 21.8     & 324.4     & $ 9.9$ &  O. Clerget \\
        2022 09 11.9  & 1.195    & 0.206  & 21.8     & 324.4     & $ 9.9$ &   D. Martin \\
        \hline
      \end{tabular}
    \end{table*}

\begin{longtable}{cccrrrl}
    \caption{\label{tab:aspect_85989}	
        Aspect data for observations of (85989) 1999~JD6.} \\
        \hline \hline 
        Date	& $r$	& $\Delta$	& $\alpha\phantom{g}$	& \multicolumn{1}{c}{$\lambda$}	& \multicolumn{1}{c}{$\beta$}	& Observatory or	\\
        & [au] 	& [au] 		& [deg] 		& \multicolumn{1}{c}{[deg]} 	& \multicolumn{1}{c}{[deg]} 	&	              Reference \\
        \hline
    \endfirsthead
    \caption{continued.}\\
     \hline \hline
        Date	& $r$	& $\Delta$	& $\alpha\phantom{g}$	& \multicolumn{1}{c}{$\lambda$}	& \multicolumn{1}{c}{$\beta$}	& Observatory or 	\\
        & [au] 	& [au] 		& [deg] 		& \multicolumn{1}{c}{[deg]} 	& \multicolumn{1}{c}{[deg]} 	&	              reference   \\
    \hline
    \endhead
    \hline
        \endfoot
        \hline
        \endlastfoot
        1999 05 22.0  & 1.289    & 0.426  & 41.8     & 231.5     & $57.7$ & D65 \\
        1999 05 24.0  & 1.299    & 0.434  & 41.0     & 228.5     & $56.2$ & D65 \\
        1999 05 24.9  & 1.304    & 0.439  & 40.7     & 227.2     & $55.5$ & D65 \\
        1999 06 13.0  & 1.384    & 0.557  & 39.2     & 213.4     & $40.6$ & D65 \\
        2000 07 06.4  & 1.321    & 0.370  & 29.8     & 261.4     & $33.9$ & (7) \\
        2000 07 07.4  & 1.316    & 0.368  & 30.6     & 259.9     & $33.5$ & (7) \\
        2000 07 09.4  & 1.306    & 0.366  & 32.4     & 257.0     & $32.6$ & (7) \\
        2000 07 10.3  & 1.301    & 0.364  & 33.3     & 255.5     & $32.1$ & (7) \\
        2004 05 15.0  & 1.303    & 0.448  & 41.3     & 261.9     & $53.7$ &  (1) \\
        2004 05 16.0  & 1.308    & 0.449  & 40.7     & 260.2     & $53.7$ &  (1) \\
        2004 05 17.0  & 1.313    & 0.450  & 40.1     & 258.4     & $53.7$ &  (1) \\
        2004 05 23.0  & 1.341    & 0.459  & 36.9     & 248.1     & $52.5$ &  (1) \\
        2004 05 27.9  & 1.361    & 0.472  & 35.2     & 240.6     & $50.5$ &  (1) \\
        2004 06 05.4  & 1.391    & 0.505  & 34.4     & 230.4     & $45.8$ & (8) \\
        2004 06 06.5  & 1.394    & 0.510  & 34.4     & 229.4     & $45.1$ & (8) \\
        2004 06 09.5  & 1.402    & 0.526  & 34.7     & 226.8     & $43.2$ & (8) \\
        2004 06 11.4  & 1.407    & 0.537  & 35.0     & 225.4     & $42.0$ & (8) \\
        2004 06 12.4  & 1.410    & 0.543  & 35.2     & 224.8     & $41.3$ & (8) \\
        2005 07 08.9  & 1.261    & 0.293  & 29.7     & 272.1     & $35.4$ &  (1) \\
        2005 07 10.0  & 1.255    & 0.289  & 30.6     & 270.1     & $35.2$ &  (1) \\
        2005 07 14.9  & 1.223    & 0.272  & 36.2     & 259.9     & $33.3$ & Modra \\
        2005 07 15.9  & 1.217    & 0.269  & 37.5     & 257.9     & $32.7$ & Modra \\
        2014 05 20.4  & 1.394    & 0.533  & 36.0     & 275.9     & $42.7$ &     (2) \\
        2014 05 21.4  & 1.397    & 0.530  & 35.4     & 274.9     & $42.9$ &     (2) \\
        2014 05 22.3  & 1.399    & 0.528  & 34.9     & 273.9     & $43.1$ &     (2) \\
        2015 06 07.4  & 1.336    & 0.575  & 45.1     & 323.1     & $22.0$ &     (3) \\ %offset
        2015 06 08.4  & 1.332    & 0.563  & 45.0     & 323.4     & $22.2$ &     (3) \\
        2015 06 11.4  & 1.317    & 0.526  & 44.9     & 324.2     & $22.8$ &     (3) \\
        2015 06 12.4  & 1.312    & 0.514  & 44.9     & 324.5     & $23.0$ &     (3) \\
        2015 06 14.4  & 1.302    & 0.490  & 44.8     & 325.1     & $23.4$ &     (3) \\
        2015 06 15.4  & 1.297    & 0.477  & 44.8     & 325.4     & $23.6$ &     (3) \\
        2018 06 01.2  & 1.181    & 0.518  & 58.8     & 167.6     & $41.2$ &     (4) \\
        2018 06 02.3  & 1.188    & 0.527  & 58.2     & 168.9     & $40.6$ &     (4) \\
        2018 06 03.3  & 1.195    & 0.536  & 57.6     & 170.1     & $39.9$ &     (4) \\
        2018 06 04.3  & 1.202    & 0.546  & 57.0     & 171.3     & $39.3$ &     (4) \\
        2019 06 03.4  & 1.437    & 0.522  & 29.3     & 269.2     & $41.1$ &     (5) \\
        2019 06 04.4  & 1.438    & 0.520  & 28.9     & 268.0     & $41.1$ &     (5) \\
        2019 06 05.4  & 1.439    & 0.518  & 28.5     & 266.8     & $41.1$ &     (5) \\
        2019 06 06.4  & 1.439    & 0.516  & 28.2     & 265.6     & $41.1$ &     (5) \\
        2019 06 07.4  & 1.440    & 0.514  & 28.0     & 264.4     & $41.0$ &     (5) \\
        2020 06 18.4  & 1.231    & 0.445  & 51.6     & 337.8     & $20.0$ &     (6) \\
        2020 06 19.4  & 1.224    & 0.433  & 51.9     & 338.6     & $20.1$ &     (6) \\
        2020 06 20.4  & 1.217    & 0.420  & 52.3     & 339.3     & $20.2$ &     (6) \\
        2020 06 21.4  & 1.211    & 0.408  & 52.6     & 340.1     & $20.3$ &     (6) \\
        2020 06 22.4  & 1.204    & 0.395  & 53.0     & 341.0     & $20.5$ &     (6) \\
        2020 06 23.4  & 1.197    & 0.383  & 53.5     & 341.8     & $20.6$ &     (6) \\
        2023 05 10.0  & 1.057    & 0.335  & 72.7     & 135.7     & $67.6$ & D65 \\
        2023 05 15.9  & 1.110    & 0.368  & 65.0     & 156.7     & $61.9$ & D65 \\
        2023 06 25.0  & 1.356    & 0.729  & 47.5     & 195.2     & $30.0$ & DK154 
    \end{longtable}
\tablebib{(1) \cite{Pol.Bro:08}; (2) \cite{War:14k}; (3) \cite{War:15r}; (4) \cite{War:18m}; (5) \cite{War.Ste:19p}; (6) \cite{War.Ste:20g}; (7) observed by Brandon Bozek, other information lost; (8) observed by Luke Dundon, other information lost}

  \begin{longtable}{cccrrrl}
    \caption{\label{tab:aspect_138852}	
      Aspect data for observations of (138852) 2000~WN10.} \\
        \hline \hline 
        Date	& $r$	& $\Delta$	& $\alpha\phantom{g}$	& \multicolumn{1}{c}{$\lambda$}	& \multicolumn{1}{c}{$\beta$}	& Observatory	\\
        & [au] 	& [au] 		& [deg] 		& \multicolumn{1}{c}{[deg]} 	& \multicolumn{1}{c}{[deg]} 	&	\\
        \hline
    \endfirsthead
    \caption{continued.}\\
     \hline \hline
        Date	& $r$	& $\Delta$	& $\alpha\phantom{g}$	& \multicolumn{1}{c}{$\lambda$}	& \multicolumn{1}{c}{$\beta$}	& Observatory	\\
        & [au] 	& [au] 		& [deg] 		& \multicolumn{1}{c}{[deg]} 	& \multicolumn{1}{c}{[deg]} 	&	\\
    \hline
    \endhead
    \hline
    \endfoot
      \hline
        \endlastfoot
        2008 11 27.8  & 1.172    & 0.190  & 11.8     &  53.5     & $ 6.9$ &     Rozhen \\
        2009 11 18.0  & 1.130    & 0.146  & 13.4     &  63.6     & $-13.2$ &     Simeiz \\
        2009 11 22.9  & 1.150    & 0.163  &  4.7     &  56.0     & $-2.8$ &     Simeiz \\
        2009 11 26.9  & 1.165    & 0.183  & 11.8     &  51.2     & $ 4.0$ &     Simeiz \\
        2010 11 13.0  & 1.104    & 0.135  & 30.1     &  69.7     & $-28.6$ &     Abastumani \\
        2010 11 25.8  & 1.157    & 0.174  & 11.9     &  49.2     & $ 0.5$ &     Maidanak \\
        2010 11 26.8  & 1.160    & 0.179  & 13.7     &  48.2     & $ 2.1$ &     Maidanak \\
        2010 12 07.7  & 1.199    & 0.254  & 29.5     &  41.1     & $14.4$ &     Maidanak \\
        2011 11 19.9  & 1.129    & 0.145  & 12.3     &  53.3     & $-13.7$ &     D65 \\
        2011 11 20.7  & 1.133    & 0.148  & 11.2     &  52.1     & $-11.6$ &     Abastumani \\
        2011 11 22.8  & 1.141    & 0.157  & 10.9     &  49.4     & $-6.9$ &     Abastumani \\
        2011 11 22.9  & 1.142    & 0.157  & 10.9     &  49.4     & $-6.8$ &     D65 \\
        2011 11 23.9  & 1.146    & 0.162  & 11.7     &  48.2     & $-4.7$ &     Kharkiv \\
        2011 11 24.9  & 1.149    & 0.167  & 12.9     &  47.2     & $-2.8$ &     Kharkiv \\
        2011 11 26.0  & 1.154    & 0.173  & 14.6     &  46.1     & $-0.8$ &     Abastumani \\
        2012 11 19.2  & 1.127    & 0.144  & 14.8     &  49.3     & $-15.0$ &     DK154 \\
        2012 11 22.8  & 1.142    & 0.160  & 14.6     &  45.1     & $-6.7$ &     Abastumani \\
        2012 11 23.8  & 1.145    & 0.165  & 15.5     &  44.2     & $-4.8$ &     Abastumani \\
        2013 10 30.4  & 1.032    & 0.157  & 71.4     & 103.2     & $-64.4$ &     DK154 \\
        2013 10 31.3  & 1.037    & 0.153  & 69.2     &  98.8     & $-63.5$ &     DK154 \\
        2013 11 05.3  & 1.061    & 0.136  & 56.1     &  77.6     & $-56.0$ &     DK154 \\
        2013 11 06.3  & 1.066    & 0.133  & 53.1     &  74.0     & $-53.8$ &     DK154 \\
        2013 11 07.3  & 1.070    & 0.131  & 50.0     &  70.5     & $-51.5$ &     DK154 \\
        2013 11 08.3  & 1.075    & 0.130  & 46.7     &  67.3     & $-48.9$ &     DK154 \\
        2013 11 09.3  & 1.080    & 0.128  & 43.5     &  64.4     & $-46.2$ &     DK154 \\
        2013 11 11.3  & 1.089    & 0.127  & 37.0     &  59.2     & $-40.4$ &     DK154 \\
        2013 11 22.8  & 1.138    & 0.159  & 17.6     &  41.8     & $-8.5$ &     Abastumani \\
        2013 11 23.3  & 1.140    & 0.161  & 17.9     &  41.5     & $-7.5$ &     DK154 \\
        2014 11 17.2  & 1.111    & 0.135  & 24.0     &  43.9     & $-25.1$ & DK154 \\
        2014 12 16.1  & 1.216    & 0.333  & 39.9     &  33.2     & $16.1$ & DK154 \\
        2015 11 14.4  & 1.094    & 0.128  & 33.6     &  42.8     & $-36.8$ & DK154 \\
        2015 11 15.3  & 1.098    & 0.130  & 31.2     &  41.5     & $-33.6$ & DK154 \\
        2015 11 16.3  & 1.102    & 0.132  & 29.1     &  40.3     & $-30.4$ & DK154 \\
        2016 10 23.3  & 0.986    & 0.188  & 87.2     & 153.9     & $-75.3$ & DK154 \\
        2016 12 01.1  & 1.162    & 0.219  & 33.1     &  29.1     & $ 2.5$ & DK154 \\
        2016 12 07.1  & 1.184    & 0.267  & 37.3     &  29.1     & $ 8.8$ & DK154 \\
        2017 11 21.1  & 1.119    & 0.157  & 31.1     &  27.0     & $-17.7$ & DK154 \\
        2017 11 24.1  & 1.131    & 0.173  & 31.3     &  26.6     & $-10.8$ & DK154 \\
        2017 11 25.2  & 1.135    & 0.180  & 31.7     &  26.5     & $-8.5$ & DK154 \\
        2018 11 29.2  & 1.147    & 0.207  & 36.0     &  23.6     & $-2.9$ & DK154 \\
        2019 11 04.3  & 1.032    & 0.145  & 70.0     &  15.0     & $-76.4$ & DK154 \\
        2019 11 05.0  & 1.035    & 0.142  & 68.3     &  15.5     & $-74.3$ & DK154 \\
        2019 11 26.1  & 1.131    & 0.187  & 36.8     &  20.8     & $-9.5$ & DK154 \\
        2019 11 27.1  & 1.135    & 0.194  & 37.0     &  21.0     & $-7.7$ & DK154 \\
        2019 11 28.0  & 1.139    & 0.200  & 37.3     &  21.2     & $-6.0$ & DK154 \\
        2019 12 02.0  & 1.154    & 0.229  & 38.8     &  22.1     & $ 0.0$ & DK154 \\
        2020 11 18.1  & 1.097    & 0.153  & 41.9     &  15.4     & $-27.8$ & DK154 \\
        2020 11 21.2  & 1.111    & 0.167  & 39.7     &  16.6     & $-19.6$ & DK154 \\
        2021 11 28.1  & 1.135    & 0.212  & 41.5     &  16.4     & $-6.8$ & DK154 \\
        2022 11 21.2  & 1.102    & 0.174  & 45.6     &   9.2     & $-22.2$ & DK154 \\
     \end{longtable}

    \begin{table*}[h]
      \caption{\label{tab:aspect_161989} 
      Aspect data for the new observations of (161989) Cacus.}
      \centering
      \begin{tabular}{cccrrrl}
        \hline \hline
        Date    & $r$   & $\Delta$      & $\alpha\phantom{g}$   & \multicolumn{1}{c}{$\lambda$} & \multicolumn{1}{c}{$\beta$}     & Observatory or        \\
        & [au]  & [au]          & [deg]                 & \multicolumn{1}{c}{[deg]}         & \multicolumn{1}{c}{[deg]}     &       Reference \\
        \hline
        2022 01 28.1  & 1.286    & 0.439  & 39.0     & 114.6     & $-54.2$ &         TS \\
        2022 01 29.1  & 1.284    & 0.435  & 39.0     & 113.6     & $-53.7$ &         TS \\
        2022 01 30.1  & 1.282    & 0.432  & 39.1     & 112.6     & $-53.2$ &         TS \\
        2022 02 15.1  & 1.242    & 0.391  & 42.2     & 100.8     & $-40.3$ &         TS \\
        2022 02 16.1  & 1.239    & 0.389  & 42.5     & 100.4     & $-39.4$ &         TS \\
        2022 02 17.1  & 1.236    & 0.388  & 42.9     &  99.9     & $-38.3$ &         TS \\
        2022 08 28.0  & 1.012    & 0.068  & 86.6     &  64.8     & $26.0$ &     (1) \\
        2022 08 30.0  & 1.019    & 0.061  & 79.6     &  59.1     & $13.6$ &     (1) \\
        2022 08 31.0  & 1.023    & 0.058  & 75.5     &  56.0     & $ 6.2$ &     (1) \\
        2022 08 31.1  & 1.023    & 0.058  & 75.2     &  55.7     & $ 5.6$ &     (1) \\
        2022 09 10.1  & 1.058    & 0.093  & 54.5     &  24.2     & $-49.5$ &         TS \\
        2022 09 12.3  & 1.066    & 0.108  & 54.0     &  17.8     & $-54.1$ &         TS \\
        2022 09 13.3  & 1.069    & 0.114  & 53.9     &  15.3     & $-55.5$ &         TS \\
    \hline
      \end{tabular}
      \tablebib{(1) \cite{Pan.Paj:23}}
    \end{table*}
    \end{appendix}
    
\end{document}